\documentclass[pdflatex,sn-nature]{sn-jnl}

\usepackage{enumitem}
\usepackage[utf8]{inputenc}
\usepackage{graphicx}%
\usepackage{multirow}%
\usepackage{amsmath,amssymb,amsfonts}%
\usepackage{amsthm}%
\usepackage{mathrsfs}%
\usepackage[title]{appendix}%
\usepackage{xcolor}%
\usepackage{textcomp}%
\usepackage{manyfoot}%
\usepackage{booktabs}%
\usepackage{algorithm}%
\usepackage{algorithmicx}%
\usepackage{algpseudocode}%
\usepackage{listings}%
\usepackage{float}

\usepackage{mathtools}
\usepackage{hyperref}
\usepackage{braket}
\usepackage{tikz}
\usepackage{graphicx}      
\usepackage{subfig}        

\theoremstyle{thmstyleone}%
\theoremstyle{thmstyletwo}%

\theoremstyle{thmstylethree}%

\begin{document}

\title[Quantum Stochastic Walks for Portfolio Optimization: Theory and Implementation on Financial Networks]{Quantum Stochastic Walks for Portfolio Optimization: Theory and Implementation on Financial Networks}


\author*[1,2]{\fnm{Yen Jui} \sur{Chang}}\email{aceest@cycu.edu.tw}

\author[3]{\fnm{Wei-Ting} \sur{Wang}}\email{d11245002@ntu.edu.tw}

\author[4]{\fnm{Yun-Yuan} \sur{Wang}}\email{yunyuanw@nvidia.com}
\author[5]{\fnm{Chen-Yu} \sur{Liu}}\email{d10245003@g.ntu.edu.tw}
 
\author[6,7]{\fnm{Kuan-Cheng} \sur{Chen}}\email{kuan-cheng.chen17@imperial.ac.uk}
\author[2,3]{\fnm{Ching-Ray} \sur{Chang}}\email{crchang@phys.ntu.edu.tw}

\affil*[1]{\orgdiv{Master Program in Intelligent Computing and Big Data}, \orgname{Chung Yuan Christian University}, \orgaddress{\street{No. 200, Zhongbei Rd., Zhongli Dist}, \city{Taoyuan City}, \postcode{320314},  \country{Taiwan}}}
\affil[2]{\orgdiv{Quantum Information Center}, \orgname{Chung Yuan Christian University}, \orgaddress{\street{No. 200, Zhongbei Rd., Zhongli Dist}, \city{Taoyuan City}, \postcode{320314},  \country{Taiwan}}}

\affil[3]{\orgdiv{Department of Physics}, \orgname{National Taiwan University}, \orgaddress{\street{No. 1, Sec. 4, Roosevelt Rd.}, \city{Taipei}, \postcode{106319}, \country{Taiwan}}}

\affil[4]{\orgdiv{NVIDIA AI Technology Center}, \orgname{NVIDIA Corp.}, \orgaddress{\city{Taipei}, \country{Taiwan}}}
\affil[5]{\orgdiv{Graduate Institute of Applied Physics}, \orgname{National Taiwan University}, \orgaddress{\street{No. 1, Sec. 4, Roosevelt Rd.}, \city{Taipei}, \postcode{106319}, \country{Taiwan}}}
\affil[6]{\orgdiv{Department of Electrical and Electronic Engineering}, \orgname{Imperial College London}, \orgaddress{\city{London}, \country{UK}}}

\affil[7]{\orgdiv{Centre for Quantum Engineering, Science and Technology (QuEST)}, \orgname{Imperial College London}, \orgaddress{\city{London}, \country{UK}}}


\abstract{%
Classical mean–variance optimization is powerful in theory but fragile in practice, often producing highly concentrated, high-turnover portfolios. Naive equal-weight (1/N) portfolios are more robust but largely ignore cross-sectional information. We propose a quantum stochastic walk (QSW) framework that embeds assets in a weighted graph and derives portfolio weights from the stationary distribution of a hybrid quantum–classical walk. The resulting allocations behave as a “smart 1/N” portfolio: structurally close to equal-weight, but with small, data-driven tilts and a controllable level of trading. On recent S\&P 500 universes, QSW portfolios match the diversification and stability of 1/N while delivering higher risk-adjusted returns than both mean–variance and naive benchmarks. A comprehensive hyper-parameter grid search shows that this behavior is structural rather than the result of fine-tuning and yields simple design rules for practitioners. A 34-year, multi-universe robustness study with rolling re-optimization further demonstrates that the QSW optimizer preserves these advantages across market regimes. Overall, the QSW framework improves risk-adjusted performance while maintaining strong diversification and moderate turnover.}

\maketitle



\maketitle

\section{Introduction}

Modern investment management operates in an arena of high-dimensional, tightly coupled, and rapidly evolving data. Conventional quantitative techniques—most notably the seminal mean–variance framework of Markowitz~\cite{markowitz1952portfolio}—encapsulate the trade-off between expected return and risk through a quadratic objective that minimizes portfolio variance for a target mean return. While mathematically elegant, this quadratic form becomes fragile when confronted with today’s realities: estimation error in large covariance matrices, non-Gaussian return distributions, time-varying correlations, and a proliferation of real-world constraints such as turnover limits, liquidity frictions, and regulatory capital charges~\cite{fabozzi2007robust,black1992global,mantegna1999introduction,jegadeesh1993returns,fletcher2013practical}.

The classical foundations of portfolio optimization are well established. The work of Markowitz~\cite{markowitz1952portfolio} introduced the mean–variance framework, which remains the cornerstone of modern portfolio theory. In this setting, expected returns $\boldsymbol{\mu}$ and the covariance matrix $\boldsymbol{\Sigma}$ jointly determine the efficient frontier through problems such as minimum-variance optimization and maximum-Sharpe optimization~\cite{tobin1958liquidity,sharpe1964capital,merton1972analytic}. Over the decades, numerous refinements have been proposed: the Black–Litterman model blends equilibrium returns with subjective views in a Bayesian framework~\cite{black1992global}; robust optimization formulations introduce uncertainty sets for $\boldsymbol{\mu}$ and $\boldsymbol{\Sigma}$~\cite{fabozzi2007robust,goldfarb2003robust,garlappi2007portfolio}; and alternative risk measures such as coherent risk and CVaR have been explored to better capture tail risks~\cite{artzner1999coherent,rockafellar2000optimization}.

Despite seven decades of refinement, however, mean–variance optimization faces severe challenges in practice. Estimation error and dimensionality are the first obstacles: the covariance matrix for $n$ assets contains $\tfrac{1}{2}n(n+1)$ parameters, and for $n=500$ this already exceeds $10^{5}$ parameters, with condition numbers often above $10^{6}$. With finite samples, small errors in $\boldsymbol{\mu}$ and $\boldsymbol{\Sigma}$ can produce extreme and unstable portfolios~\cite{mantegna1999introduction,demiguel2009optimal}. Markets are also non-stationary: correlations spike toward unity during crises~\cite{longin2001extreme}, precisely when diversification is most needed, and correlation structures exhibit asymmetric behavior between up and down markets~\cite{ang2002asymmetric}. Empirical returns display heavy tails, skewness, and volatility clustering~\cite{cont2001empirical}, making variance alone an incomplete proxy for risk. Real-world implementation adds further complexity: cardinality constraints, transaction costs, market impact, and liquidity constraints turn the convex quadratic program into a mixed-integer nonlinear program that is NP-hard~\cite{chang2000heuristics,fletcher2013practical}.
\label{sec:classical:limits} 

Perhaps most critically, small perturbations in expected returns can cause dramatic rebalancing; errors in means dominate those in covariances by roughly an order of magnitude~\cite{best1991sensitivity,chopra1993effect}, leading to concentrated, counterintuitive portfolios that often underperform simple equal-weight (1/N) strategies out of sample~\cite{demiguel2009optimal}.

To mitigate these issues, a growing body of work has turned to neural networks (NNs). One line of research follows a predict-then-optimize paradigm: deep learning models are trained to forecast inputs such as expected returns or risk premia more accurately than classical econometric models~\cite{gu2020empirical}, with the forecasts then fed into a conventional optimizer. A second, more recent line uses end-to-end deep reinforcement learning (DRL) to learn portfolio policies directly: agents map market states to portfolio weights to maximize cumulative reward functions such as the Sharpe ratio or risk-budgeting objectives~\cite{jiang2017deep,zhang2021deep,han2023risk}. These approaches can, in principle, bypass some of the sensitivity of mean–variance optimization to input estimates. However, NN-based approaches introduce their own challenges. The resulting models are typically highly non-linear and opaque, making it difficult to understand or audit how they respond to changes in market conditions~\cite{arrieta2020explainable}. This lack of interpretability is a serious concern for risk management, regulatory review, and practitioner trust. Moreover, deep models are data-hungry and prone to overfitting noisy, non-stationary financial time series~\cite{gu2020empirical}, and DRL agents can suffer from training instability and sensitivity to reward specification. In practice, these methods can behave as powerful but opaque “black boxes”, which limits their adoption in settings where transparency and stability are as important as raw performance.

Parallel to these developments, the econophysics literature has highlighted a complementary perspective based on financial networks. Early work showed that a covariance matrix can be filtered into sparse structures such as minimum-spanning trees (MST) or planar maximally filtered graphs (PMFG), whose topology reveals hidden economic sectors and contagion channels~\cite{mantegna1999hierarchical,onnela2003dynamic,TUMMINELLO201040}. Subsequent studies demonstrated that risk propagation and systemic fragility travel along these edges much like a diffusion process~\cite{pozzi2013spread,kenett2012networks}. From this viewpoint, the quadratic risk term $\mathbf{w}^{\top}\boldsymbol{\Sigma}\mathbf{w}$ can be interpreted as a diffusion of capital on a weighted network, where assets are vertices, covariances set edge weights, and self-loops reward each asset’s own risk-adjusted return. Classical solvers glimpse this structure only indirectly through dense linear algebra; they do not embed diversification \emph{in the dynamics} of the flow.

Recent progress in quantum information science enables this network view to be dynamic and adaptive. Quantum stochastic walks (QSWs)~\cite{whitfield2010quantum,attal2012open} and quantum network-ranking algorithms~\cite{Eduardo2012quantum,paparo2013quantum,wang2022continuous,chruscinski2017gkls} merge coherent quantum evolution with classical random walks on the same graph. In this framework, a coherent channel governed by a Hermitian Hamiltonian explores the network through superposition and interference, while a stochastic channel, implemented via a Google-type generator, ensures ergodicity and convergence. By tuning a quantum–classical mixing parameter, a QSW interpolates between fully quantum exploration and purely classical diffusion, providing an adaptive mechanism that can respond to both structured dependencies and stochastic shocks. The coherent channel can survey multiple highly covariant clusters in parallel, while the stochastic channel damps cyclic probability within tight clusters and nudges capital towards under-represented regions of the network.

Building on these insights, we develop a quantum graph–theoretic framework for portfolio optimization with four main contributions. First, we map expected returns and covariances to a weighted, directed graph in which self-loops encode asset-specific Sharpe information and inter-node edges penalize correlation, providing a network-based representation of the mean–risk trade-off. Second, we introduce a dual-channel QSW whose density matrix evolves on this graph and whose diagonal converges to a steady-state weight vector; the classical and quantum channels jointly realize a “smart 1/N” allocation, structurally close to equal-weight but with small, data-driven tilts and a controllable level of trading. Third, we provide a practical implementation of the QSW framework (including GPU acceleration for matrix operations), making it feasible to optimize and rebalance portfolios of order 100 assets in a realistic time. Fourth, we conduct a comprehensive empirical evaluation, benchmarking QSW-based portfolios against maximum-Sharpe mean–variance solutions, naive 1/N portfolios, and the S\&P 500 index. The tests span fixed parameter presets, a 625-point hyper-parameter grid search, and a 34-year multi-universe robustness study across 30 randomly sampled 100-stock S\&P 500 universes. Across these experiments, QSW-based portfolios consistently retain 1/N-like diversification while improving risk-adjusted returns and maintaining moderate turnover. The grid search shows that these gains are structural, not the result of fine-tuning, and yields simple design rules that link the QSW parameters to diversification, turnover, and efficiency.

The remainder of the paper proceeds as follows. We first present the empirical results in the Results section, followed by a Discussion of implications. Finally, we describe the QSW formulation and experimental design in the Methods section.

\section{Results}

This section presents a comprehensive empirical validation of the QSW framework. The experimental design has two phases: (1) a recent-period parameter exploration (2018–2024) to understand and tune the model, and (2) a long-horizon, multi-universe robustness study (1990–2024) to test generalizability across regimes and universes. The first phase uses a fixed “top-100-by-market-cap’’ S\&P 500 universe and includes both a preset-based analysis and a systematic grid search. The second phase runs 30 independent trials on dynamically maintained, point-in-time S\&P 500 universes to guard against selection and survivorship bias. Across all experiments, QSW-based portfolios are compared against three benchmarks: the maximum-Sharpe MPT portfolio, the naive 1/N portfolio, and the S\&P 500 index.

\subsection{Experimental overview}
Our experimental framework is designed to evaluate the QSW methodology under both controlled and realistic conditions.

In Phase~1, parameter exploration (2018–2024), we use daily data from 2018 to 2024 on the top 100 S\&P 500 companies by market capitalization as a fixed universe. This phase consists of two experiments. Experiment 1 (preset analysis) defines six QSW strategy presets that represent different investment philosophies, from ultra-diversified to high-activity trading styles. Each preset is evaluated across five quantum-mixing values $\omega \in \{0.2, 0.4, 0.6, 0.8, 1.0\}$ to study the impact of the quantum–classical spectrum. Experiment 2 (comprehensive grid search) systematically explores a 625-point grid over the four QSW parameters $(\alpha,\beta,\lambda,\omega)$ on the same top-100 universe. This grid search identifies robust high-performing regions of the parameter space and clarifies how the four control knobs interact.

In Phase~2, robustness validation (1990–2024), we test the generalizability of the approach over a 34-year backtest with rolling 2-year training windows. Experiment 3 (multi-universe testing) runs 30 independent trials on distinct 100-stock universes. As described in the Methods section (“Data and experimental setup”), each universe is constructed point-in-time by sampling from contemporaneous S\&P 500 constituents and is dynamically maintained to avoid survivorship bias. This design validates that the QSW results are not dependent on a single, favorable asset selection.

In all experiments, QSW-based portfolios are evaluated against three benchmarks: the maximum-Sharpe MPT portfolio (classical mean–variance optimizer), the naive 1/N portfolio on the same universe, and the S\&P 500 index as a market-cap-weighted passive benchmark.

\subsection{Experiment 1: preset-based analysis}

\begin{table}[htbp]
\centering
\caption{QSW strategy presets used in Phase~1.}
\label{tab:scenario_parameters}
\begin{tabular}{lcccp{6cm}}
\hline
\textbf{Preset} & \textbf{$\alpha$} & \textbf{$\beta$} & \textbf{$\lambda$} & \textbf{Investment philosophy} \\
\hline
Ultra-diversified & 1.0   & 100.0 & 10.0  & Prioritizes maximum diversification \\
Moderate-balanced & 10.0  & 10.0  & 10.0  & Balanced emphasis on risk, return, and turnover \\
Stability-focused & 1.0   & 10.0  & 100.0 & Emphasizes position stability \\
Balanced-active   & 10.0  & 1.0   & 100.0 & Moderate return focus with stable positions \\
Sharpe-maximizer  & 100.0 & 1.0   & 10.0  & Aggressive pursuit of high-Sharpe assets \\
High-activity     & 100.0 & 10.0  & 1.0   & Maximum rebalancing frequency \\
\hline
\end{tabular}
\end{table}

This experiment evaluates 30 distinct configurations, which are derived by crossing the six parameter presets in Table~\ref{tab:scenario_parameters} with five quantum-classical mixes ($\omega \in \{0.2, 0.4, 0.6, 0.8, 1.0\}$). To test the model's sensitivity to the length of historical data, we evaluate every configuration twice: (i) using a 1-year ("short-memory") training window, and (ii) using a 2-year ("long-memory") training window. Both sets of models are then out-of-sample back-tested from 2018-01-02 to 2024-12-31 with quarterly rebalancing.

\begin{figure}[htbp]
  \centering
  \includegraphics[width=\textwidth]{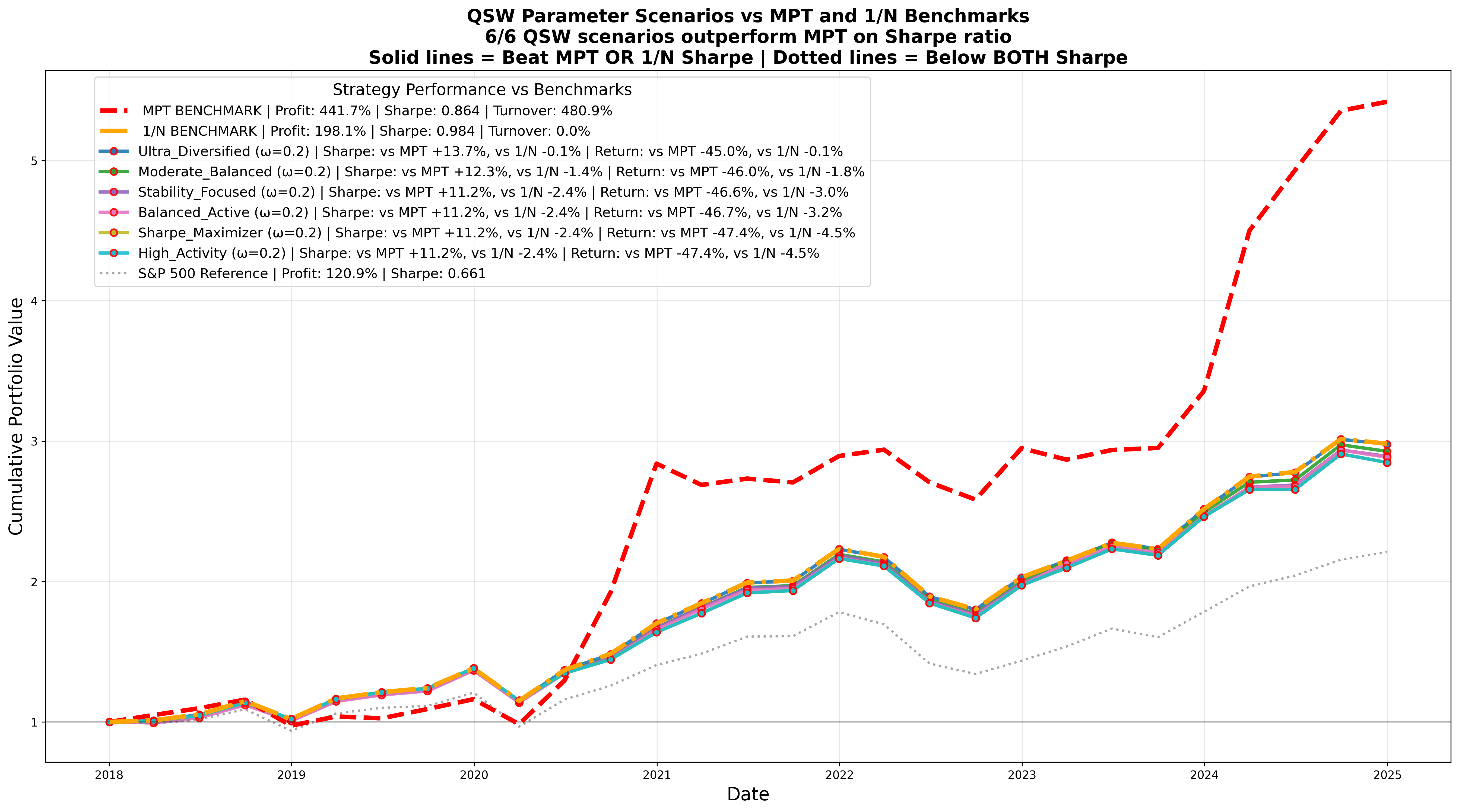}
  \includegraphics[width=\textwidth]{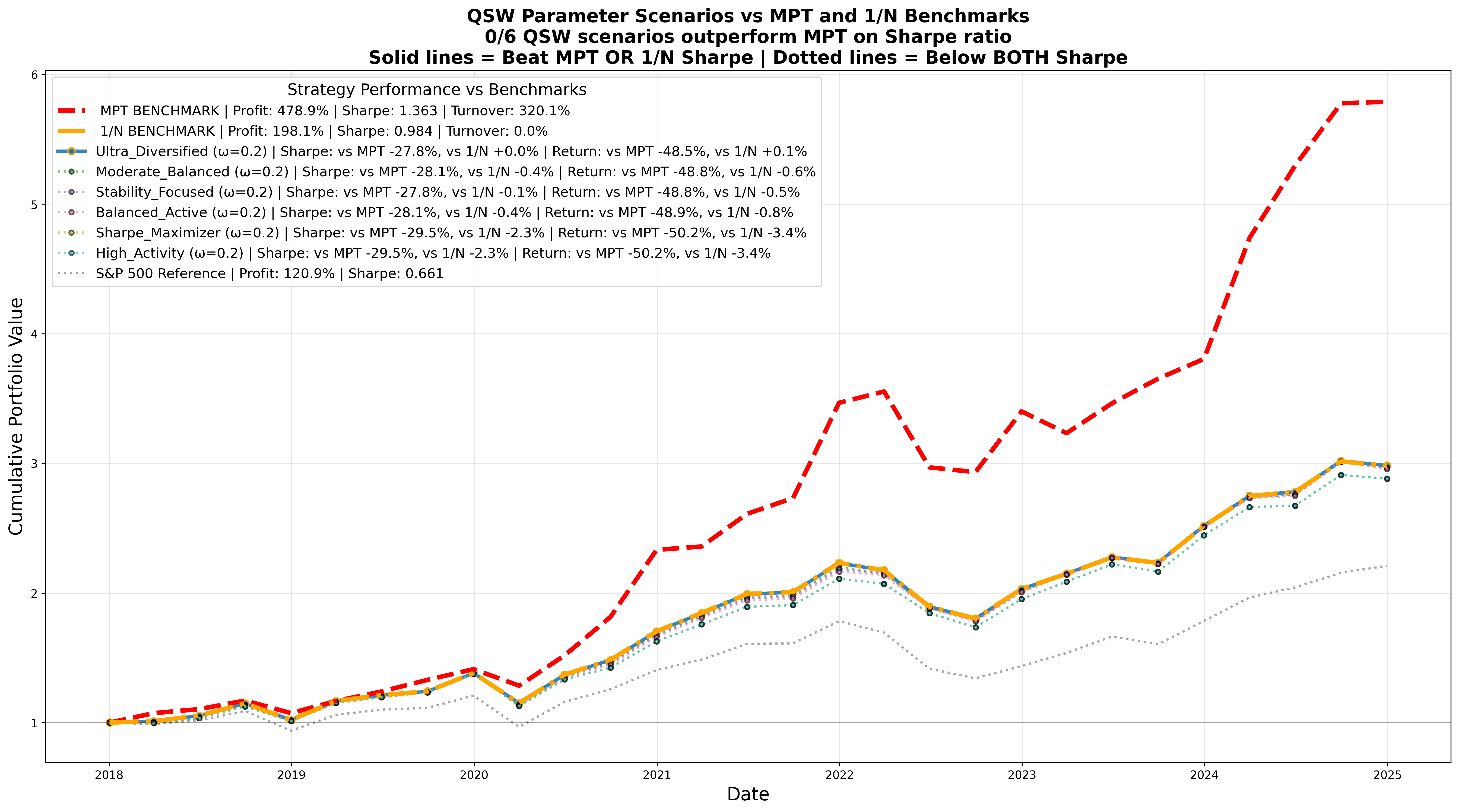}
  \caption{Cumulative value of \$1 invested in 2018. \textbf{Top:} 1-year training. \textbf{Bottom:} 2-year training. Red dashed line = MPT benchmark; dotted grey = S\&P 500. Solid QSW lines denote configurations that \emph{beat} MPT on Sharpe ratio, dotted lines underperform.}
  \label{fig:scenarioTrajectories}
\end{figure}

A combined analysis of the cumulative wealth trajectories (Figure~\ref{fig:scenarioTrajectories}) and the detailed performance metrics (Table~\ref{tab:scenario_metrics}) reveals several critical insights into the QSW model's behavior relative to the benchmarks.

First, regarding model stability, the "two starkly different regimes" visible in Figure~\ref{fig:scenarioTrajectories} are driven almost entirely by the instability of the MPT benchmark, not the QSW model. This highlights MPT's extreme sensitivity to estimation error, a key limitation we identified in the Introduction. With 1-Year Training, MPT produces a poor portfolio (Sharpe 0.86) with extremely high volatility (31.61\%). The short, noisy training window likely results in a highly unstable covariance matrix, leading to a flawed optimization. In contrast, all QSW strategy presets (Sharpe $\approx$ 0.96--0.98) are stable and easily outperform MPT. With 2-Year Training, with a more stable window, MPT's volatility drops to 20.94\% and its Sharpe ratio jumps by 58\% to 1.36. The QSW and 1/N models, however, are almost completely unaffected by the change in window length, with their Sharpe (0.98) and Volatility ($\approx$ 17.2\%) remaining rock-solid. This demonstrates the robustness of the QSW method; its performance is not dependent on a "lucky" or perfectly stable estimation window.

Second, the QSW model converges to a "Smart 1/N". The most striking finding is that the QSW model, across all QSW strategy presets, produces a risk and diversification profile that is nearly identical to the naive 1/N benchmark. As seen in Table~\ref{tab:scenario_metrics}, the QSW's Sharpe (0.96--0.98), Volatility ($\approx$ 17\%), Max Drawdown ($\approx$ -19\%), and HHI ($\approx$ 0.01) are all in lockstep with the 1/N portfolio. This suggests the QSW's quantum-graph dynamics naturally find the highly diversified, stable 1/N state. The key difference is that QSW acts as a "smart 1/N", initiating a very low (2\%--35\%) but non-zero turnover, in contrast to MPT's hyper-active 320\%--480\% turnover.

Third, while MPT achieves the highest "paper" Sharpe ratio (1.36) in the 2-year case, Table~\ref{tab:scenario_metrics} shows this is operationally unfeasible. This performance is achieved by making an extreme, concentrated bet (HHI $\geq$ 0.25, or $\approx$ 4-5 effective stocks) and by turning over the entire portfolio 3-5 times per year (320\%--480\% turnover). The transaction costs from this churn would completely erase its "paper" alpha. QSW, by contrast, generates its alpha with minimal turnover and maximum diversification (HHI $\approx$ 0.01, $\approx$ 90-100 effective stocks), making it a practical and cost-effective strategy.

Finally, Figure~\ref{fig:scenarioTrajectories} shows that in both 1-year and 2-year setups, all QSW configurations (final wealth $\approx$ 3.0$\times$) and the 1/N benchmark ($\approx$ 3.0$\times$) consistently and significantly outperform the passive S\&P 500 index (final wealth $\approx$ 2.2$\times$).

\begin{table}[htbp]
\centering
\caption{Headline metrics for the six QSW strategy presets (best mix \(\omega=0.2\)).}
\label{tab:scenario_metrics}
\begin{tabular}{lccccccc}
\hline
\multicolumn{8}{c}{\textbf{(a) 1-year training}}\\
\hline
Preset & Sharpe & Vol.\,[\%] & MDD\,[\%] & Turn.\,[\%] & Eff. & HHI & Eff.\#\,stk \\
\hline
Ultra-Divers. & 0.9846  & 17.17  & -19.27 & 4.00   & 19.64 & 0.0100 & 99.96 \\
Moderate-Bal. & 0.9800  & 17.10  & -19.15 & 31.92  & 2.95  & 0.0102 & 98.81 \\
Stability-F. & 0.9837  & 17.05  & -19.19 & 35.28  & 2.65  & 0.0102 & 97.37 \\
Balanced-Act. & 0.9803  & 17.03  & -19.12 & 50.49  & 1.87  & 0.0105 & 94.84 \\
Sharpe-Max. & 0.9615  & 16.80  & -19.56 & 79.71  & 1.19  & 0.0107 & 92.98 \\
High-Activity  & 0.9615  & 16.80  & -19.56 & 79.71  & 1.19  & 0.0107 & 92.98 \\
\hline
\textbf{1/N benchmark} & \textbf{0.98} & \textbf{17.17} & \textbf{-19.27} & \textbf{0} & \textbf{$\infty$} & \textbf{0.01}  & \textbf{100} \\
MPT benchmark  & 0.86  & 31.61 & -16.03 & 480.85 (--) & 0.18 & 0.268 & 3.7 \\
\hline
\multicolumn{8}{c}{\textbf{(b) 2-year training}}\\
\hline
Preset & Sharpe & Vol.\,[\%] & MDD\,[\%] & Turn.\,[\%] & Eff. & HHI & Eff.\#\,stk \\
\hline
Ultra-Divers.  & 0.98  & 17.18  & -19.20  & 2.11   & 31.68 & 0.0100 & 99.98 \\
Moderate-Bal. & 0.98  & 17.15  & -18.59  & 17.66  & 5.25 & 0.0101 & 98.80 \\
Stability-F. & 0.98  & 17.10  & -18.13  & 26.06  & 3.64 & 0.0102 & 97.59 \\
Balanced-Act. & 0.98  & 17.10  & -17.75  & 34.39  & 2.77 & 0.0104 & 95.95 \\
Sharpe-Max. & 0.96  & 16.99  & -18.25  & 71.24  & 1.33 & 0.0109 & 91.99 \\
High-Activity  & 0.96  & 16.99  & -18.25  & 71.24  & 1.33 & 0.0109 & 92.00 \\
\hline
\textbf{1/N benchmark} & \textbf{0.98 } & \textbf{17.17} & \textbf{-19.27} & \textbf{0} & \textbf{$\infty$} & \textbf{0.01} & \textbf{100} \\
MPT benchmark  & 1.36 & 20.94  & -17.5  & 320.08  & 0.42 & 0.254 & 4.65 \\
\hline
\multicolumn{8}{l}{\scriptsize{Vol.\,= annualized volatility; MDD\,= max drawdown; Turn.\,= annual turnover; Eff.\,= Sharpe / Turnover.}}
\end{tabular}
\end{table}

Comparing the 1-year (Table~\ref{tab:scenario_metrics}a) and 2-year (Table~\ref{tab:scenario_metrics}b) results reveals the core weakness of MPT and the core strength of QSW. The MPT benchmark is fundamentally unstable. When the training window changes from 1-year to 2-years, MPT's Sharpe ratio leaps 58\% (from 0.86 to 1.36) and its volatility drops from 31.6\% to 20.9\%. This demonstrates that MPT is extremely sensitive to the estimation error in its inputs, a key classical limitation. In stark contrast, the QSW and 1/N models are rock-solid. Their key metrics (Sharpe $\approx$ 0.98, Vol $\approx$ 17.2\%) are almost completely unaffected by the change in training data. This provides strong empirical evidence that our QSW framework, like the 1/N benchmark, is structurally robust and successfully insulated from the parameter instability that plagues classical MPT.

To understand the behavior within each preset, we analyze the effect of the quantum-classical mixing parameter, $\omega$, and the portfolio's concentration over the 2018–2024 backtest period. We present the results for the 1-year training window here; the corresponding analysis for the 2-year window, which confirms the same structural conclusions, is provided in the Supplementary Information. Figure~\ref{fig:omega_sensitivity} shows the performance metrics for all six QSW strategy presets as a function of the quantum-classical balance, $\omega$. The results reveal a clear and powerful trend: the $\omega$ parameter acts primarily as a "turnover dial." In all six QSW strategy presets, increasing $\omega$ (making the model more "classical") leads to a near-linear increase in portfolio turnover. For example, in the "Ultra-Diversified" preset, turnover rises from 4\% to 30\% as $\omega$ goes from 0.2 to 1.0. Crucially, this increase in turnover has almost no impact on the Sharpe Ratio or Volatility. These metrics remain remarkably flat across the entire $\omega$ spectrum, and always superior to MPT's 1-year performance. Most importantly, $\omega$ does not break the model's diversification. The HHI remains at its 1/N floor ($\approx 0.01$) in all but the most active presets, and even then, it stays an order of magnitude more diversified than MPT. This demonstrates that the QSW model's core benefits (high Sharpe, low volatility, and extreme diversification) are structurally inherent and robust, while the $\omega$ parameter provides a simple, interpretable knob to control the "cost" (i.e., turnover) of the strategy.

\begin{figure}[htbp]
    \centering
    \caption{Impact of the quantum-classical mixing parameter ($\omega$) on key metrics for all six QSW strategy presets (1-year training, 2018–2024 backtest). $\omega$ primarily functions as a "turnover dial" without degrading the portfolio's Sharpe Ratio or diversification.}
    \label{fig:omega_sensitivity}
    \includegraphics[width=0.49\textwidth]{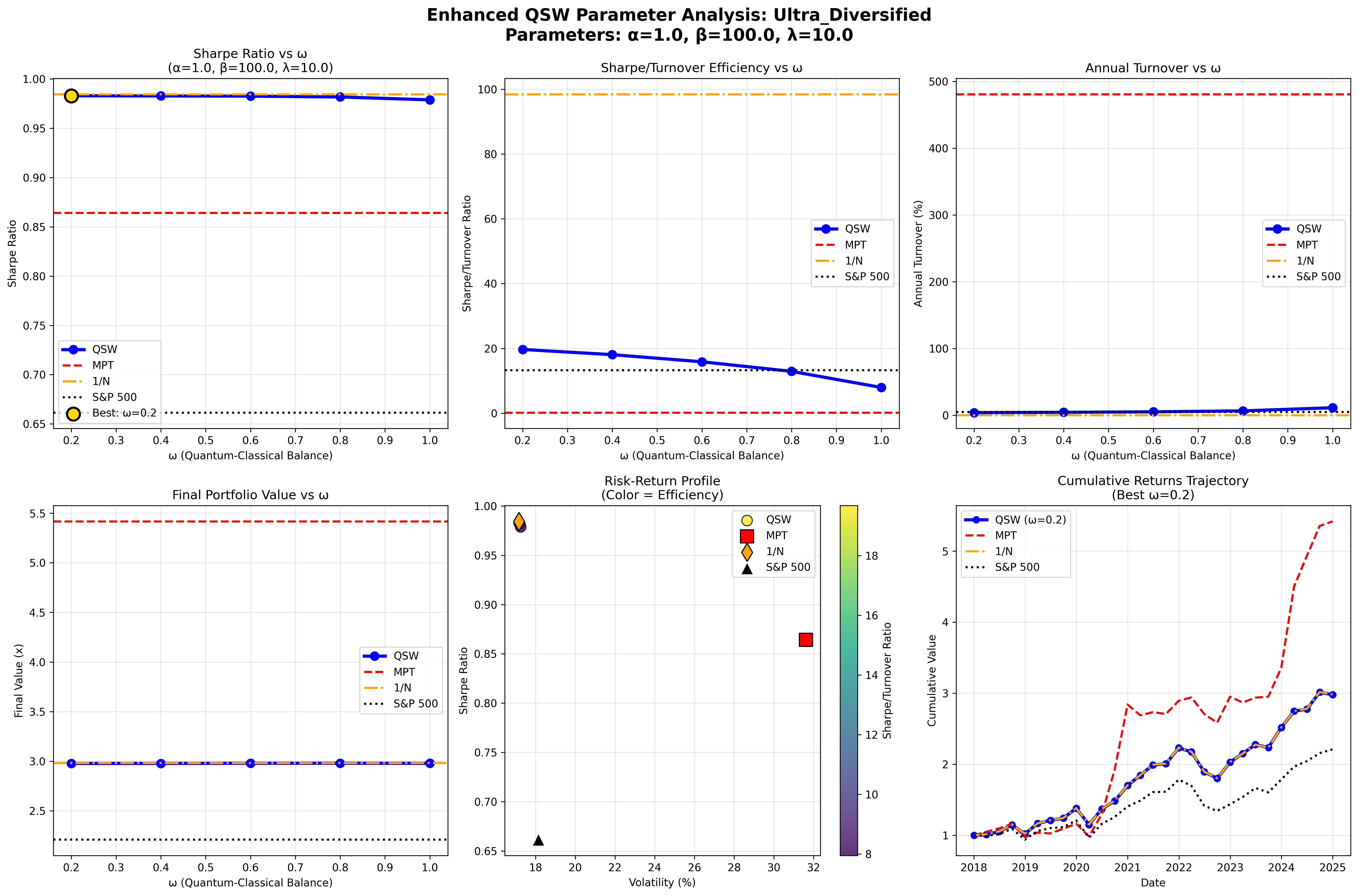}
    \includegraphics[width=0.49\textwidth]{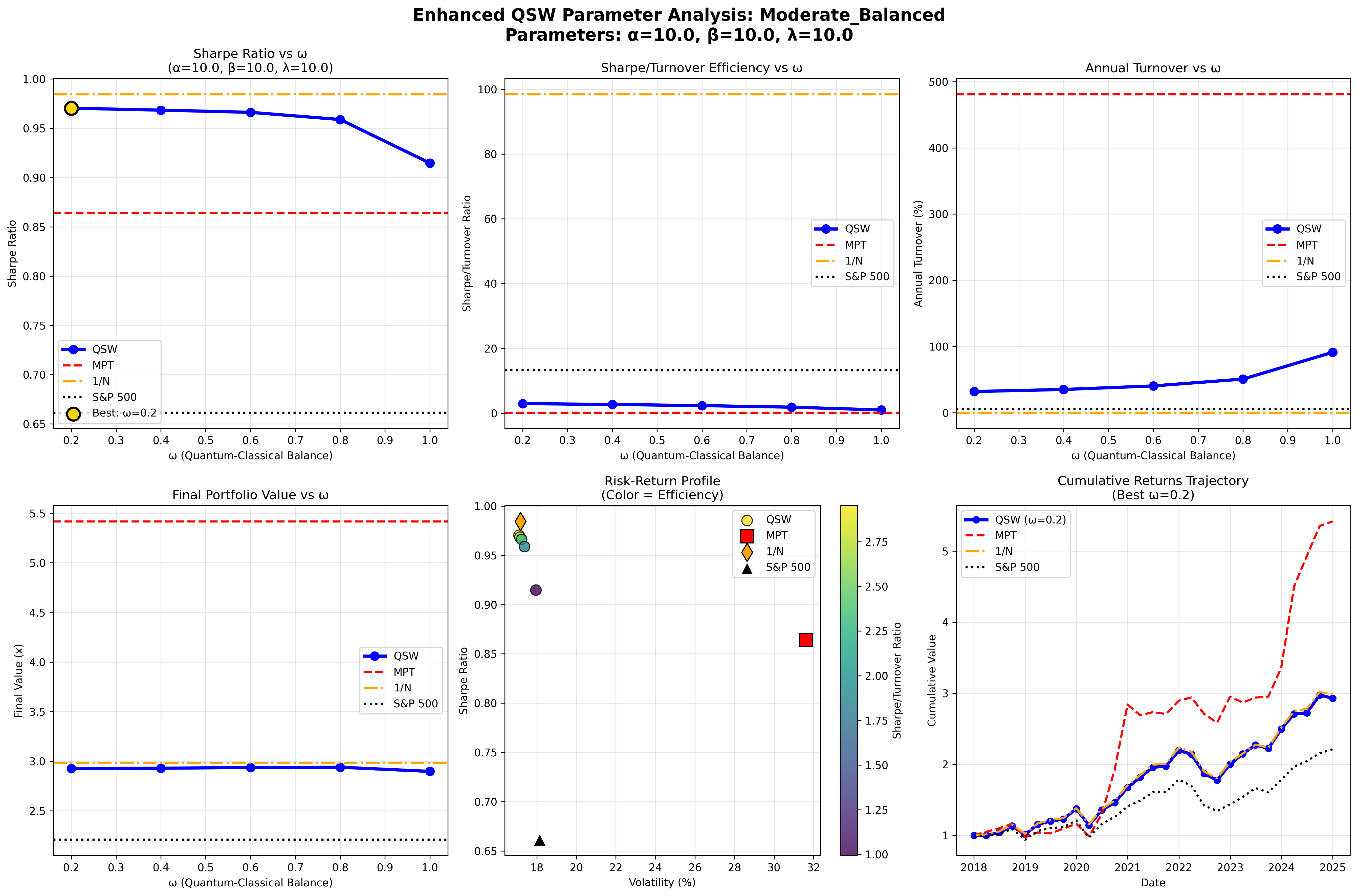}
    \includegraphics[width=0.49\textwidth]{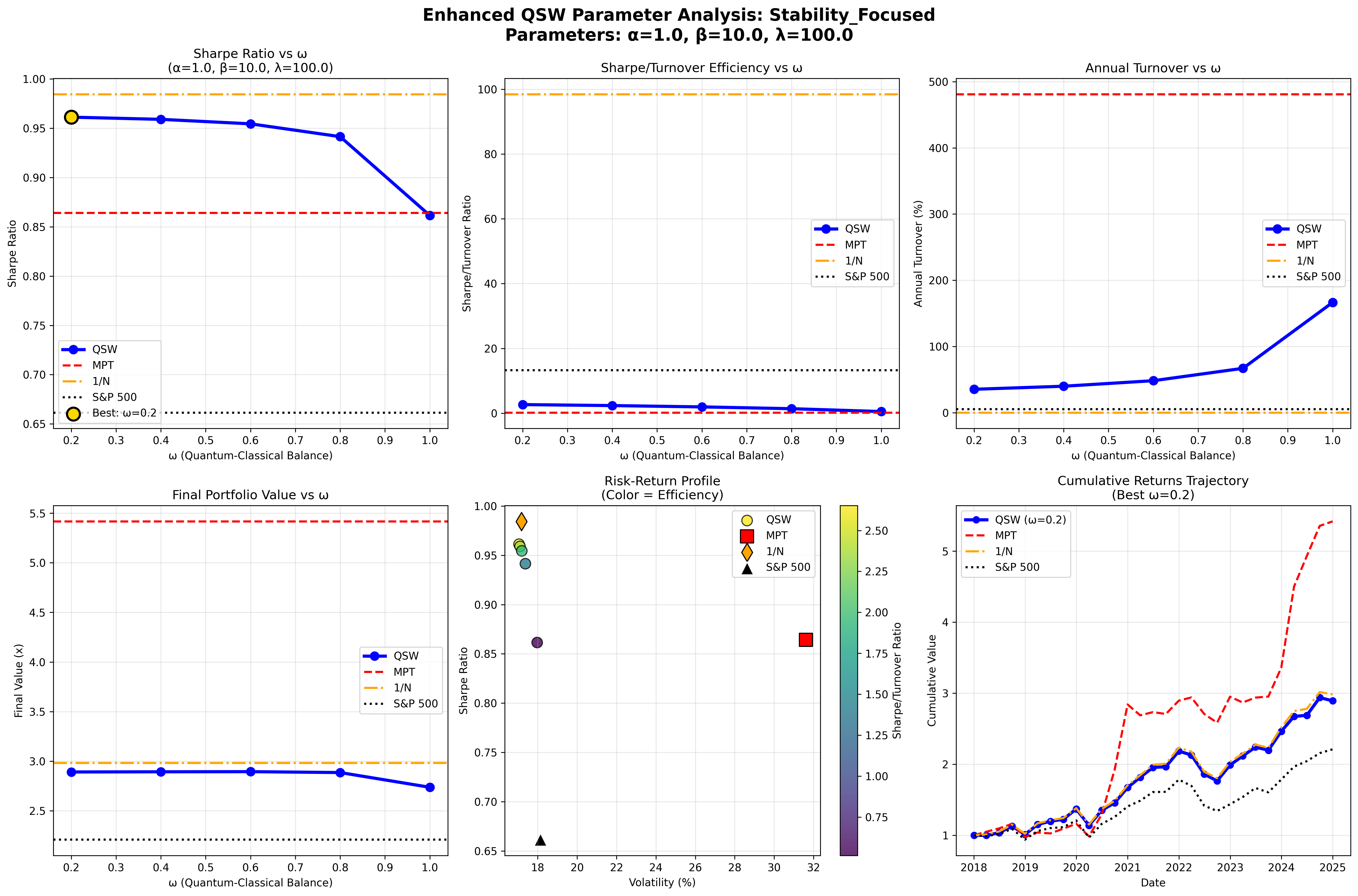}
    \includegraphics[width=0.49\textwidth]{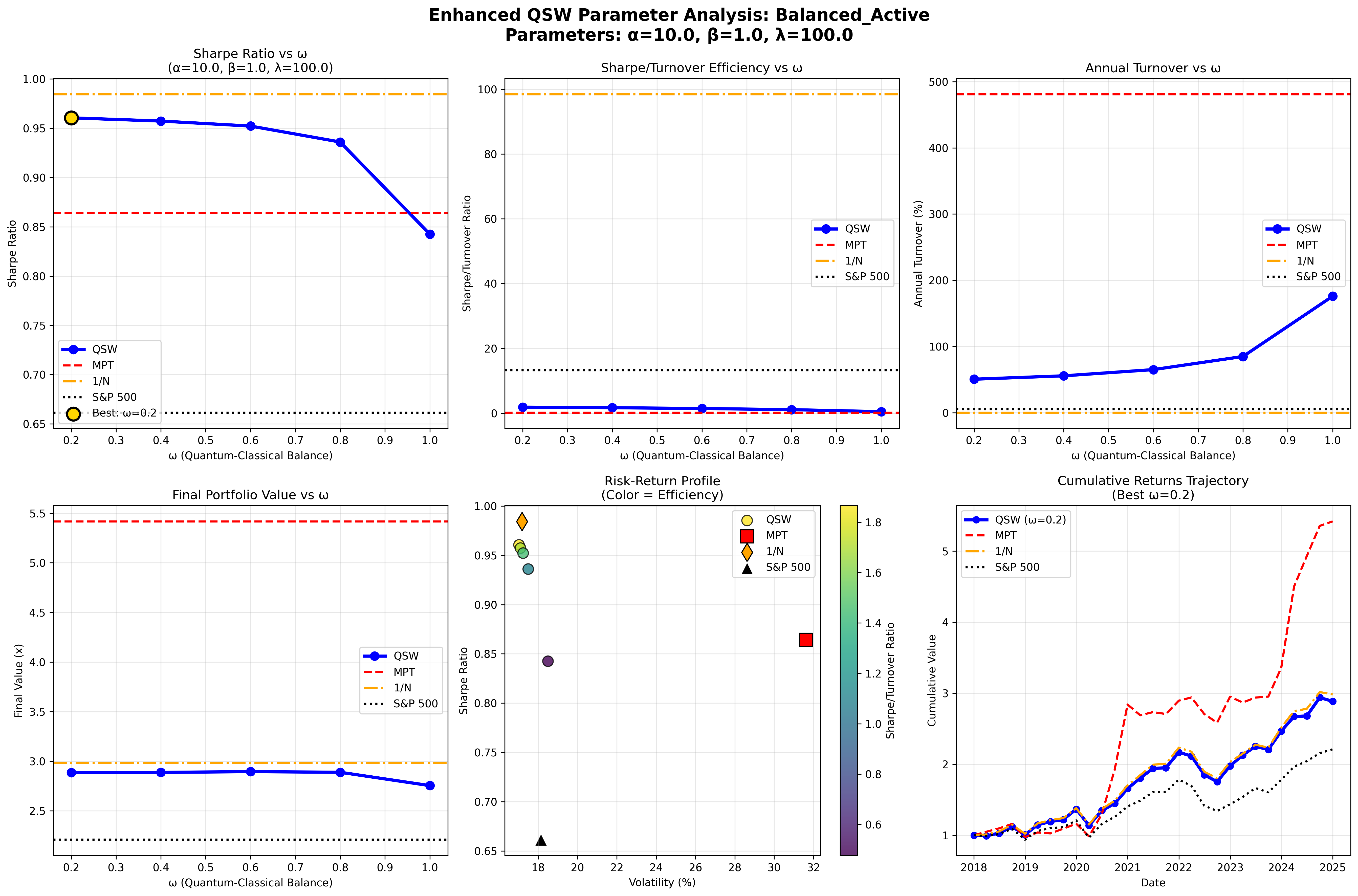}
    \includegraphics[width=0.49\textwidth]{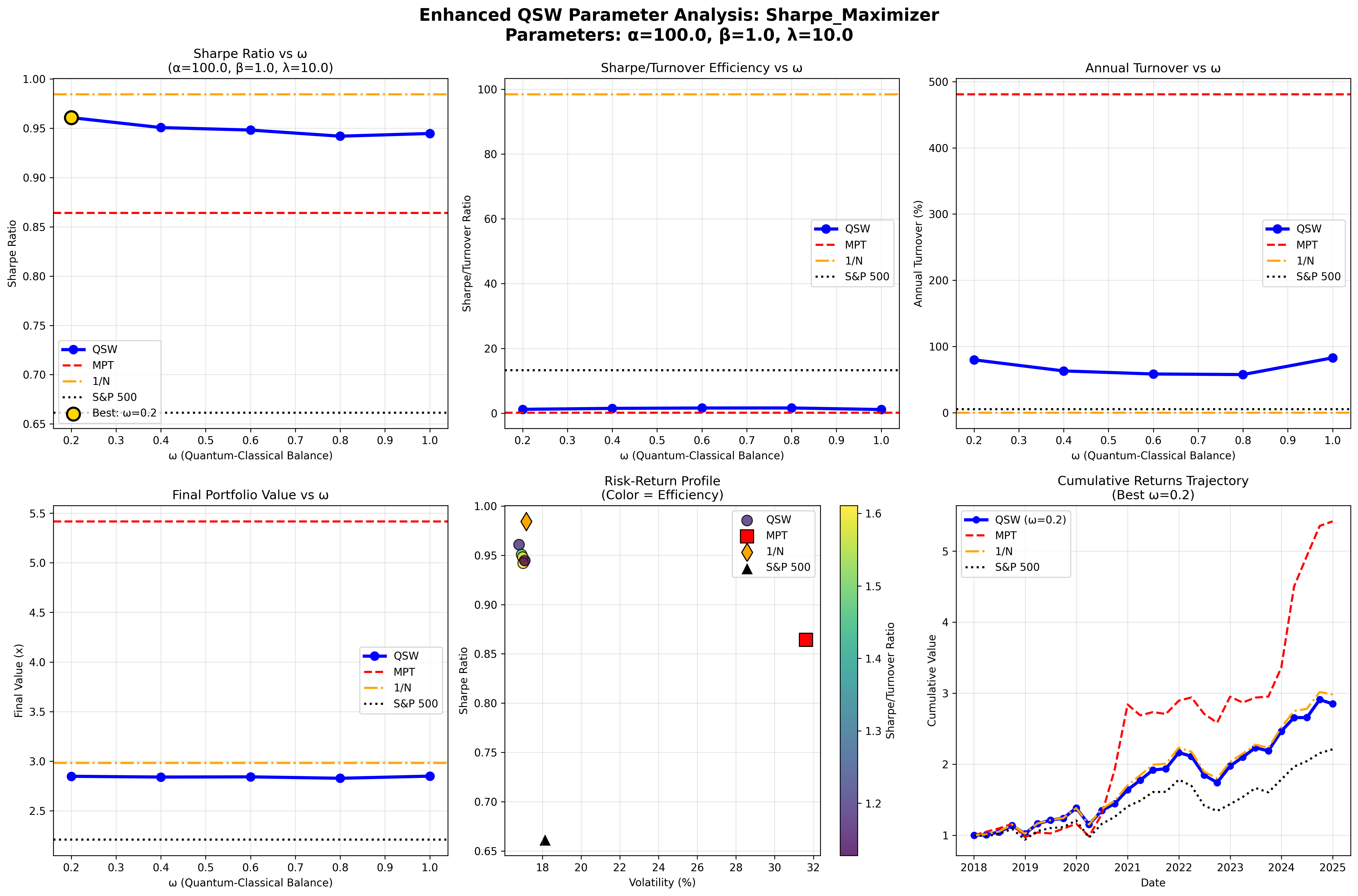}
    \includegraphics[width=0.49\textwidth]{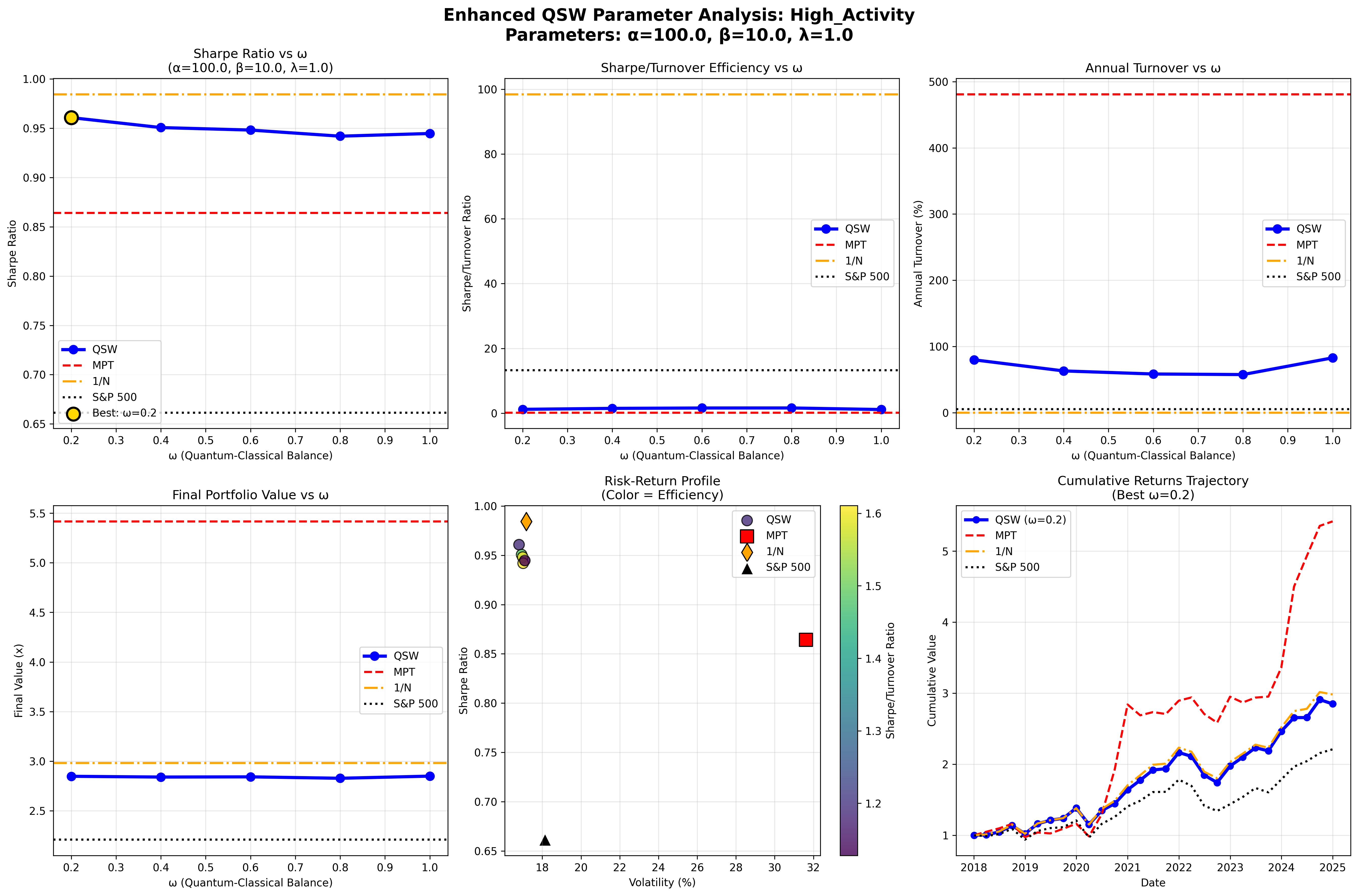}
\end{figure}

\begin{figure}[htbp]
\centering
\includegraphics[width=0.9\textwidth]{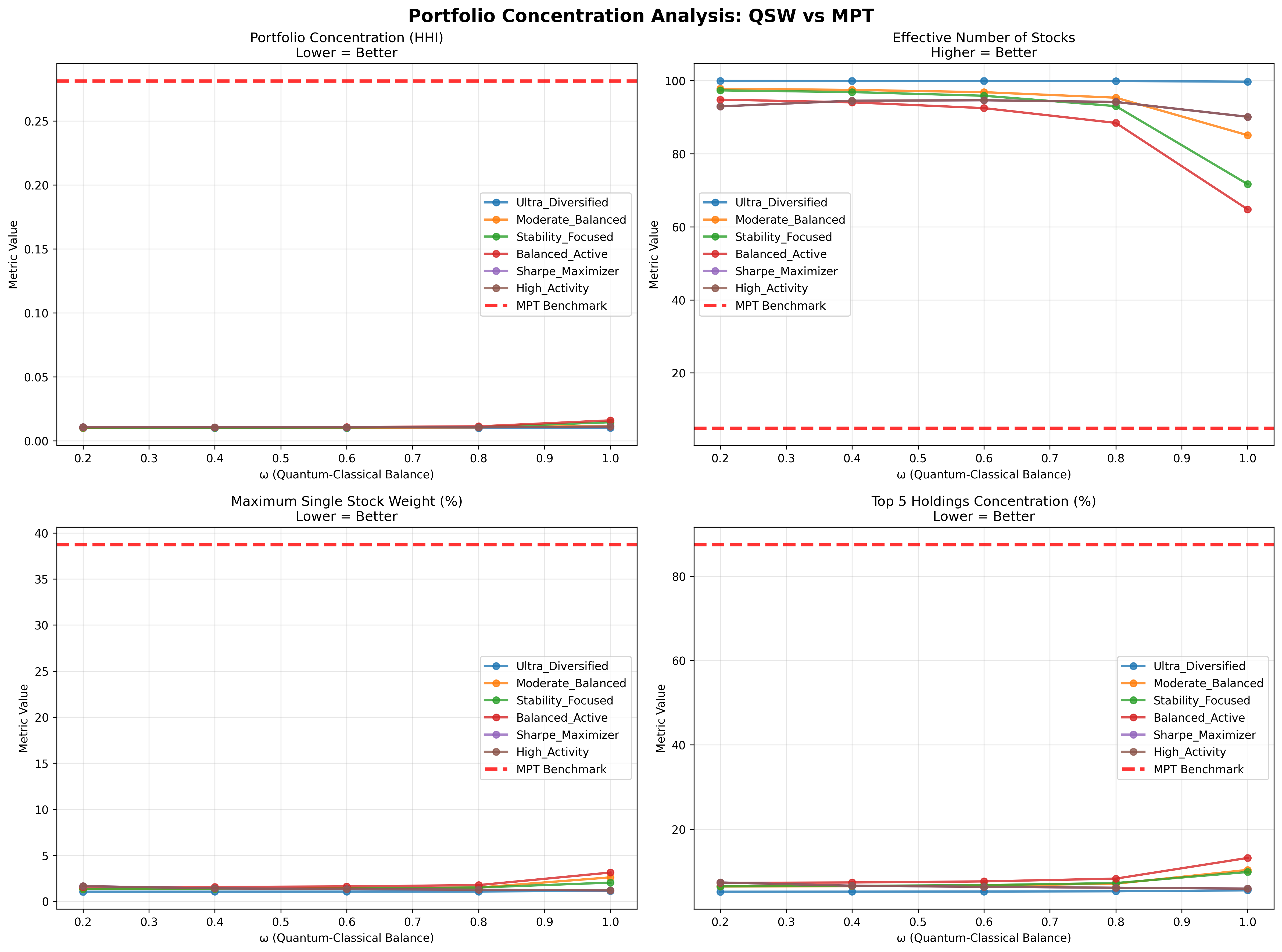}
\caption{Evolution of portfolio concentration (Herfindahl–Hirschman Index, HHI) for QSW-based portfolios (blue and purple) compared with the classical MPT benchmark (red) during the 2018–2024 backtest period. The MPT portfolio exhibits structural instability, alternating between highly concentrated and diffuse allocations, whereas all QSW configurations maintain stable, near-uniform diversification (HHI $\approx 0.01$) even through periods of market stress such as the 2020 COVID-19 crash.}
\label{fig:concentration_analysis}
\end{figure}

A core structural difference between QSW and MPT is revealed by analyzing their portfolio concentration. We compare the models over the 2018–2024 backtest using four standard metrics, as shown in Figure~\ref{fig:concentration_analysis}. The QSW model, across all presets, behaves as a "smart 1/N". Its concentration is tunable and directly linked to the quantum-classical mix, $\omega$. As seen in Figure~\ref{fig:concentration_analysis}, when $\omega$ is low (more quantum), the model's HHI is locked at around the theoretical floor of 0.01. As $\omega$ increases, the model is permitted to become slightly more concentrated, and the HHI, Effective Number of Stocks, Max Single Stock Weight, and Top 5 Holdings all show a small, controlled deviation from the 1/N state. However, even at its most "classical" ($\omega=1.0$), the QSW model remains exceptionally well-diversified. The MPT benchmark (red line) is defined by its extreme and erratic concentration. Its HHI is dangerously unstable. While its average HHI is $\approx$ 0.28, the Top 5 Holdings chart shows that MPT's typical state is to hold $\approx$ 87–89\% of the entire portfolio in just five names. This confirms MPT is a structurally unstable and highly concentrated strategy, while QSW is robustly diversified by design.

Portfolio annual turnover directly translates into implementation costs that can significantly erode theoretical performance gains. These costs compound with trading frequency, making turnover control a critical factor in real-world portfolio management. Institutional studies such as \cite{frazzini2018trading} place the \emph{all-in} round-trip cost of trading cash equities at: large-cap stocks (5–15 bp), mid-/small-cap (15–30 bp), and market impact (a further 10–50 bp). In this work, we apply a conservative 20 bp all-in round-trip cost for large-cap equities (10 bp commission \& spread + 10 bp market impact). Table \ref{tab:costComparison} converts realized turnover into \emph{ex-ante implementation drag} using this rate. Even under the more forgiving 2-year training calibration, MPT forfeits \textbf{$\sim$64 bp} of alpha to dealing costs versus \textbf{$\sim$7.4 bp} for the QSW preset mean.

\begin{table}[htb]
    \centering
    \caption{Illustrative annual implementation drag assuming a 20\,bp round-trip cost. Costs are computed as $\text{Turnover}\times 0.20\%$.\protect\\ \qquad \footnotesize\textit{Note.} 1 basis point (bp) = 0.01 percentage points.}
    \label{tab:costComparison}
    \begin{tabular}{lcccc} \toprule & \multicolumn{2}{c}{\textbf{1-year training}} & \multicolumn{2}{c}{\textbf{2-year training}}\\ \cmidrule(r){2-3}\cmidrule(l){4-5} \textbf{Strategy / Preset} & Turn.\,[\%] & Cost\,[bp] & Turn.\,[\%] & Cost\,[bp] \\ \midrule \textit{QSW presets}\\ \quad Ultra-Diversified & 4.0 & 0.8 & 2.1 & 0.4 \\ \quad Moderate-Balanced & 31.9 & 6.4 & 17.7 & 3.5 \\ \quad Stability-Focused & 35.3 & 7.1 & 26.1 & 5.2 \\ \quad Balanced-Active & 50.5 & 10.1 & 34.4 & 6.9 \\ \quad Sharpe-Maximizer & 79.7 & 15.9 & 71.2 & 14.2 \\ \quad High-Activity & 79.7 & 15.9 & 71.2 & 14.2 \\ \midrule \textbf{QSW preset mean} & \textbf{46.9} & \textbf{9.4} & \textbf{37.1} & \textbf{7.4} \\ \addlinespace \textbf{1/N (naive)} & \textbf{0.0} & \textbf{0.0} & \textbf{0.0} & \textbf{0.0} \\ \textbf{MPT (max-Sharpe)}& \textbf{480.9} & \textbf{96.2} & \textbf{320.1} & \textbf{64.0} \\ \bottomrule \end{tabular} \end{table}

\medskip\noindent \textit{Is a 64 bp vs.\ 7 bp gap economically meaningful?} Yes. Compare each strategy’s “paper’’ alpha (CAGR in excess of the S\&P 500) with, and without, implementation drag (Table~\ref{tab:alpha_retention}).

\begin{table}[htbp]
    \centering \caption{After-cost alpha retention\protect\\ \footnotesize\textit{Note.} 1 basis point (bp) = 0.01 percentage points.} \label{tab:alpha_retention} \vspace{4pt}
    \begin{tabular}{lccc}
    \toprule & Paper $\alpha$ [bp] & Cost drag [bp] & Net $\alpha$ [bp] \\
    \midrule MPT (2-yr calibration) & 1\,220 & 65 & 1\,155 \\
    QSW (6-preset mean) & 600 & 8 & 592 \\
    \bottomrule
    \end{tabular}
\end{table}

\[
\text{Retention}_{\text{MPT}} = \frac{1\,155}{1\,220} \approx 95\%, \qquad
\text{Retention}_{\text{QSW}} = \frac{592}{600} \approx 99\%.
\]

Even after deducting realistic dealing costs, QSW preserves virtually all of its back-tested edge, whereas MPT forfeits about 5 \% of its “paper’’ alpha. Because the classical maximum-Sharpe solution trades \emph{eight to thirteen times more} than QSW, its cost drag—while “only’’ 64 bp—is still an order of magnitude larger in proportional terms. QSW’s subdued trading footprint additionally mitigates price pressure, timing risk, and information leakage, making its low-turnover profile a \emph{structural} implementation advantage. In Experiment~3 we show that, even when the QSW parameters are re-optimized every quarter over 1990–2024, the dynamic QSW optimizer still trades substantially less than the long-memory MPT benchmark while preserving its risk-adjusted edge.

\begin{figure}[htbp]
\centering
\includegraphics[width=0.32\textwidth]{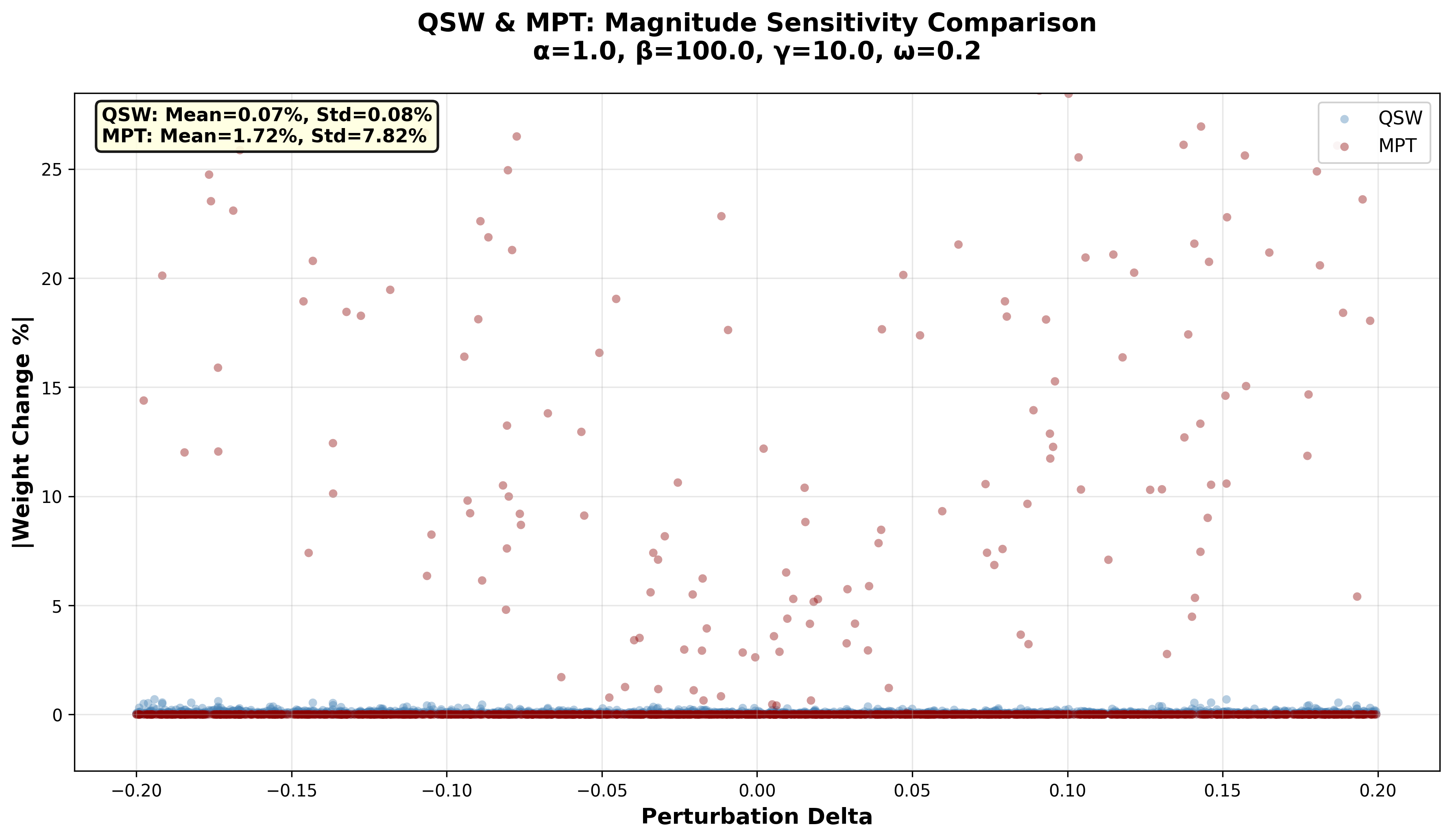}
\includegraphics[width=0.32\textwidth]{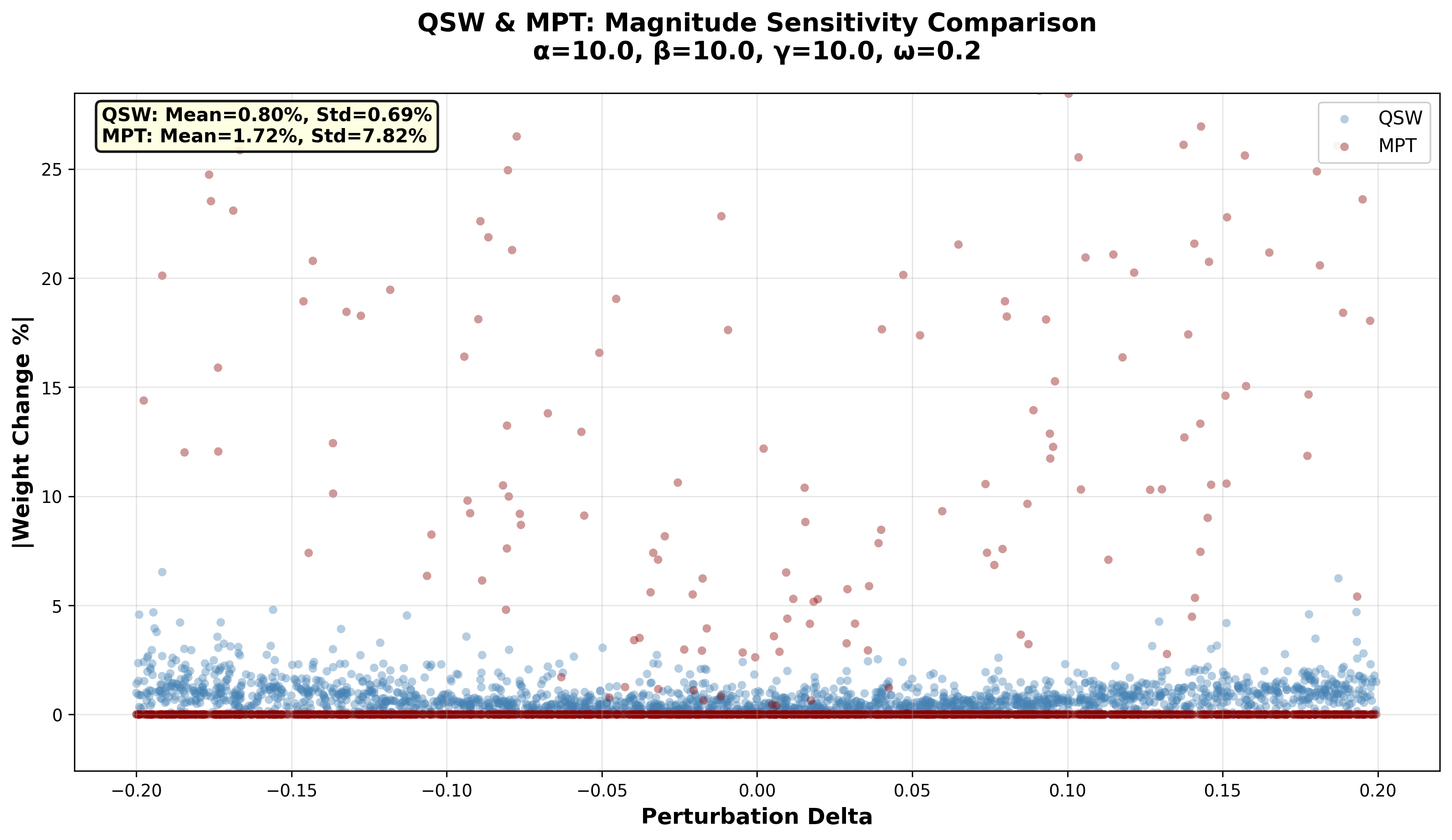}
\includegraphics[width=0.32\textwidth]{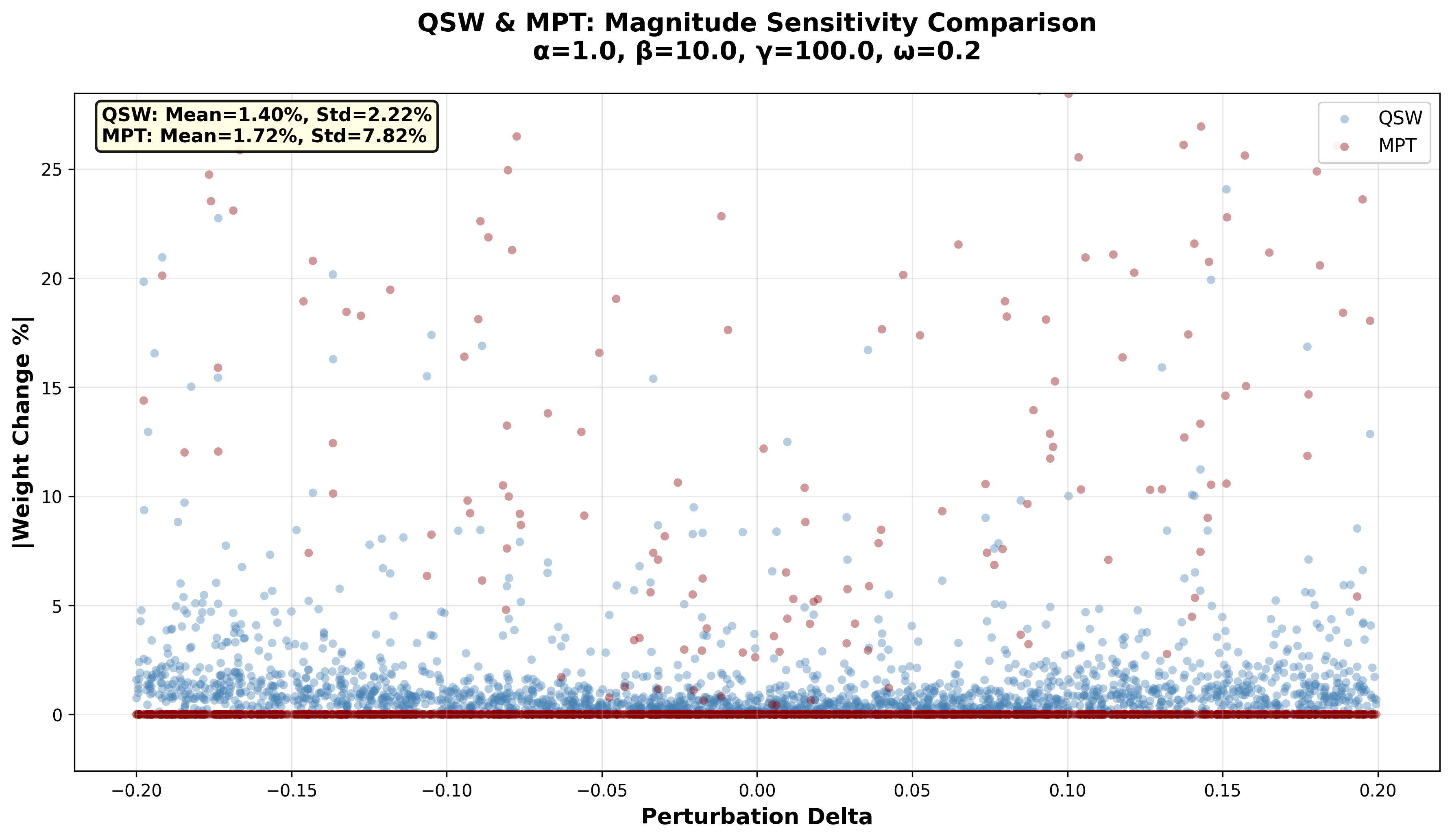}
\includegraphics[width=0.32\textwidth]{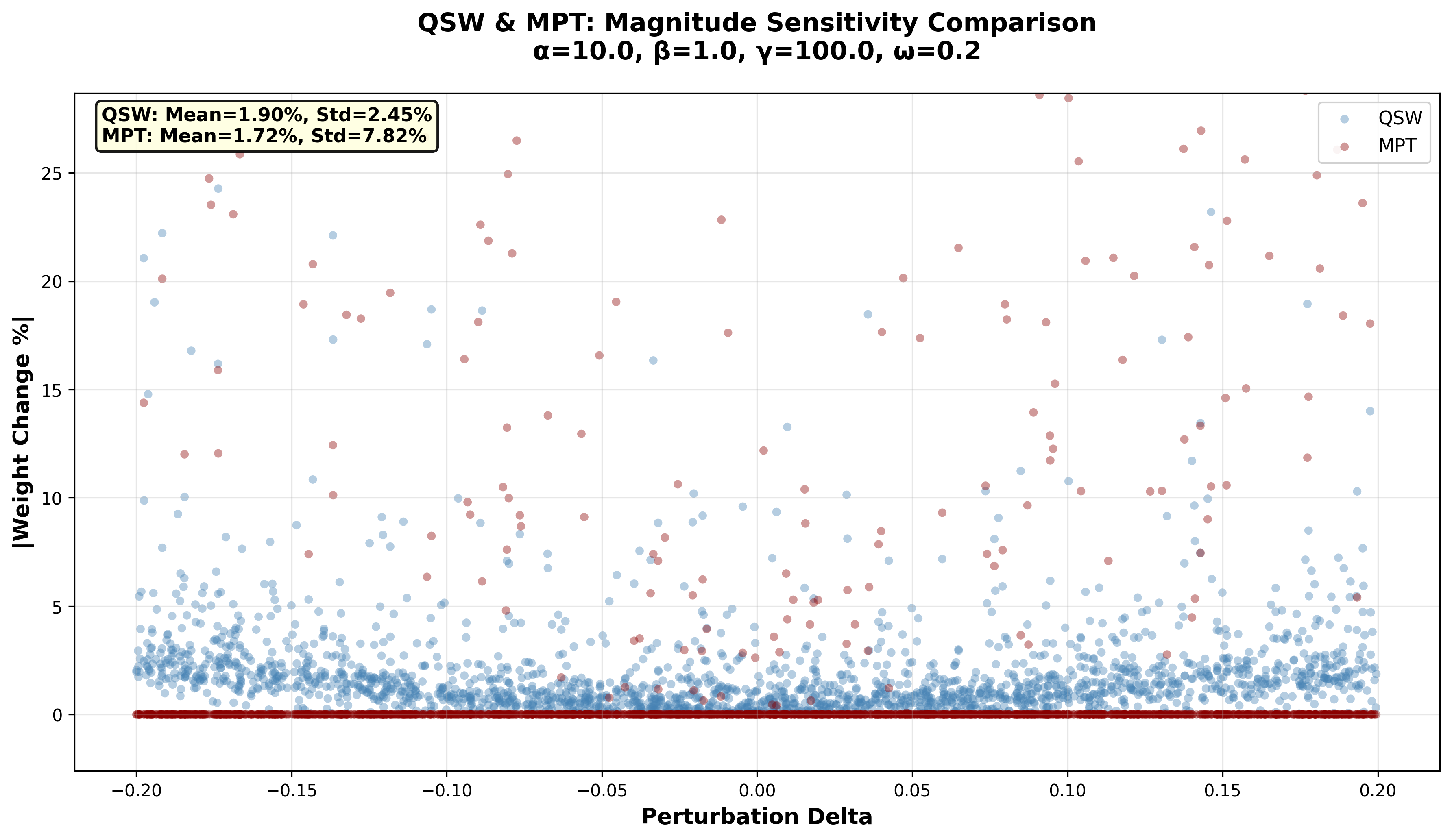}
\includegraphics[width=0.32\textwidth]{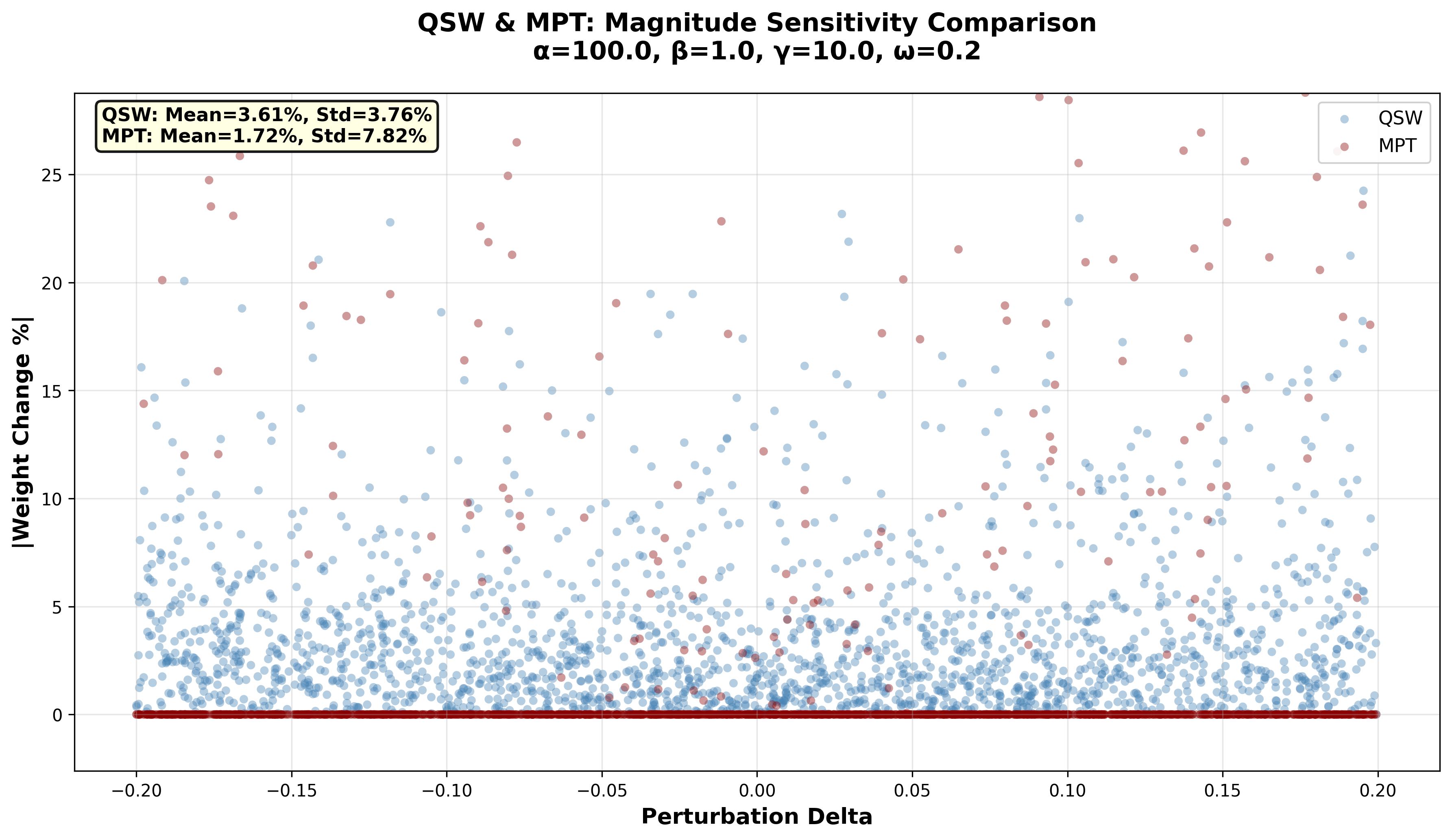}
\includegraphics[width=0.32\textwidth]{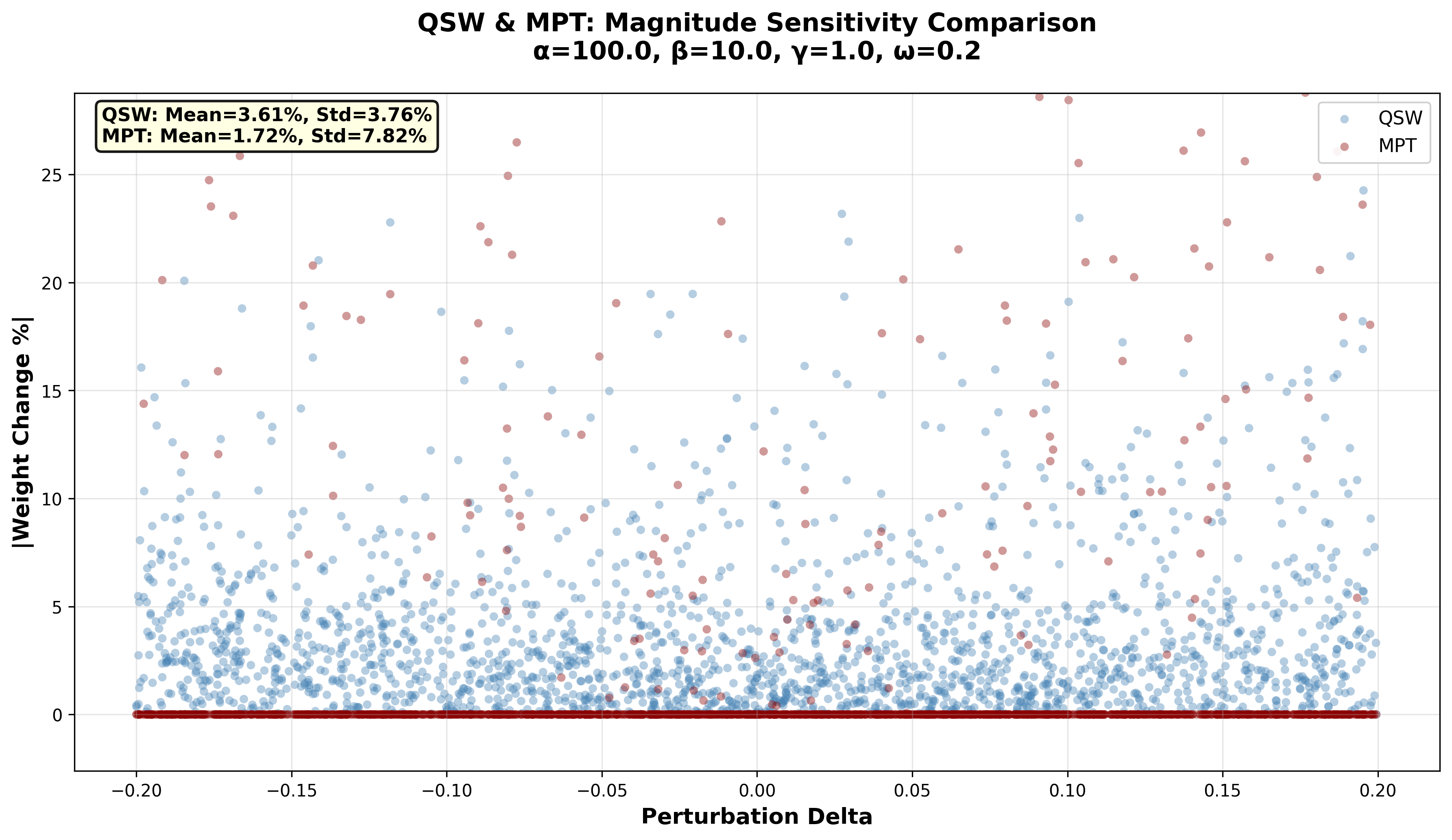}
\caption{Portfolio stability under a 20\% random shock to a single asset's return, repeated \textbf{1000} times. Each plot corresponds to a QSW preset (blue) vs. the MPT benchmark (red).}
\label{fig:shock_test}
\end{figure}

In the previous analysis, we demonstrated that MPT is structurally unstable at the macro level: changing the training window from 1 to 2 years led to dramatic changes in performance and risk profile (a 58\% jump in Sharpe). This section's Monte Carlo test confirms this fragility on a micro level. A fundamental flaw of classical MPT is its extreme sensitivity to input parameters. A small, statistically insignificant change in the expected return vector $\boldsymbol{\mu}$ can cause the optimizer to produce a radically different portfolio, leading to massive, unforced turnover. To test the robustness of our QSW model to this, we conduct a Monte Carlo sensitivity analysis. We take a single period's data and apply a 20\% random "shock" to the returns of a single, randomly chosen asset (i.e., $r_i \times \text{random}(0.8, 1.2)$). We then re-run the optimization for both QSW and MPT and measure the \textbf{total absolute change in portfolio weights} (i.e., $\sum_i |w_{i, \text{new}} - w_{i, \text{old}}|$). This process is repeated \textbf{1000} times for each of the six QSW strategy presets. Figure~\ref{fig:shock_test} plots the 1000-run distribution of this weight change, while Table~\ref{tab:shock_test_summary} quantifies the results.

\begin{table}[htbp]
    \centering
    \caption{Portfolio stability under a Monte Carlo shock test (1000 runs).}
    \label{tab:shock_test_summary}
    \begin{tabular}{lrrrcc} \toprule \textbf{Strategy / Preset} & \textbf{$\alpha$} & \textbf{$\beta$} & \textbf{$\lambda$} & \textbf{Mean \% Wt. Change} & \textbf{Std. Dev. \% Wt. Change} \\ \midrule \textit{QSW presets} \\ \quad Ultra-Diversified & 1.0 & 100.0 & 10.0 & 0.07\% & 0.08\% \\ \quad Moderate-Balanced & 10.0 & 10.0 & 10.0 & 0.80\% & 0.69\% \\ \quad Stability-Focused & 1.0 & 10.0 & 100.0 & 1.40\% & 2.22\% \\ \quad Balanced-Active & 10.0 & 1.0 & 100.0 & 1.90\% & 2.45\% \\ \quad Sharpe-Maximizer & 100.0 & 1.0 & 10.0 & 3.61\% & 3.76\% \\ \quad High-Activity & 100.0 & 10.0 & 1.0 & 3.61\% & 3.76\% \\ \midrule \textbf{MPT (max-Sharpe)} & -- & -- & -- & \textbf{1.72\%} & \textbf{7.82\%} \\ \bottomrule
    \end{tabular}
\end{table}

The results are clear and highly suggestive. The MPT benchmark (red distributions in Figure~\ref{fig:shock_test}) exhibits extreme fragility. As quantified in Table~\ref{tab:shock_test_summary}, while MPT's \textit{mean} weight change (1.72\%) is moderate, its \textbf{standard deviation (7.82\%) is massive}—over 2x to 100x larger than any QSW preset. This highlights how MPT amplifies minor input errors into significant portfolio changes. The QSW model, by contrast, is both structurally robust and \textbf{tunable}. Its stability is directly governed by its parameters, which function as an accelerator and a brake: presets with high $\alpha$ ("Sharpe-Maximizer", "High-Activity") show the highest mean weight change (3.61\%), as they are designed to chase returns, while presets with a high diversification penalty ($\beta$) or holding coefficient ($\lambda$) (such as "Ultra-Diversified") show exceptionally low mean weight changes (0.07\%–1.90\%). This test provides clear empirical evidence that the QSW framework not only overcomes the fragility of MPT but also provides an interpretable, tunable mechanism to control the model's sensitivity to input noise. The next experiment broadens the evidence: a full 625-point grid search pinpoints globally robust parameter regions, and a 30-universe robustness study confirms that these findings generalize beyond this fixed top-100 universe.

\subsection{Experiment 2: comprehensive grid search}

Having established in Experiment 1 that all six QSW strategy presets behave robustly and outperform MPT's 1-year model, we now search the full four-dimensional parameter space \(\boldsymbol{\theta} = (\alpha,\beta,\lambda,\omega)\) to answer three key questions: (i) Does an even better configuration exist than the presets? (ii) How sensitive is performance to each hyper-parameter? (iii) Can we extract simple design rules for practitioners?

This experiment performs a systematic grid search over the 625 parameter combinations defined by: $\alpha,\beta,\lambda \in \{0.1,\,5,\,50,\,100,\,500\}$ and $\omega \in \{0.2,\,0.4,\,0.6,\,0.8,\,1.0\}$, yielding $625$ unique combinations. All 625 combinations are evaluated using the Parameter Exploration Setup described in the Methods (2018–2024 backtest, top-100 universe, and quarterly rebalancing). The entire 625-point grid search is performed twice: once using the \textbf{1-year} ("short-memory") rolling training window, and a second time using the \textbf{2-year} ("long-memory") window. This results in a total of $625 \times 2 = 1{,}250$ full, independent back-tests. For each of the 1,250 back-tests, we record the full set of performance metrics (Sharpe ratio, volatility, MDD, turnover, efficiency $\mathcal{E}$, and HHI).

\begin{table}[htbp] 
\centering
\caption{Descriptive statistics across the full $625$-point grid.}
\label{tab:grid_summary} 
\begin{tabular}{lrrrrrr}
\toprule
\multirow{2}{*}{\textbf{Metric}} &
\multicolumn{3}{c}{\textbf{1-year window}} &
\multicolumn{3}{c}{\textbf{2-year window}} \\
\cmidrule(r){2-4}\cmidrule(l){5-7}
& Min & Median & Max & Min & Median & Max \\
\midrule
Sharpe                    & 0.85 & 0.96 & 0.99 & 0.86 & 0.95 & 0.99 \\
Turnover (\%)             & 0.03 & 56.42 & 248.42 & 0.01 & 48.56 & 212.27 \\
Efficiency (Sharpe/Turn.) & 0.35 & 1.71  & \textbf{95.72} & 0.41 & 1.94 & \textbf{97.50} \\
\bottomrule 
\end{tabular} 
\end{table}

\begin{table}[htbp]
\centering
\caption{Top 10 QSW configurations from the 625-point grid search, compared against benchmarks.}
\label{tab:top_sharpe_configs}
\begin{tabular}{lccccccccc}
\hline
\multicolumn{10}{c}{\textbf{(a) 1-year training (sorted by Sharpe ratio)}} \\
\hline
\textbf{Rank} & \textbf{$\alpha$} & \textbf{$\beta$} & \textbf{$\lambda$} & \textbf{$\omega$} & \textbf{Sharpe} & \textbf{Final Val.} & \textbf{HHI} & \textbf{Turn.(\%)} & \textbf{Efficiency} \\
\hline
1  & 0.1 & 500 & 0.1  & 1.0 & 0.989++ & 3.02+ & 0.010 & 2.9 & 25.6 \\
2  & 0.1 & 500 & 5.0  & 1.0 & 0.989++ & 3.02+ & 0.010 & 3.3 & 23.0 \\
3  & 0.1 & 500 & 0.1  & 0.8 & 0.988++ & 3.01+ & 0.010 & 1.9 & 33.8 \\
4  & 0.1 & 500 & 0.1  & 0.6 & 0.988++ & 3.01+ & 0.010 & 1.7 & 36.3 \\
5  & 0.1 & 500 & 5.0  & 0.6 & 0.988++ & 3.01+ & 0.010 & 1.9 & 34.5 \\
6  & 0.1 & 500 & 0.1  & 0.4 & 0.988++ & 3.00+ & 0.010 & 1.6 & 37.6 \\
7  & 0.1 & 500 & 5.0  & 0.8 & 0.988++ & 3.01+ & 0.010 & 2.1 & 31.5 \\
8  & 0.1 & 500 & 5.0  & 0.4 & 0.988++ & 3.00+ & 0.010 & 1.7 & 36.3 \\
9  & 0.1 & 500 & 0.1  & 0.2 & 0.988++ & 3.00+ & 0.010 & 1.5 & 39.6 \\
10 & 0.1 & 500 & 5.0  & 0.2 & 0.988++ & 3.00+ & 0.010 & 1.6 & 38.6 \\
\hline
1/N & -- & -- & -- & -- & 0.980 & 2.98 & 0.010 & 0.0 & $\infty$ \\
MPT & -- & -- & -- & -- & 0.860 & 4.56 & 0.268 & 480.9 & 0.18 \\
\hline
\\
\multicolumn{10}{c}{\textbf{(b) 2-year training (sorted by Sharpe ratio)}} \\
\hline
\textbf{Rank} & \textbf{$\alpha$} & \textbf{$\beta$} & \textbf{$\lambda$} & \textbf{$\omega$} & \textbf{Sharpe} & \textbf{Final Val.} & \textbf{HHI} & \textbf{Turn.(\%)} & \textbf{Efficiency} \\
\hline
1  & 0.1 & 500 & 0.1   & 1.0 & 0.989+ & 3.01+ & 0.010 & 1.4 & 41.6 \\
2  & 0.1 & 500 & 50.0  & 0.2 & 0.989+ & 2.99+ & 0.010 & 10.4 & 8.7 \\
3  & 0.1 & 500 & 5.0   & 1.0 & 0.989+ & 3.01+ & 0.010 & 1.6 & 38.4 \\
4  & 0.1 & 500 & 0.1   & 0.8 & 0.988+ & 3.00+ & 0.010 & 0.9 & 50.7 \\
5  & 0.1 & 500 & 5.0   & 0.8 & 0.988+ & 3.00+ & 0.010 & 1.1 & 47.9 \\
6  & 0.1 & 500 & 0.1   & 0.6 & 0.988+ & 3.00+ & 0.010 & 0.9 & 53.4 \\
7  & 0.1 & 500 & 5.0   & 0.6 & 0.988+ & 3.00+ & 0.010 & 0.9 & 51.6 \\
8  & 0.1 & 500 & 0.1   & 0.4 & 0.988+ & 2.99+ & 0.010 & 0.8 & 54.2 \\
9  & 0.1 & 500 & 100.0 & 0.2 & 0.988+ & 2.98+ & 0.010 & 25.2 & 3.8 \\
10 & 0.1 & 500 & 5.0   & 0.4 & 0.988+ & 2.99+ & 0.010 & 0.9 & 53.1 \\
\hline
1/N & -- & -- & -- & -- & 0.980 & 2.98 & 0.010 & 0.0 & $\infty$ \\
MPT & -- & -- & -- & -- & 1.360 & 5.84 & 0.254 & 320.1 & 0.42 \\
\hline
\end{tabular}
\end{table}

The grid search results provide a powerful and conclusive answer to our three research questions by revealing the QSW's "smart" hybrid nature. The model is not a sequential "two-step" process, but a \textbf{simultaneous dual-channel optimizer} that intelligently balances its classical and quantum components to achieve a robust, efficient portfolio. As noted, Tables~\ref{tab:grid_summary} and \ref{tab:top_sharpe_configs} provide a strong and conclusive answer to our first research question. First, QSW is structurally robust. The descriptive statistics in Table~\ref{tab:grid_summary} show that the QSW framework is inherently stable. In the 1-year window, the \emph{worst-performing} QSW configuration (Sharpe 0.85) is still comparable to the MPT benchmark (0.86). Furthermore, the \emph{median} QSW Sharpe ($\approx$ 0.96) is vastly superior, proving that the model delivers strong performance across the majority of its parameter space, in stark contrast to the brittle MPT solution. Second, QSW converges to a smart 1/N-like portfolio. The most critical finding comes from Table~\ref{tab:top_sharpe_configs}. In \emph{both} the 1-year and 2-year windows, the Top-10 best-performing QSW configurations are effectively identical: they all have minimal return-chasing ($\alpha=0.1$) and maximal diversification penalty ($\beta=500$). The QSW model, when allowed to search the entire parameter space, consistently identifies an almost infinitesimally traded, 1/N-like portfolio as the most robust solution in this universe. Its performance (Sharpe 0.989, HHI 0.010, Turnover $<$ 3\%) is a near-perfect match for the 1/N benchmark. Third, QSW correctly identifies MPT's "paper" victory. The difference between the 1-year and 2-year results perfectly frames the QSW's value. In the 1-Year case, MPT (0.86) fails due to estimation error. QSW (0.989) discovers that 1/N (0.980) is the superior strategy and outperforms \emph{both} benchmarks. In the 2-Year case, MPT (1.36) stabilizes and finds a high-turnover, high-concentration "paper" victory. QSW, faced with the same data, \emph{still} identifies the 1/N-like state as its optimal solution. It correctly avoids MPT's operationally unfeasible (320\% turnover) solution and selects the \textbf{most practical and efficient} strategy.

\begin{figure}[htbp]
\centering
\caption{Hyper-parameter heatmaps (1-year training). These plots reveal the "design rules" of the QSW model, showing how $\alpha$, $\beta$ and $\lambda$ control the portfolio's behavior.}
\label{fig:param_heatmaps}
\subfloat[Sharpe ratio]{
    \includegraphics[width=0.31\textwidth]{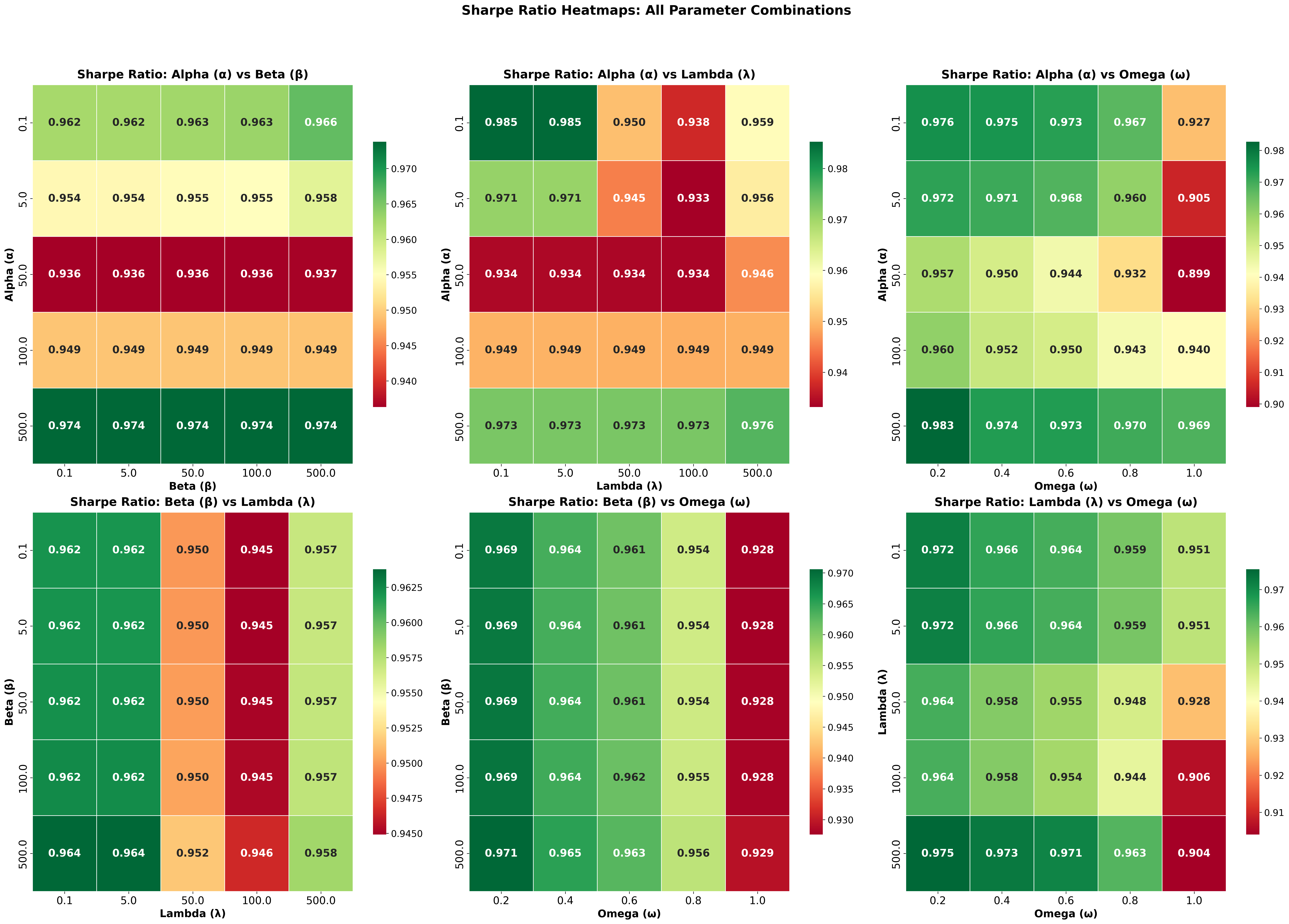}
    \label{fig:heatmap_sharpe}
}
\hfill
\subfloat[HHI (concentration)]{
    \includegraphics[width=0.31\textwidth]{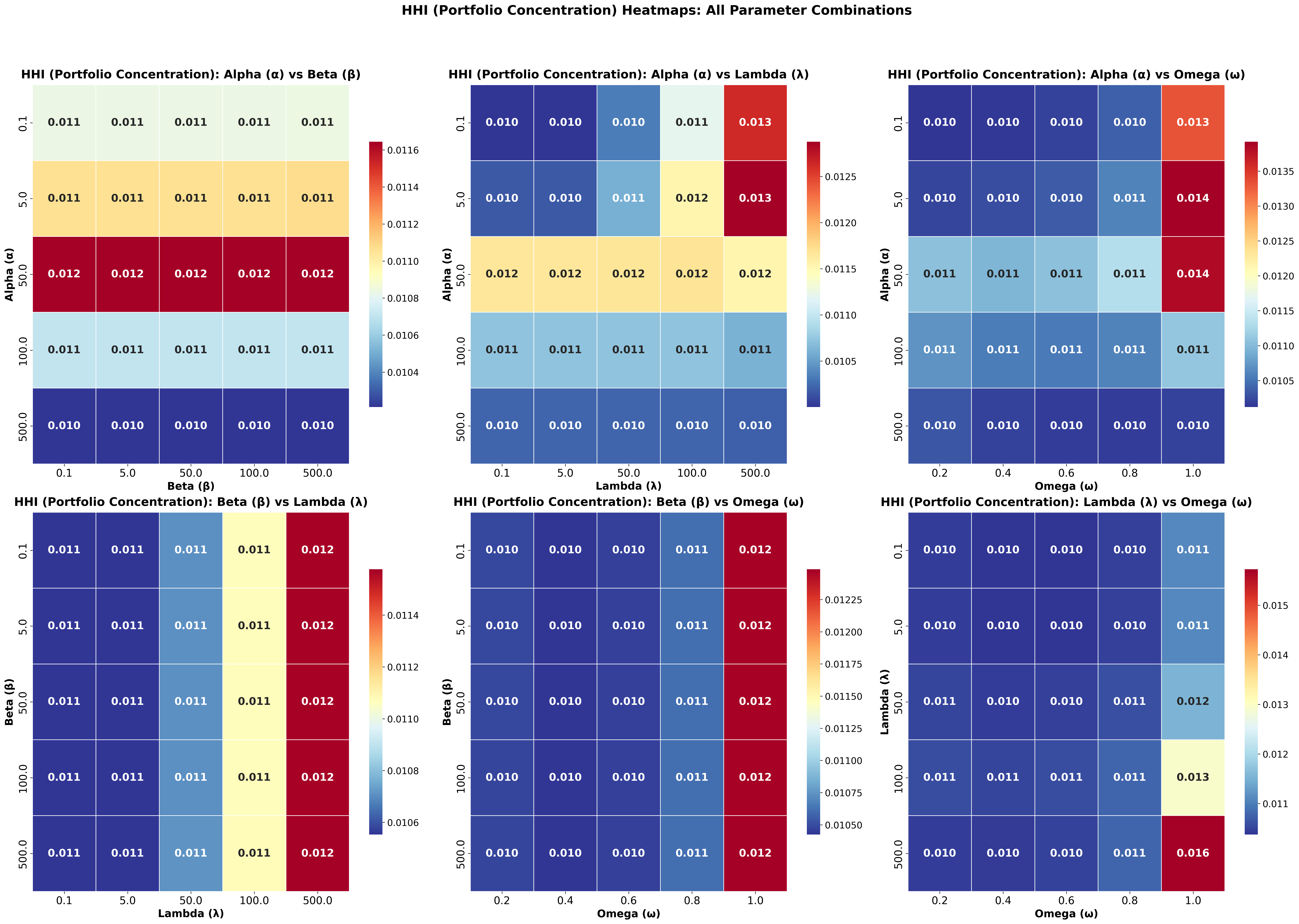}
    \label{fig:heatmap_hhi}
}
\hfill
\subfloat[Efficiency (Sharpe/Turn.)]{
    \includegraphics[width=0.31\textwidth]{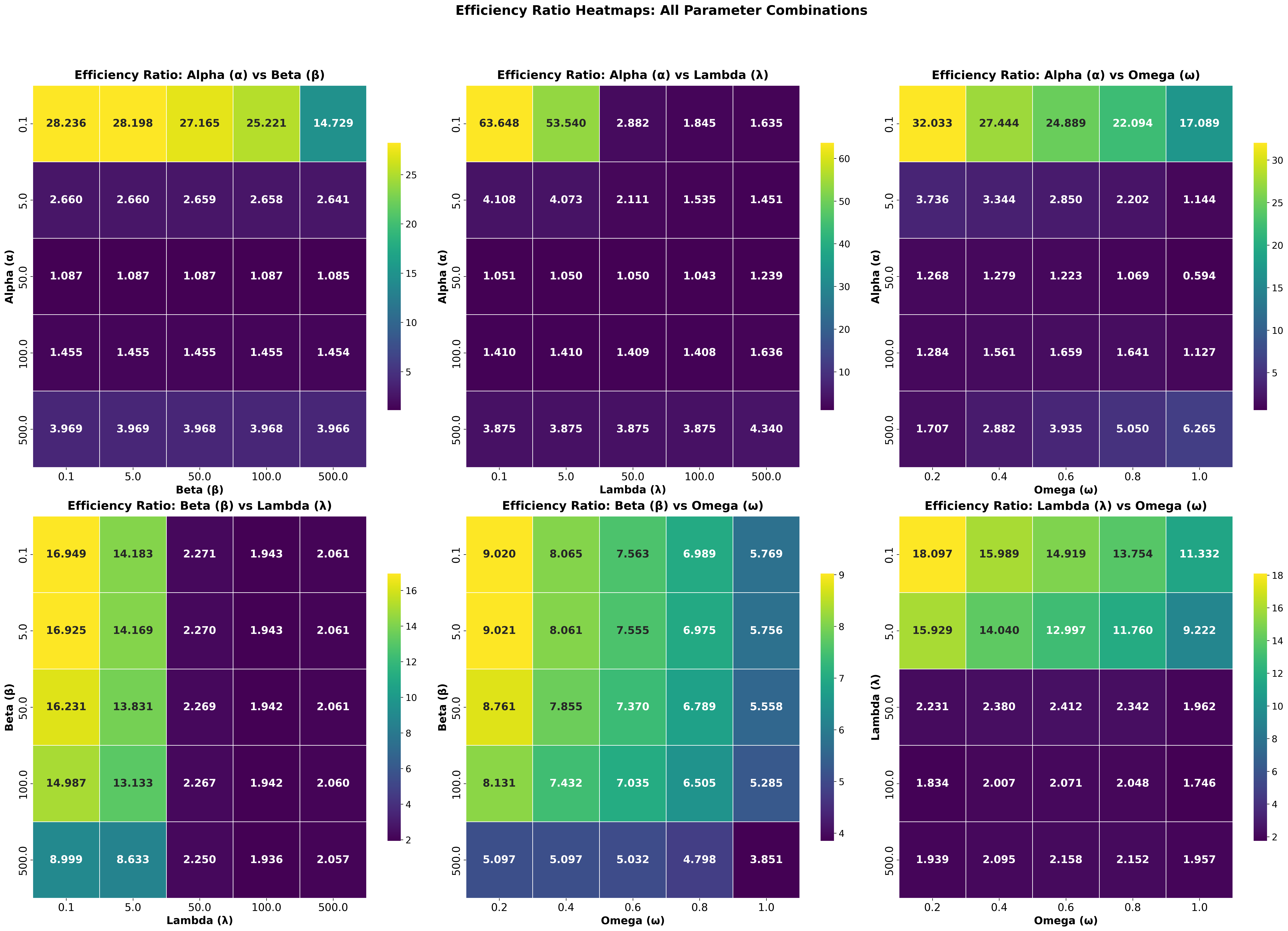}
    \label{fig:heatmap_efficiency}
}
\end{figure}

\begin{figure}[htbp] 
\centering 
\includegraphics[width=0.6\textwidth]{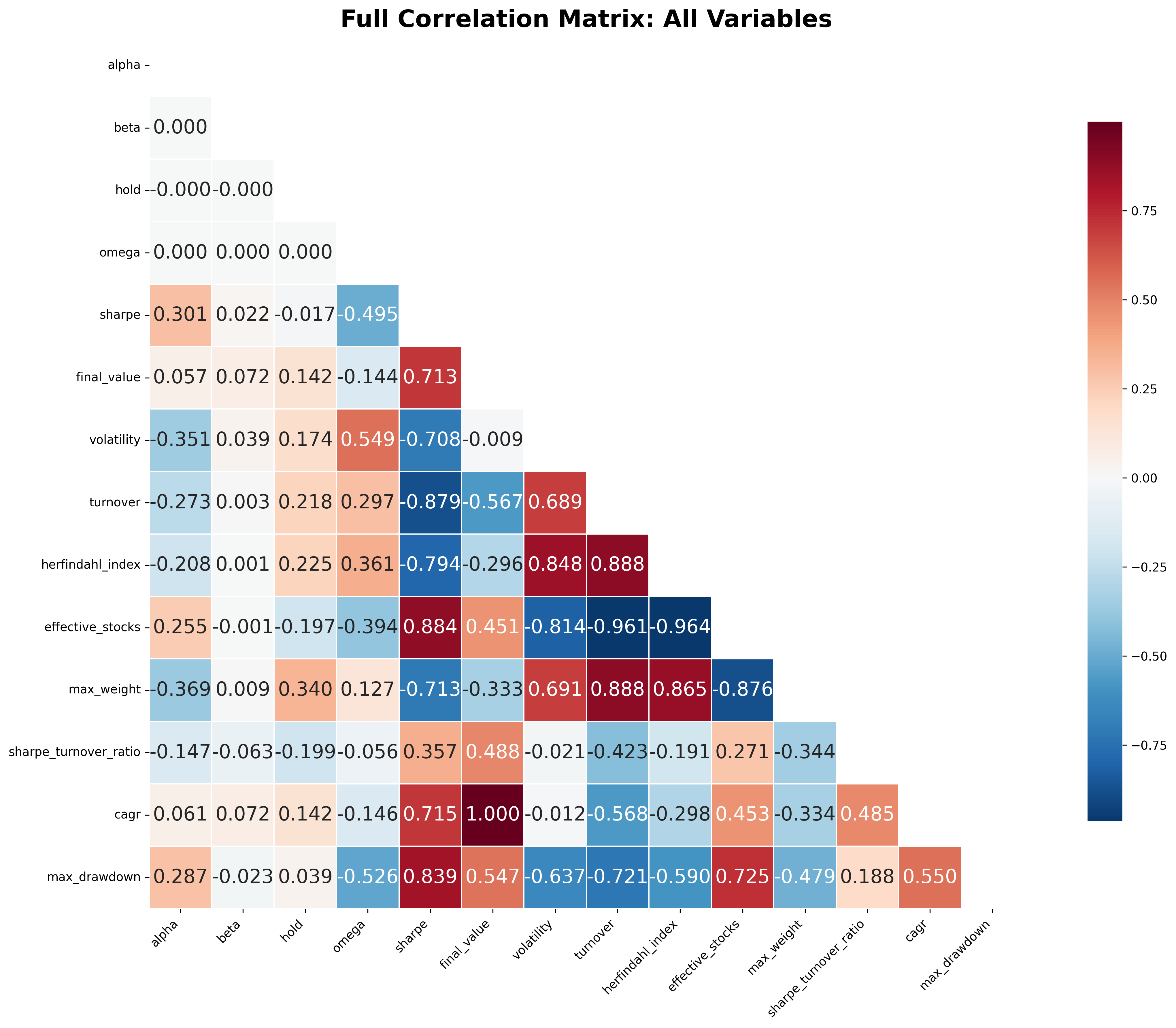} 
\caption{Full hyper-parameter and metric correlation matrix (1-year training). Note the near-zero correlation for Sharpe, confirming its robustness, and the strong, opposing correlations for $\alpha$ and $\beta$ on HHI and Efficiency.}
\label{fig:param_correlation} 
\end{figure}

Our analysis of the tables has answered our first question: the optimal strategy is the "smart 1/N" state. To visualize the global impact of each hyper-parameter, we plot the full 625-run parameter space. The 2D heatmaps (Figure~\ref{fig:param_heatmaps}) and the full correlation matrix (Figure~\ref{fig:param_correlation}) answer our second and third questions, showing how the model finds this state by using its dual-channel design. The 2-year data, which are nearly identical, are provided in the Supplementary Information and confirm that these conclusions are robust to the training window choice.

Figure~\ref{fig:param_heatmaps} makes the QSW’s structural behavior immediately visible. Most strikingly, the concentration panel in Figure~\ref{fig:heatmap_hhi} shows that, across the \emph{entire} 625-run grid, the HHI remains confined to an extremely narrow band: from a floor of 0.010 (dark blue) to a maximum of 0.014 (bright green). In other words, even the ``worst-case'' combination of $(\alpha,\beta,\lambda,\omega)$ never produces anything close to a concentrated portfolio. By comparison, the MPT benchmark sits at HHI $=0.268$, more than $20\times$ higher. This gap is not a tuning artefact but a structural property: the QSW framework is \emph{incapable} of generating the kind of brittle, single-bet allocations that MPT routinely produces. The Sharpe and Efficiency panels (Figures~\ref{fig:heatmap_sharpe} and \ref{fig:heatmap_efficiency}) complete this picture by revealing how the hyper-parameters act as \emph{behavioral knobs} rather than sources of instability. Sharpe ratios remain tightly clustered in the $0.94$--$0.99$ range over most of the grid, indicating that performance is remarkably insensitive to moderate perturbations of $(\alpha,\beta,\lambda,\omega)$. Efficiency, by contrast, varies much more strongly along the $\lambda$ and $\omega$ axes: higher values of $\lambda$ (the ``holding'' parameter) and $\omega$ systematically reduce $\mathcal{E}$ by lowering Sharpe and/or increasing turnover, while leaving the underlying diversification almost untouched.

These visual impressions are quantified by the full correlation matrix in Figure~\ref{fig:param_correlation}. Within our tested range $\omega\in[0.2,1.0]$, $\omega$ exhibits a strong negative correlation with Sharpe and a positive correlation with HHI, confirming that it primarily acts as a \emph{noise} parameter: smaller $\omega$ values lead, on average, to more diversified, higher-quality portfolios, whereas larger $\omega$ values inject additional stochasticity that slightly raises concentration and erodes risk-adjusted returns. In contrast, $\lambda$ shows a pronounced negative correlation with Efficiency while having only a mild effect on Sharpe, confirming its role as a trade-off knob between turnover and implementability rather than a driver of raw performance. Together, the heatmaps and the correlation matrix show that QSW’s smart 1/N solution is not a single fine-tuned point, but a broad, structurally robust region of the parameter space.

\begin{figure}[htbp]
\centering
\caption{Hyper-parameter correlation bar charts for 1-year (top row) and 2-year (bottom row) training. These plots reveal the "design rules" of the QSW model and how its behavior adapts to the input data.}
\label{fig:param_correlations}
\subfloat[Sharpe correlations (1Y)]{
    \includegraphics[width=0.31\textwidth]{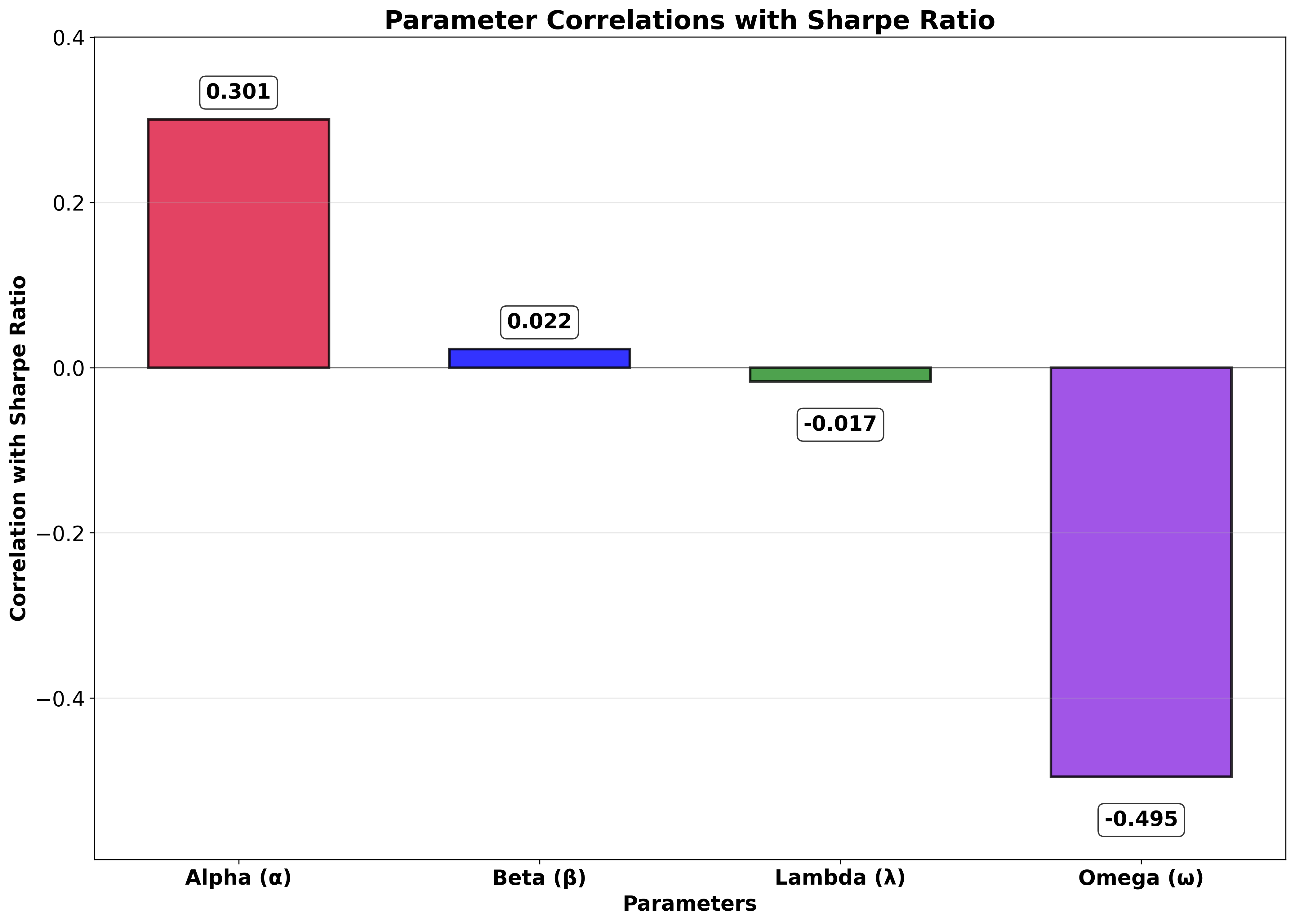}
    \label{fig:corr_sharpe_1y}
}
\hfill
\subfloat[HHI correlations (1Y)]{
    \includegraphics[width=0.31\textwidth]{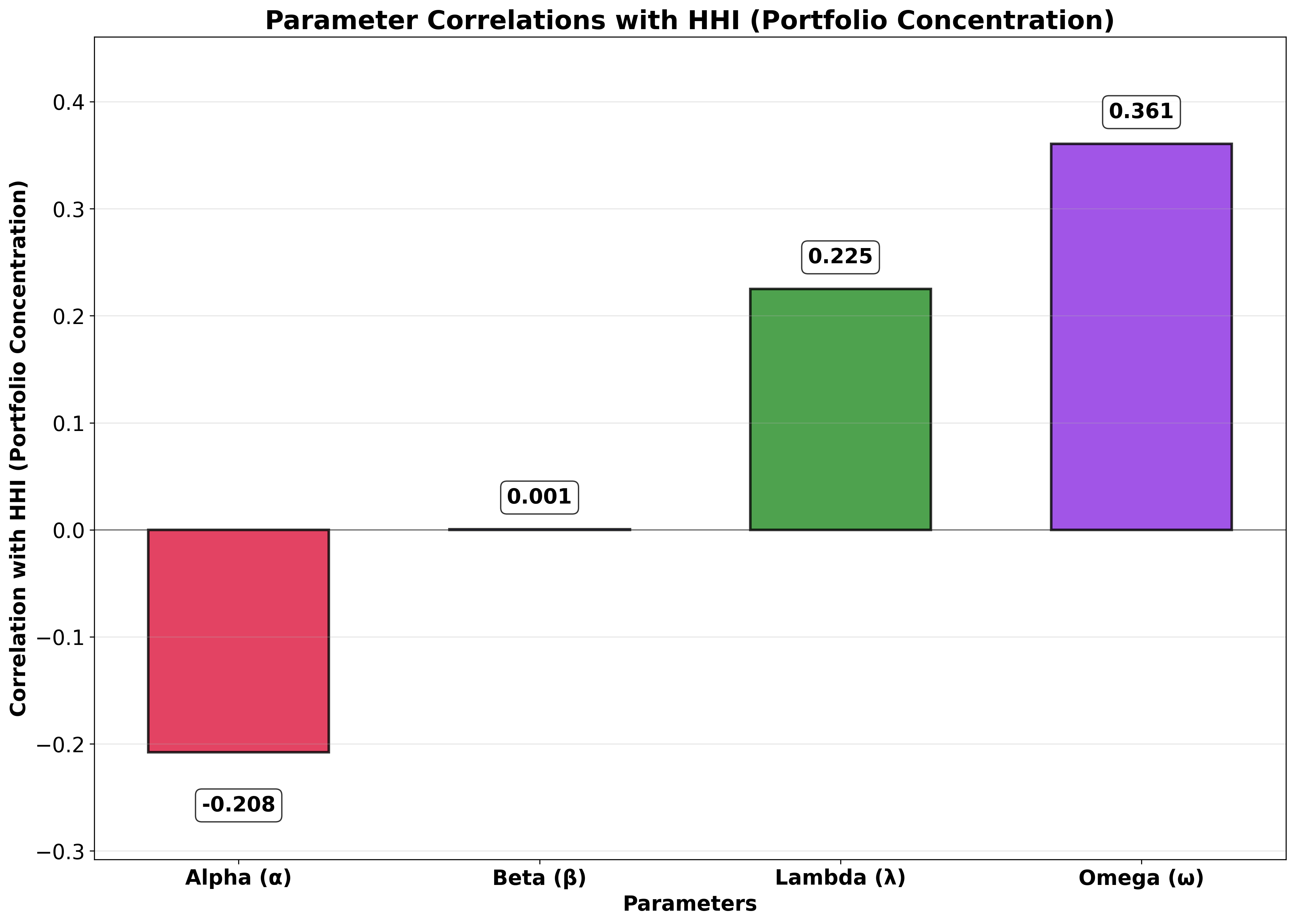}
    \label{fig:corr_hhi_1y}
}
\hfill
\subfloat[Efficiency correlations (1Y)]{
    \includegraphics[width=0.31\textwidth]{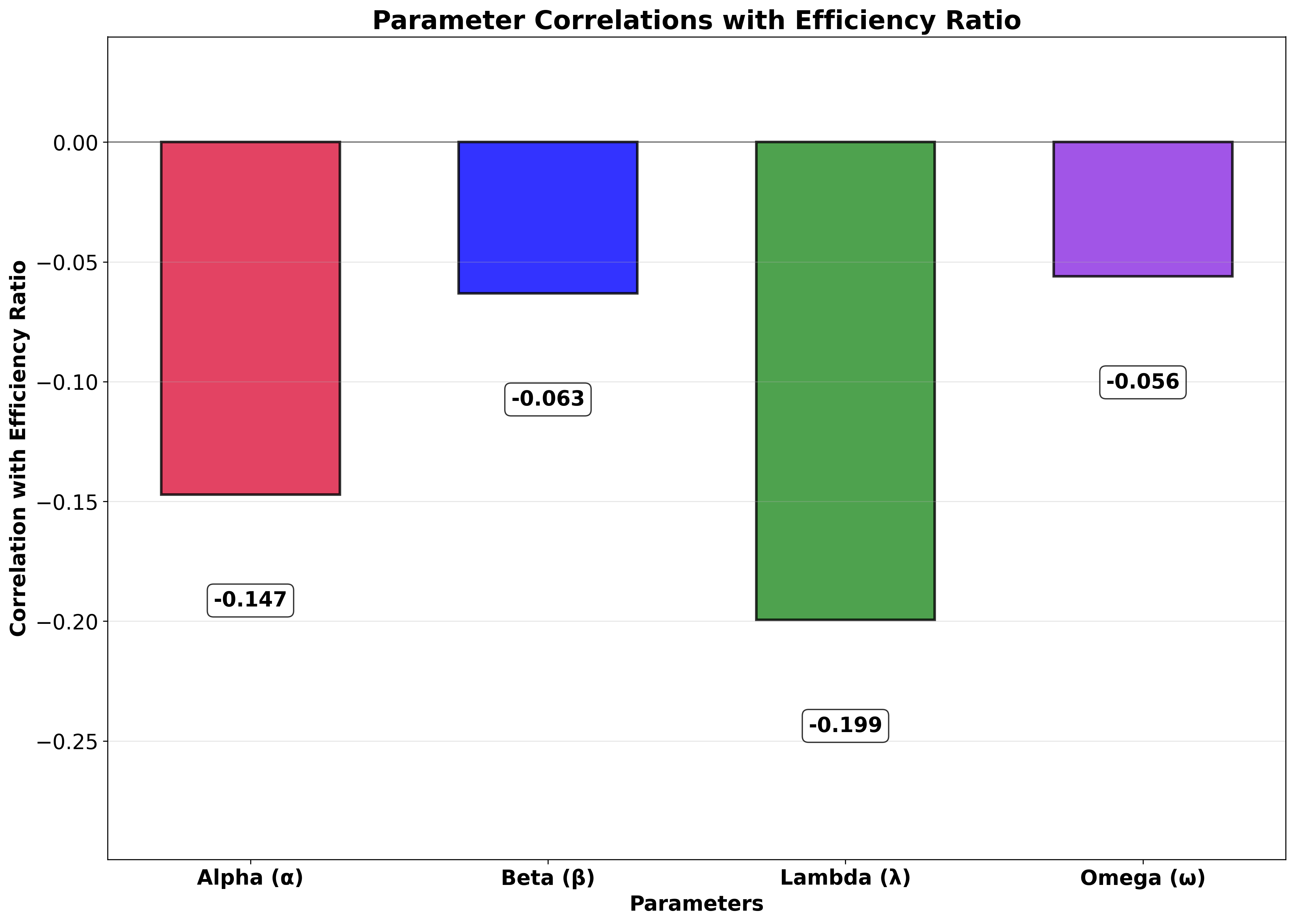}
    \label{fig:corr_efficiency_1y}
}
\\
\subfloat[Sharpe correlations (2Y)]{
    \includegraphics[width=0.31\textwidth]{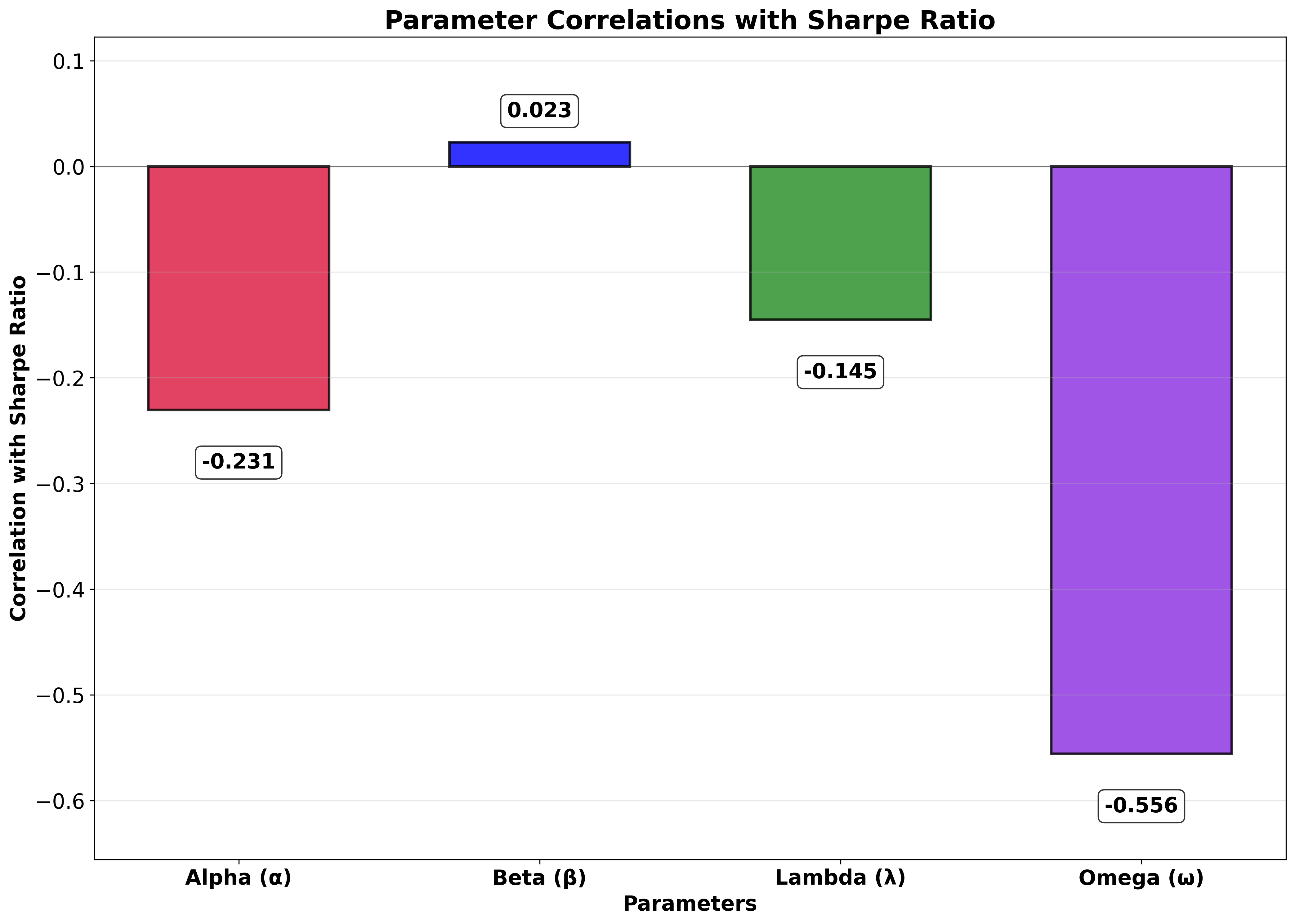}
    \label{fig:corr_sharpe_2y}
}
\hfill
\subfloat[HHI correlations (2Y)]{
    \includegraphics[width=0.31\textwidth]{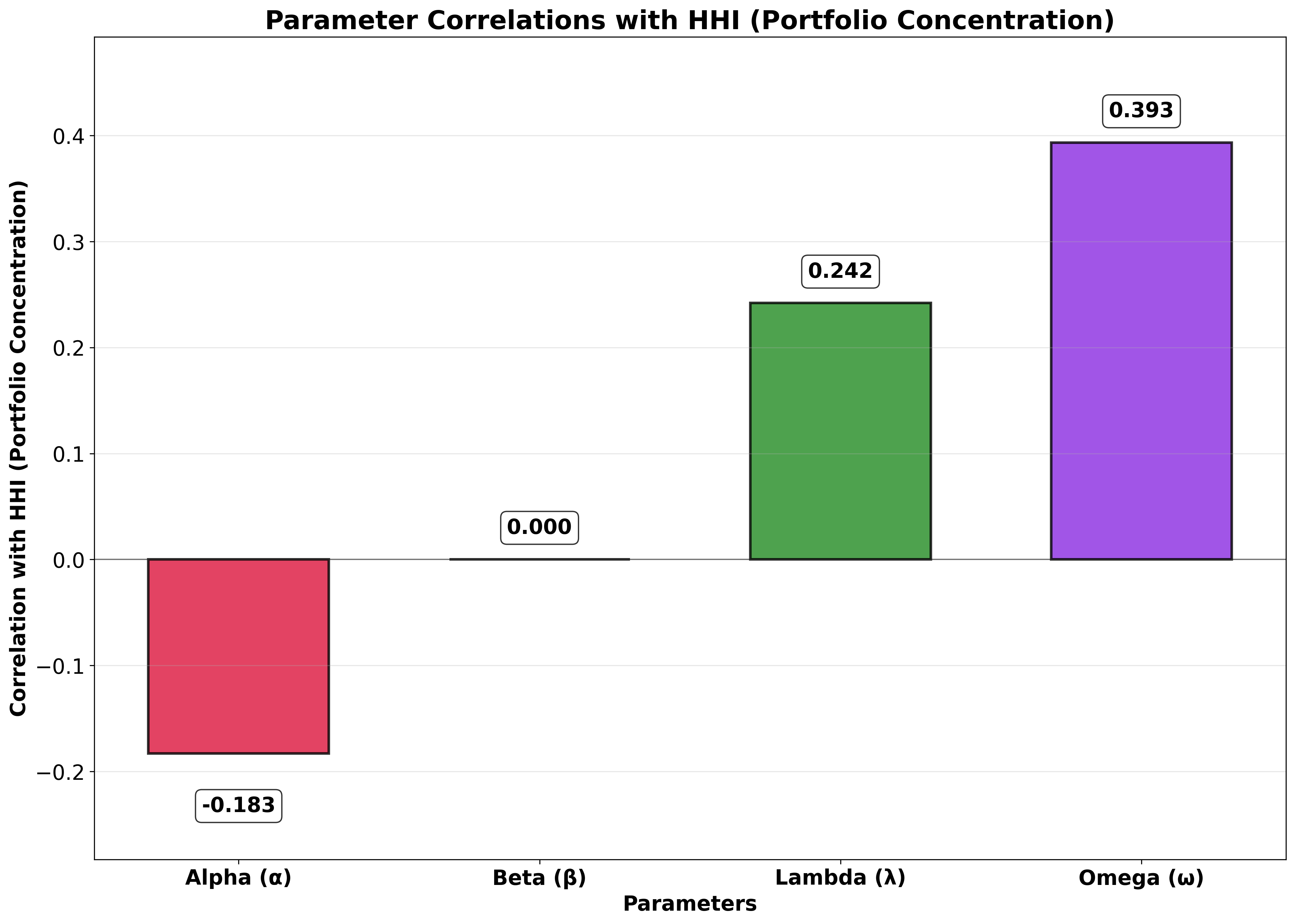}
    \label{fig:corr_hhi_2y}
}
\hfill
\subfloat[Efficiency correlations (2Y)]{
    \includegraphics[width=0.31\textwidth]{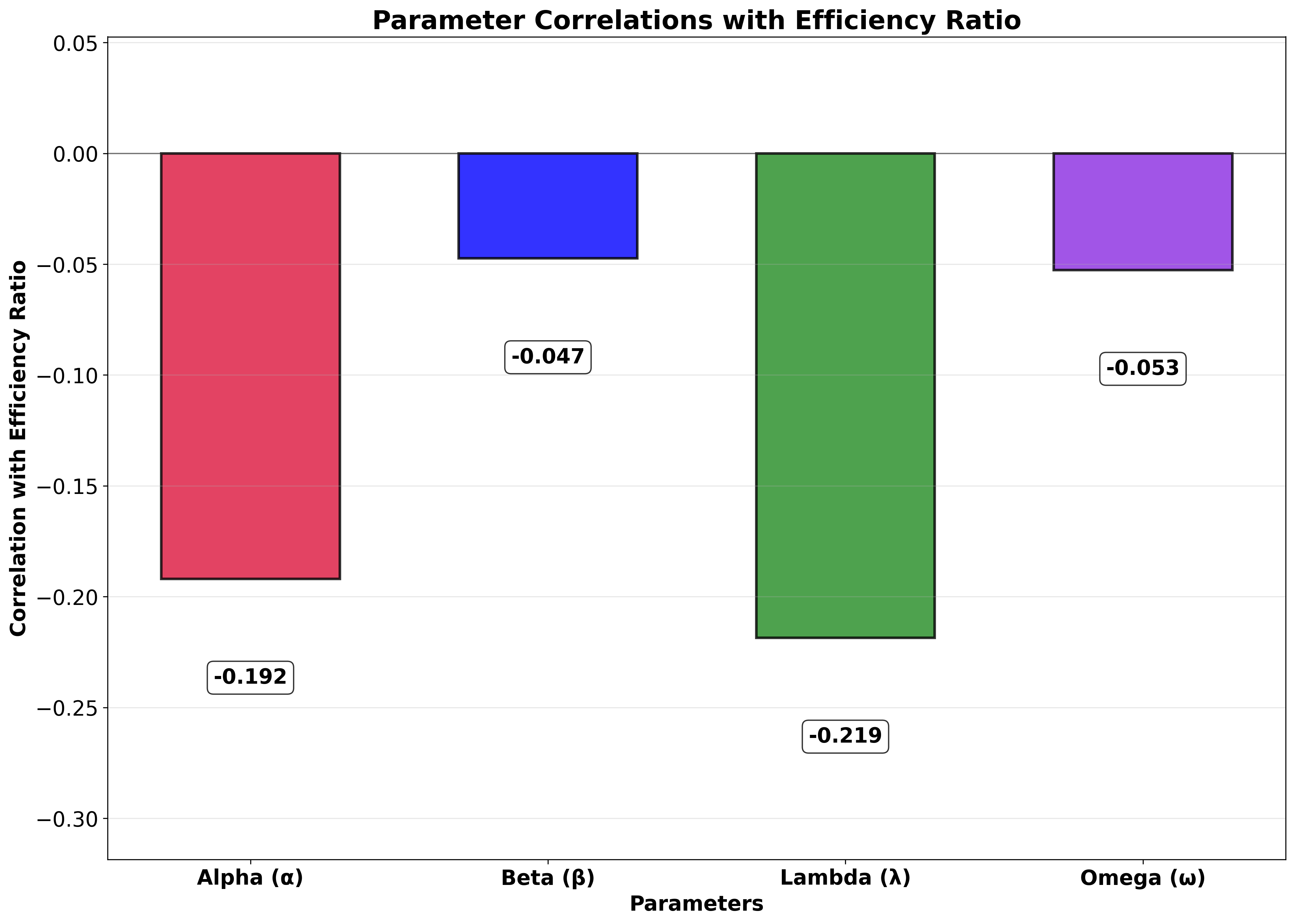}
    \label{fig:corr_efficiency_2y}
}
\end{figure}

To visually confirm these findings and answer our second research question, we analyze the global impact of each hyper-parameter. Figure~\ref{fig:param_correlations} plots the Pearson correlations between the four control knobs ($\alpha, \beta, \lambda, \omega$) and our key metrics. We present the plots for both the 1-year "noisy" regime and the 2-year "stable" regime, as their comparison provides insight into the QSW model's adaptive, structural behavior.

The correlation plots provide four practical design rules for practitioners. First, $\omega$ is the main ``noise'' dial, with a mild impact on concentration. Among the four knobs, $\omega$ shows the strongest and most stable pattern across both horizons. Within our tested range $\omega\in[0.2,1.0]$, it is \emph{strongly negatively correlated} with Sharpe (1Y: $-0.495$, 2Y: $-0.556$), and \emph{positively correlated} with HHI (1Y: $0.225$, 2Y: $0.393$). In other words, higher $\omega$ injects additional stochastic noise into the walk, which slightly increases concentration and systematically erodes risk-adjusted returns. Conversely, lower $\omega$ values tend, on average, to produce more diversified, higher-quality portfolios, although the magnitude of the HHI effect is small relative to the structural diversification enforced by the QSW dynamics. Second, $\lambda$ is the ``turnover'' knob, trading off implementability against efficiency. As expected from its construction, $\lambda$ behaves as a holding/inertia parameter. Across both windows, it exhibits a \emph{moderate positive correlation} with HHI (1Y: $0.225$, 2Y: $0.242$) and a \emph{moderate negative correlation} with Efficiency (1Y: $-0.199$, 2Y: $-0.219$), while its correlation with Sharpe is weak. Forcing the model to ``stay where it is'' (large $\lambda$) suppresses trading and reduces turnover, but at the cost of higher concentration and lower efficiency. In other words, $\lambda$ does not change the core performance level so much as it controls the trade-off between practical implementability and how aggressively the strategy rebalances back toward its preferred low-HHI state. Third, $\alpha$ is a regime-dependent return bias, not a standalone source of alpha. The return-preference parameter $\alpha$ shows only weak and \emph{inconsistent} correlations with Sharpe (positive in the 1-year window, slightly negative in the 2-year window), while maintaining a tendency to \emph{reduce} HHI and Efficiency when increased. This indicates that, over the 2018–2024 sample, classical return signals behave more like a noisy, regime-dependent bias than a robust source of outperformance: raising $\alpha$ tilts the walk toward past winners and can slightly improve Sharpe in the short-memory regime, but it does not produce a reliable, horizon-independent Sharpe uplift. Fourth, $\beta$ looks neutral globally, but is decisive in the winning region. At first glance, the global bar charts in Figure~\ref{fig:param_correlations} suggest that the diversification penalty $\beta$ has almost \emph{no} correlation with HHI or Sharpe (correlations near zero). Taken in isolation, this would seem to imply that $\beta$ does not matter. However, the Top-10 list in Table~\ref{tab:top_sharpe_configs} tells a very different story: \emph{every} optimal configuration sets $\beta=500$, its maximum value. This is not a contradiction but a selection effect: once $\alpha$ and $\omega$ are in a favorable regime (low return-chasing, low noise), the QSW is already close to fully diversified (HHI $\approx 0.01$), so further changes in $\beta$ have little marginal impact and its global correlation is washed out. Conditioning on the high-Sharpe region, however, reveals the true design rule: \textbf{strong diversification pressure (large $\beta$) is a prerequisite for ending up in the structurally robust, “smart 1/N’’ regime.}

The grid search results provide empirical support for the dual-channel framework introduced earlier. Taken together, the Top-10 list, the full-grid statistics, and the visual diagnostics reveal a genuinely hybrid design: a classical channel that pins the portfolio to a structurally robust baseline, and a quantum channel that fine-tunes how this baseline is implemented in practice. The Top-10 list in Table~\ref{tab:top_sharpe_configs} is the most direct piece of evidence for the classical channel acting as a stabilizer. Across both the 1-year and 2-year windows, \emph{every} optimal configuration chooses the same classical settings: a minimal return-chasing coefficient ($\alpha=0.1$) and the strongest possible diversification penalty ($\beta=500$). In other words, the model “learns’’ that the 2018–2024 market is effectively noisy from a classical perspective and responds by turning the accelerator $\alpha$ almost off, while slamming the diversification brake $\beta$ to its maximum. The 2D heatmaps in Figure~\ref{fig:param_heatmaps} and the correlation plots in Figure~\ref{fig:param_correlations} confirm this picture: once $\alpha$ is kept small and $\beta$ is large, the portfolio is structurally forced into an almost perfectly diversified “smart 1/N’’ state (HHI $\approx 0.010$) across a broad region of the grid. The classical channel therefore acts as a stabilizer that locks in a robust risk–return profile by overpowering unstable return signals. Given this stabilized baseline, the quantum channel, controlled by $\omega$, is then used to fine-tune how the same robust allocation is implemented. This can be seen most clearly in Table~\ref{tab:top_sharpe_configs}(a) when we fix $(\alpha,\beta,\lambda) = (0.1, 500, 0.1)$ and only vary $\omega$: as $\omega$ decreases from $1.0$ to $0.2$, the Sharpe ratio remains essentially unchanged (0.989 $\rightarrow$ 0.988); turnover decreases (from 2.9\% to 1.5\%); efficiency increases (from 25.6 to 39.6); while HHI stays fixed at 0.010 in all cases. In other words, once the classical channel has pinned the portfolio to a highly diversified, smart 1/N-like state, the quantum channel does not search for a different target allocation; instead, it \emph{searches for cheaper ways} to reach and maintain that allocation. Consistent with the global correlations, lower $\omega$ reduces noisy fluctuations and turnover, improving cost-efficiency without sacrificing Sharpe or diversification. This is the practical “quantum exploration’’ advantage: the QSW simultaneously explores multiple paths to the same robust state and selects the one with the most attractive implementation profile.

Experiment~2 shows that, for a fixed period (2018–2024), the QSW can reliably discover a structurally robust, smart 1/N-like regime and fine-tune its implementation cost through its hybrid quantum–classical channels. The natural next question is whether this behavior is truly \emph{adaptive}: can the same mechanism re-learn an appropriate balance between the classical channel $(\alpha,\beta,\lambda)$ and the quantum channel $(\omega)$ as market conditions change? Experiment~3 addresses this question by moving from a static to a fully dynamic setting. Instead of identifying a single “best’’ parameter set over 2018–2024, we run the entire 625-point grid search at \emph{every} quarterly rebalancing date from 1990 to 2024. At each quarter, the QSW optimizer re-estimates its optimal mix of classical and quantum parameters for that specific market regime. This 34-year, rolling experiment serves as the ultimate test of the QSW’s robustness and “smartness’’ over multiple cycles, crises, and structural shifts.

\subsection{Experiment 3: robustness validation (1990--2024)}

This Phase~2 experiment stress-tests the QSW as a dynamic hybrid optimizer over a 34-year period (1990--2024) that spans multiple major crises, including the Dot-com bubble, the 2008 global financial crisis, and the 2020 COVID crash. Building directly on the insights from Phase~1, we now ask a harder question: can the same dual-channel mechanism \emph{continuously re-learn} an appropriate balance between its classical channel $(\alpha,\beta,\lambda)$ and quantum channel $(\omega)$ as market conditions change, and can it do so \emph{robustly across different equity universes}? To answer this, Experiment~3 combines a dynamic dual-channel strategy with multi-universe testing.

The experimental setup involves a dynamic adaptive model that re-optimizes channels every quarter. At each quarterly rebalancing date from 1990 to 2024, the model runs a full 625-point grid search on the most recent 2-year training window. It then selects the parameter set that best balances the classical and quantum channels $(\alpha,\beta,\lambda,\omega)$ under that specific market regime. In contrast to Experiment~2, which identified a single “best-fit’’ configuration for 2018–2024, this experiment allows the QSW optimizer to re-learn its settings every quarter over 34 years. We perform multi-universe testing using 30 independent trials. To ensure that the results are not an artefact of a single, “lucky’’ 100-stock universe, we repeat the entire 34-year dynamic backtest 30 times. As described in the Methods (“Data and experimental setup”), each trial starts from a different, randomly sampled 100-stock subset of the point-in-time S\&P 500 constituents in 1990Q1 and is dynamically maintained: at each quarterly date, delisted stocks are removed and immediately replaced by new names drawn from the contemporaneous S\&P 500 membership. This multi-universe design ensures that our robustness conclusions do not rely on a particular asset selection or survivorship bias.

\begin{table}[htbp]
\centering
\caption{Summary statistics for Experiment 3 (1990--2024), averaged over 30 random S\&P~500 universes.}
\label{tab:exp3_summary}
\begin{tabular}{lcccccc}
\toprule
\textbf{Strategy} & \textbf{CAGR} & \textbf{Volatility} & \textbf{Sharpe} & \textbf{Calmar} & \textbf{Max DD} & \textbf{Turnover} \\
\midrule
QSW      & $23.3\% \pm 3.4\%$ & $20.6\% \pm 1.9\%$ & $0.979 \pm 0.082$ & $0.63 \pm 0.20$ & $38.0\% \pm 5.0\%$ & $201.1\% \pm 11.2\%$ \\
MPT      & $14.6\% \pm 2.4\%$ & $25.8\% \pm 9.7\%$ & $0.480 \pm 0.114$ & $0.28 \pm 0.07$ & $55.1\% \pm 9.7\%$ & $333.1\% \pm 13.8\%$ \\
1/N      & $17.2\% \pm 3.5\%$ & $19.0\% \pm 2.2\%$ & $0.742 \pm 0.083$ & $0.41 \pm 0.19$ & $43.8\% \pm 5.5\%$ & $0.1\% \pm 0.1\%$  \\
S\&P 500 & $8.3\% \pm 0.0\%$  & $18.0\% \pm 0.0\%$ & $0.368 \pm 0.000$ & $0.17 \pm 0.00$ & $47.7\% \pm 0.0\%$ & N/A                \\
\bottomrule
\end{tabular}
\end{table}

Table~\ref{tab:exp3_summary} reports the cross-universe means and standard deviations. Three patterns stand out. First, QSW delivers the highest average CAGR and Sharpe, with markedly lower dispersion than MPT. Second, this is not achieved by taking more risk: QSW’s volatility lies between 1/N and MPT, its average maximum drawdown is smaller than both MPT and the S\&P 500, and its Calmar ratio is materially higher than all benchmarks. Third, QSW attains these outcomes with \emph{less} turnover than MPT, indicating that the QSW optimizer is not simply “over-trading’’ its way to higher returns but improves the overall risk–return–cost profile.

\begin{figure}[htbp]
\centering
\caption{Distribution of risk-adjusted performance and final wealth across 30 random S\&P~500 universes (1990--2024). Each violin summarizes the cross-universe distribution for one strategy; dots show individual universes, and horizontal bars mark the mean.}
\label{fig:exp3_violin_perf}
\subfloat[Sharpe ratio]{%
    \includegraphics[width=0.7\textwidth]{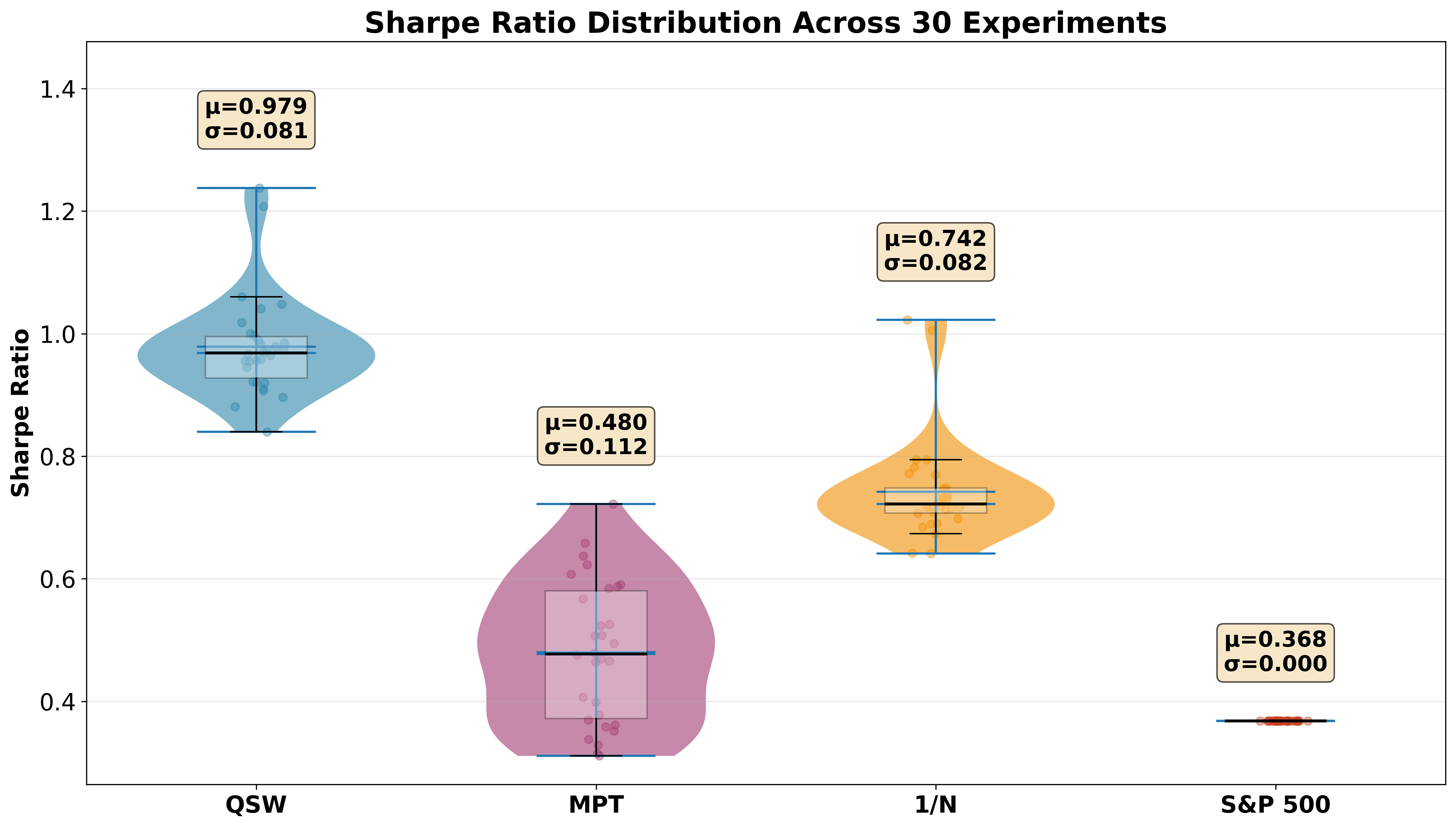}%
}
\hfill
\subfloat[Calmar ratio (CAGR/Max~DD)]{%
    \includegraphics[width=0.7\textwidth]{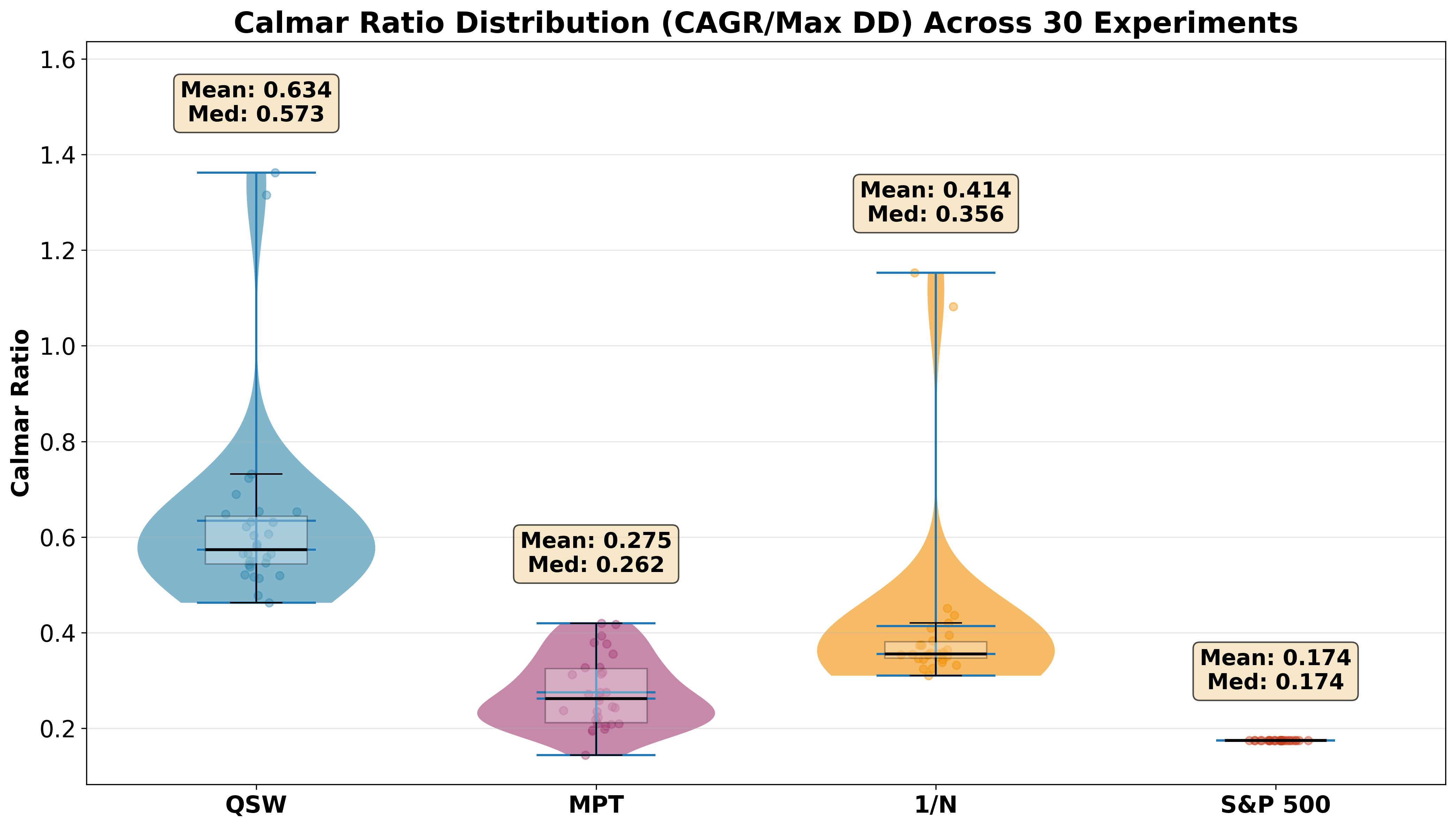}%
}
\hfill
\subfloat[Final portfolio value (log scale)]{%
    \includegraphics[width=0.7\textwidth]{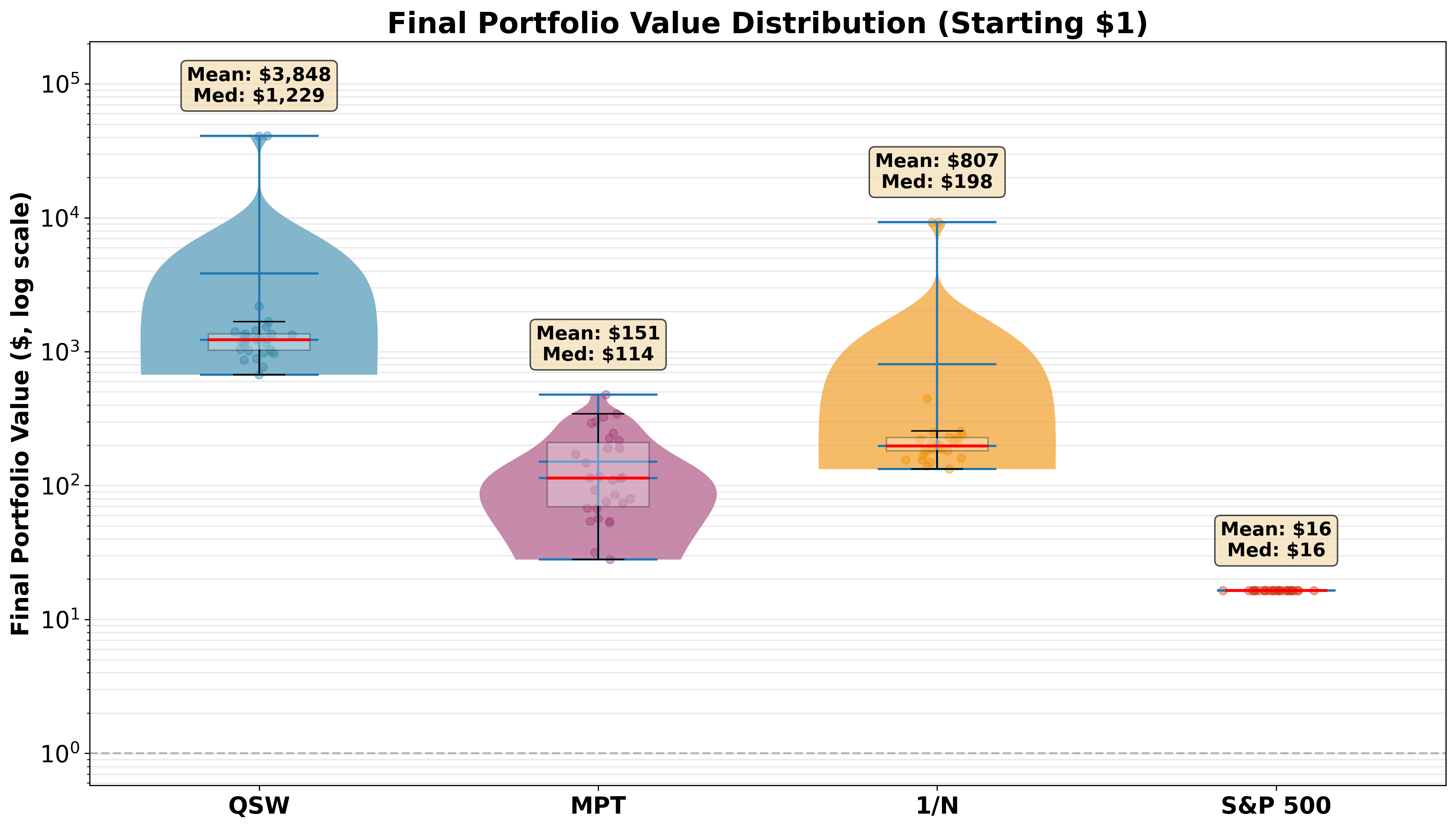}%
}
\end{figure}

Figure~\ref{fig:exp3_violin_perf} complements Table~\ref{tab:exp3_summary} by showing the full cross-universe distributions. The Sharpe violins in Figure~\ref{fig:exp3_violin_perf}a show that QSW dominates not only on average but in most individual universes: the bulk of its mass lies above both 1/N and MPT, with little overlap. Figure~\ref{fig:exp3_violin_perf}b shows the same ordering for Calmar ratios, confirming that the advantage persists after normalizing by maximum drawdown. Finally, the final-value distribution in Figure~\ref{fig:exp3_violin_perf}c illustrates how these differences compound over 34 years: starting from \$1, QSW typically reaches several times the terminal wealth of 1/N and an order of magnitude more than MPT or the index. The medians and interquartile ranges closely track the means, indicating that the outperformance is systematic rather than driven by a few outlier universes.

\begin{figure}[htbp]
\centering
\caption{Risk and trading-cost profile across 30 universes. QSW achieves higher risk-adjusted returns than 1/N and MPT while keeping volatility and drawdowns at moderate levels and trading substantially less than the MPT optimizer.}
\label{fig:exp3_violin_risk}
\subfloat[Annual turnover]{%
    \includegraphics[width=0.7\textwidth]{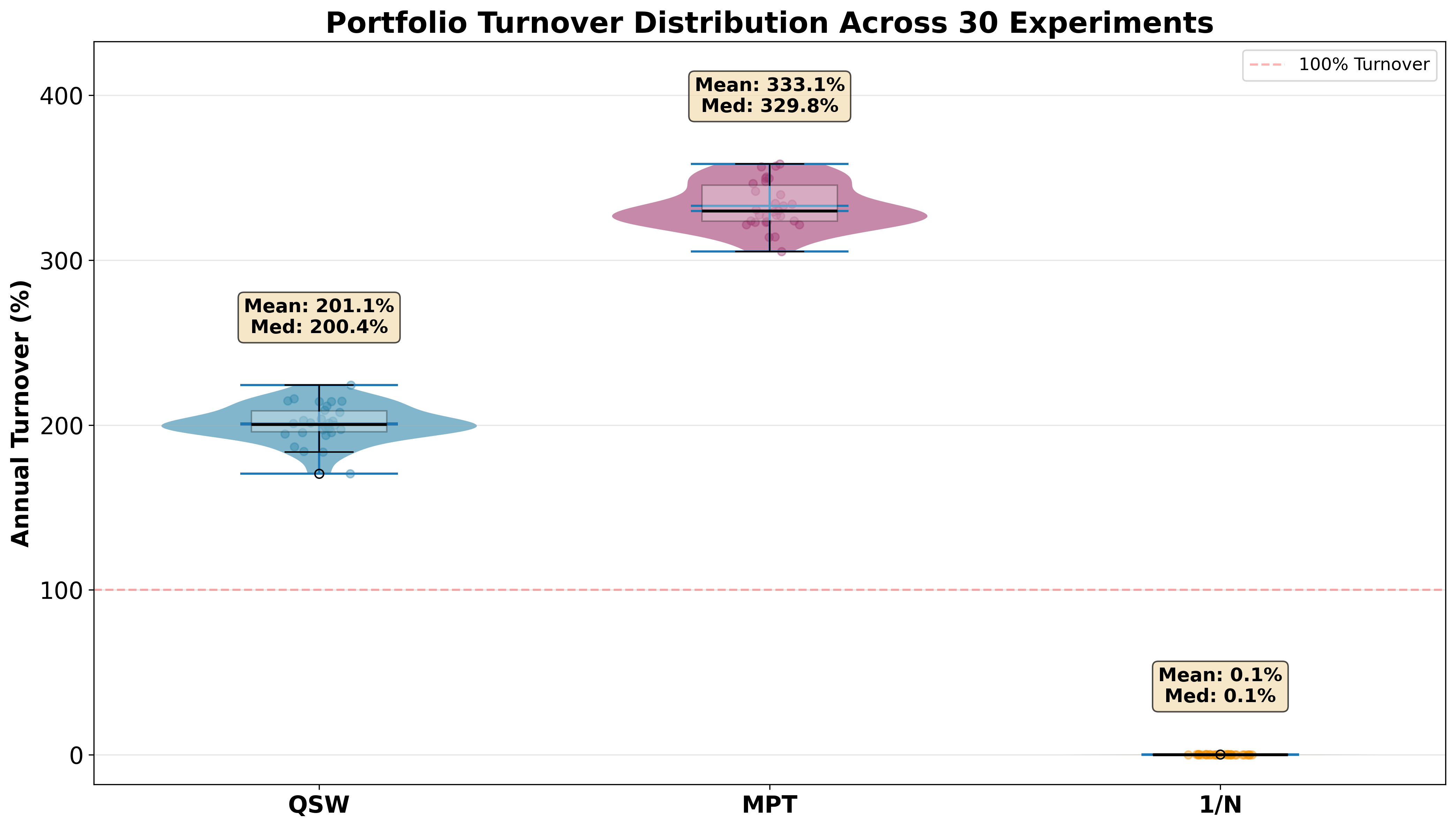}%
}
\hfill
\subfloat[Annual volatility]{%
    \includegraphics[width=0.7\textwidth]{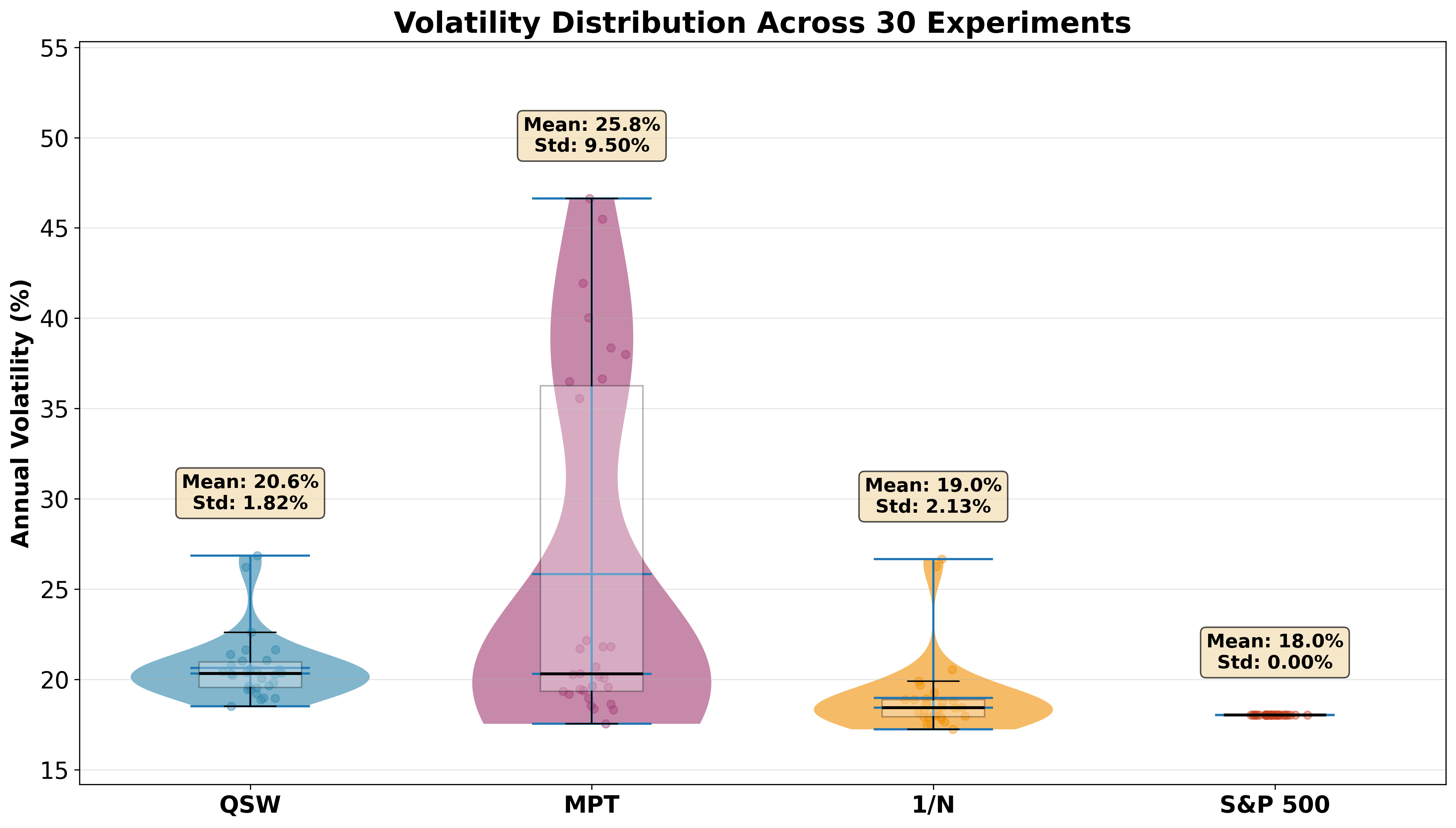}%
}
\hfill
\subfloat[Maximum drawdown]{%
    \includegraphics[width=0.7\textwidth]{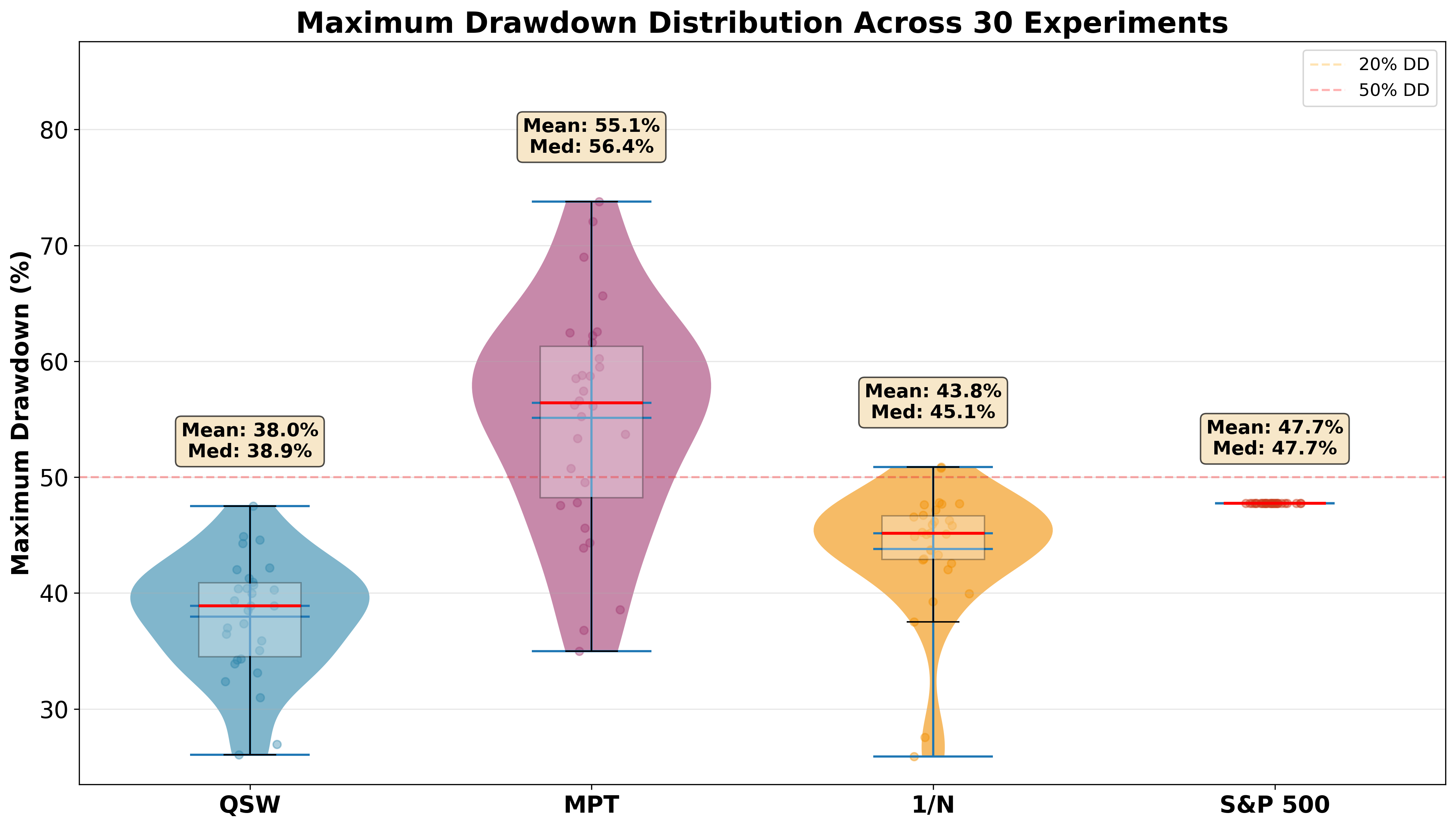}%
}
\end{figure}

Figure~\ref{fig:exp3_violin_risk} examines how QSW earns its outperformance. On volatility, QSW sits between 1/N and MPT, indicating that its higher Sharpe is driven by better risk pricing rather than simply taking more overall risk. On drawdowns, QSW again occupies the favorable middle ground: its maximum drawdowns are consistently shallower than those of MPT and the index, and are either slightly better or comparable to 1/N. Finally, QSW strikes a pragmatic balance on trading activity: it trades substantially less than MPT while, of course, trading more than the nearly buy-and-hold 1/N portfolio. Taken together, these distributions show that the QSW optimizer improves the overall risk–return–cost trade-off rather than relying on extreme leverage, concentration, or turnover.

It is worth noting that the average QSW turnover in Experiment~3 (around 200\% per year) is higher than in the static Phase~1 experiments (typically 20–80\% for the best presets in Experiment~1 and a preset mean of about 40\% in Experiment~2). This is an expected consequence of moving from a fixed parameter configuration to a fully dynamic hybrid optimizer. At each quarter, the model not only updates the portfolio weights in response to new returns, but may also switch to a different point in the $(\alpha,\beta,\lambda,\omega)$ grid. Even if the selected configurations are individually low- or moderate-turnover, the act of re-optimizing and jumping between configurations introduces an additional layer of trading. Crucially, however, QSW still trades substantially less than the MPT benchmark (about 200\% vs.\ 330\% annual turnover on average) while delivering markedly higher Sharpe and Calmar ratios. Hence, the higher turnover in Experiment~3 reflects the cost of adaptivity rather than a loss of control.

\begin{figure}[htbp]
\centering
\caption{Consistency of QSW outperformance across 30 universes. Left: fraction of universes in which QSW outperforms each benchmark on each metric. Right: risk–return scatter plot, with each point corresponding to one universe.}
\label{fig:exp3_win_scatter}
\subfloat[QSW win rates by metric]{%
    \includegraphics[width=0.8\textwidth]{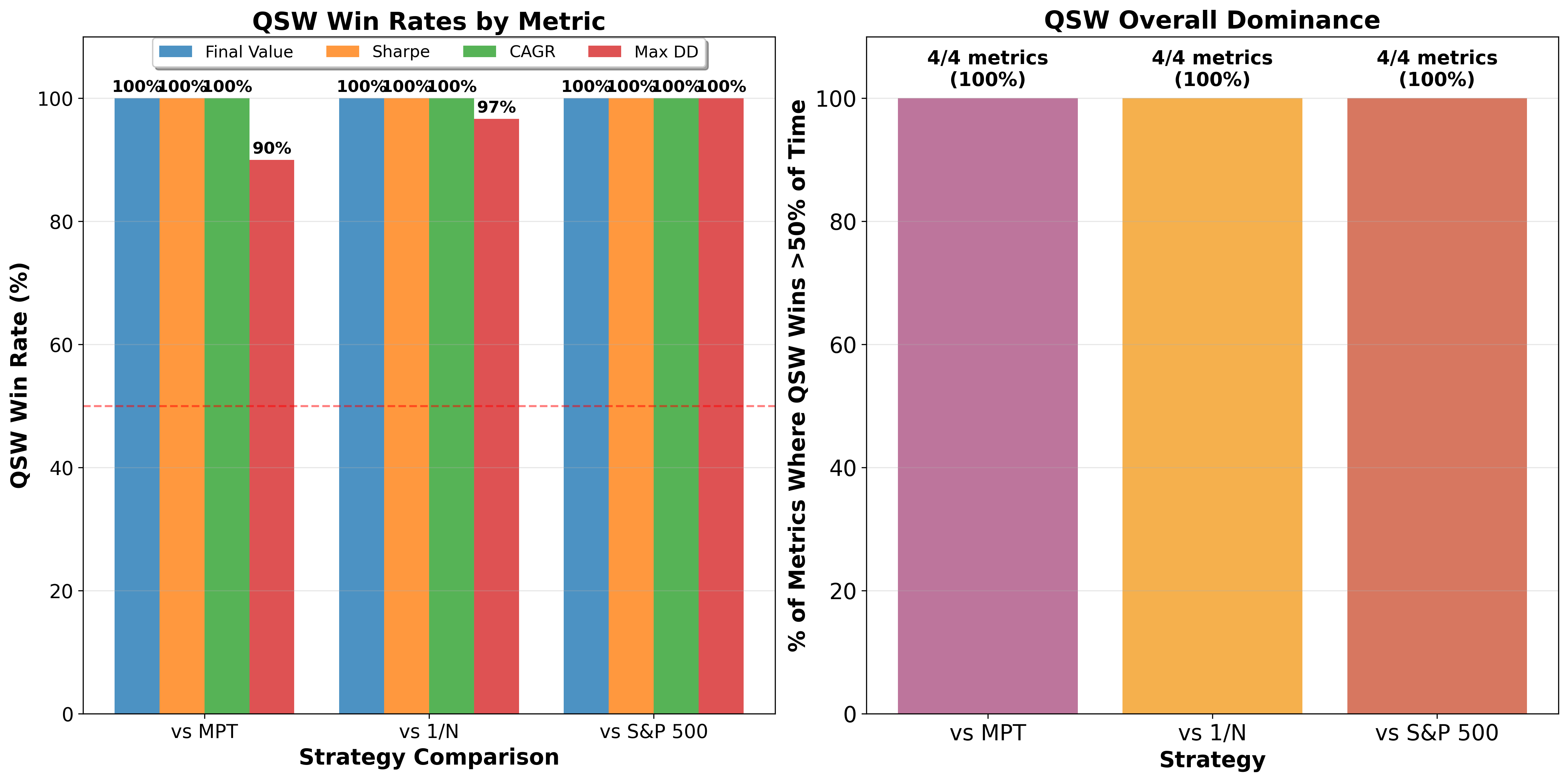}%
}
\hfill
\subfloat[Risk–return scatter (CAGR vs volatility)]{%
    \includegraphics[width=0.8\textwidth]{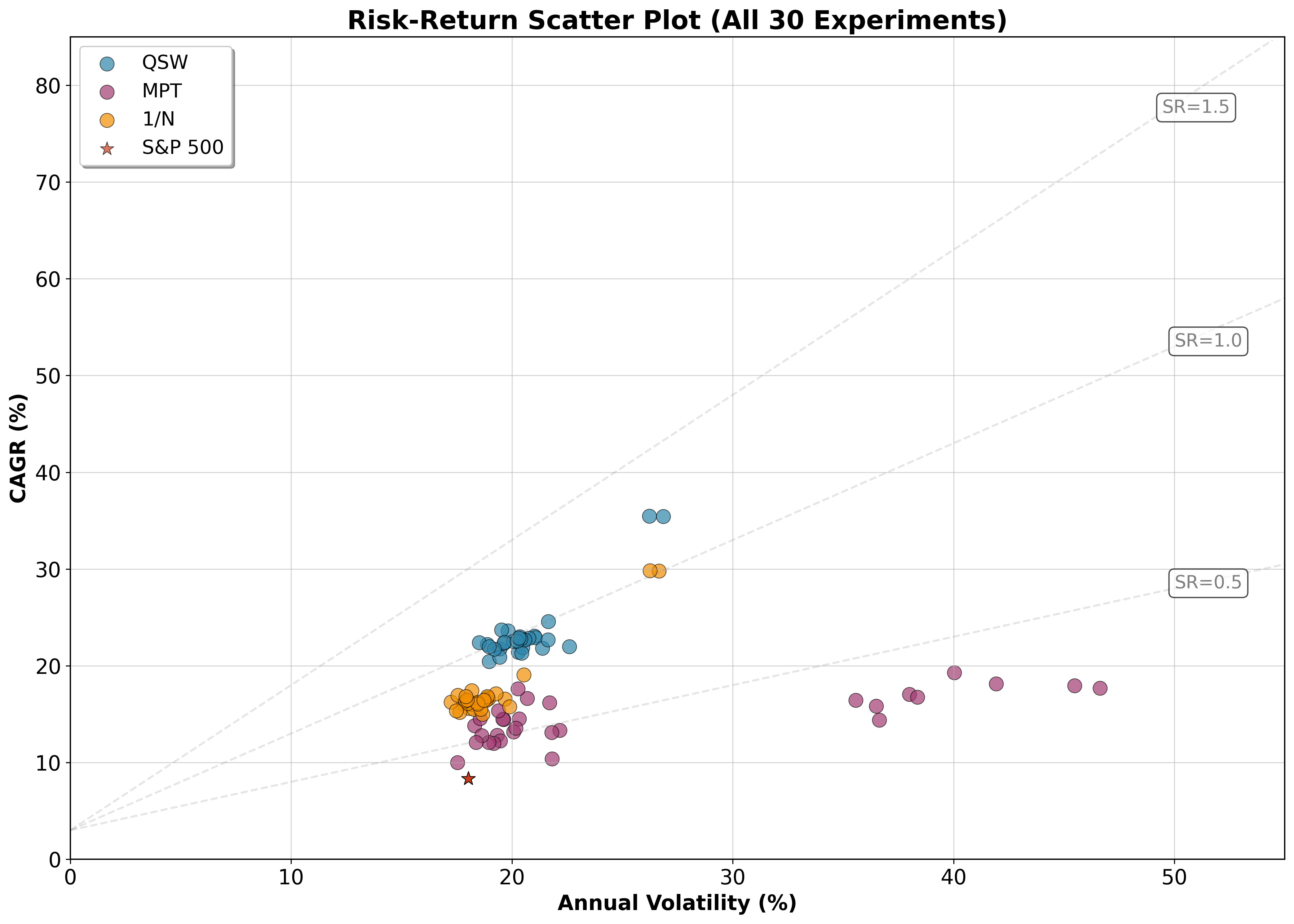}%
}
\end{figure}

The win-rate bars in Figure~\ref{fig:exp3_win_scatter}a summarize the robustness of these results. QSW outperforms MPT, 1/N, and the S\&P 500 in the vast majority of universes across all key metrics (Sharpe, Calmar, CAGR, maximum drawdown, and final value), with win rates close to 100\% in most comparisons. Figure~\ref{fig:exp3_win_scatter}b shows the same message in risk–return space: QSW outcomes form a tight cluster along a high-Sharpe ridge, while 1/N sits on a lower ridge and MPT outcomes are both more dispersed and systematically below the QSW cloud. The index lies at the bottom-left corner, with the lowest CAGR and Sharpe. This confirms that QSW’s advantage is not the result of a few extreme outliers, but a stable shift of the entire risk–return distribution.

\begin{figure}[htbp]
\centering
\includegraphics[width=0.95\textwidth]{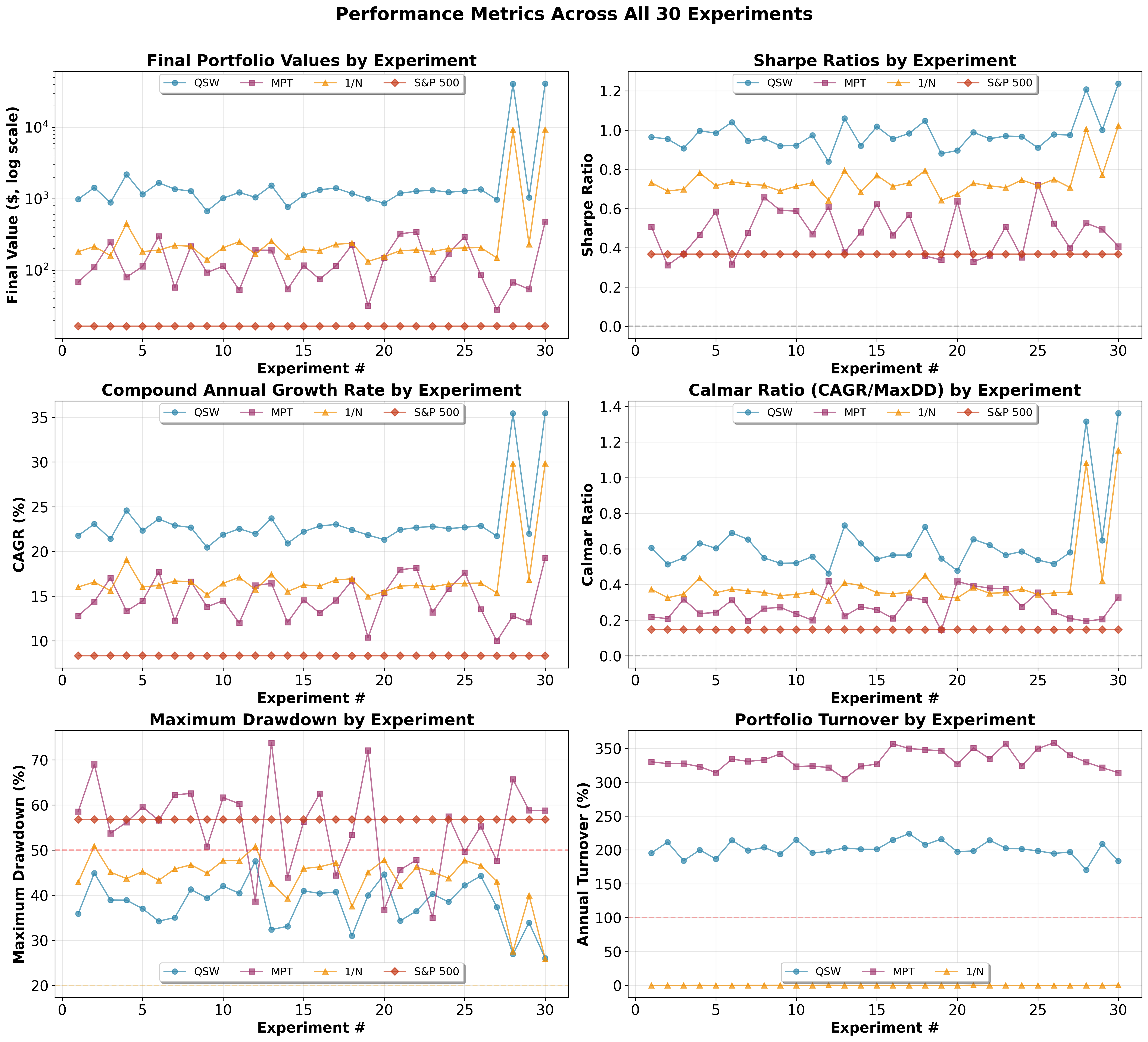}
\caption{Performance metrics across all 30 universes (1990--2024). Each panel plots one metric (final value, Sharpe, CAGR, Calmar ratio, maximum drawdown, and annual turnover) for QSW, MPT, 1/N, and the S\&P~500 as a function of the universe index.}
\label{fig:exp3_by_experiment}
\end{figure}

Figure~\ref{fig:exp3_by_experiment} shows that the \emph{shape} of the QSW curves closely tracks that of the 1/N benchmark across all 30 universes. Whenever the equal-weight portfolio delivers a higher final wealth, Sharpe, or Calmar ratio in a given universe, the QSW strategy tends to sit slightly above it, with a very similar risk and drawdown profile and a moderate increase in turnover. In contrast, the MPT series is both noisier and structurally different, with large swings in volatility, drawdown, and trading activity. This behavior is exactly what one would expect from the “smart 1/N’’ structure identified in Experiment~2: the dynamic QSW optimizer repeatedly re-discovers an almost equal-weight allocation—driven by small $\alpha$, large $\beta$, and low-to-moderate $\omega$—and then adds a thin, data-driven tilt on top of that baseline. Across universes, QSW therefore behaves not as an unstable maximizer, but as a robust, smart version of 1/N: it inherits the diversification and resilience of the naive equal-weight portfolio, while systematically lifting its risk-adjusted performance.

\begin{figure}[htbp]
\centering
\includegraphics[width=0.95\textwidth]{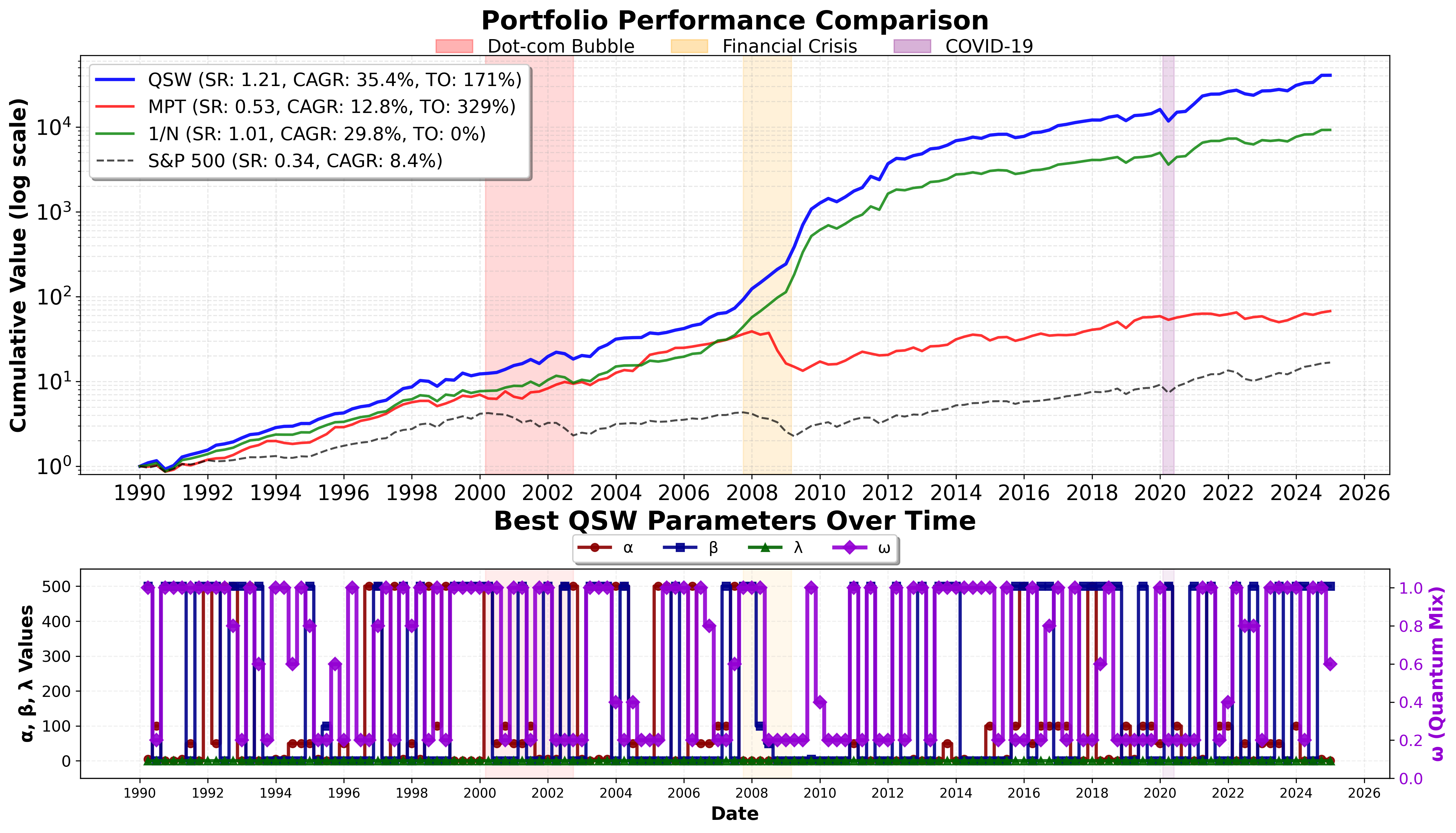}
\caption{Single-universe case study (1990--2024). \textbf{Top:} Cumulative portfolio values (log scale) for QSW, MPT, 1/N, and the S\&P~500, with major crisis periods highlighted. \textbf{Bottom:} Evolution of the QSW parameters $(\alpha,\beta,\lambda,\omega)$ selected by the rolling 2-year grid search at each quarter.}
\label{fig:exp3_case_study}
\end{figure}

While the cross-universe results establish statistical robustness, it is also instructive to inspect a single, representative universe in detail. Figure~\ref{fig:exp3_case_study} illustrates the time evolution of the dynamic QSW strategy on a randomly sampled 100-stock universe, together with the corresponding path of the optimal hyper-parameters.

Several features are worth noting. First, the dynamic QSW strategy consistently stays ahead of both MPT and 1/N, with the performance gaps widening after each major crisis episode (Dot-com, 2008, and COVID-19). This confirms that the gains observed in the cross-universe statistics are not confined to a particular bull market regime, but persist across markedly different environments. Second, the bottom panel shows that the optimizer does not lock into a single parameter setting. Instead, it repeatedly re-learns a familiar pattern: $\alpha$ is kept low most of the time, $\beta$ is frequently driven to its upper bound, and $\omega$ toggles between more classical and more quantum regimes depending on market conditions. During volatile or stressed periods, the model tends to reinforce the diversification channel (large $\beta$, low-to-moderate $\omega$), while in calmer regimes it allows slightly more classical exposure. This behavior is fully consistent with the design rules inferred from the 2018–2024 grid search in Phase~1.

\begin{figure}[htbp]
\centering
\caption{Risk path for the single-universe case study. Left: cumulative drawdowns over time; Right: rolling 8-quarter volatility.}
\label{fig:exp3_case_study_risk}
\subfloat[Drawdown over time]{%
    \includegraphics[width=0.48\textwidth]{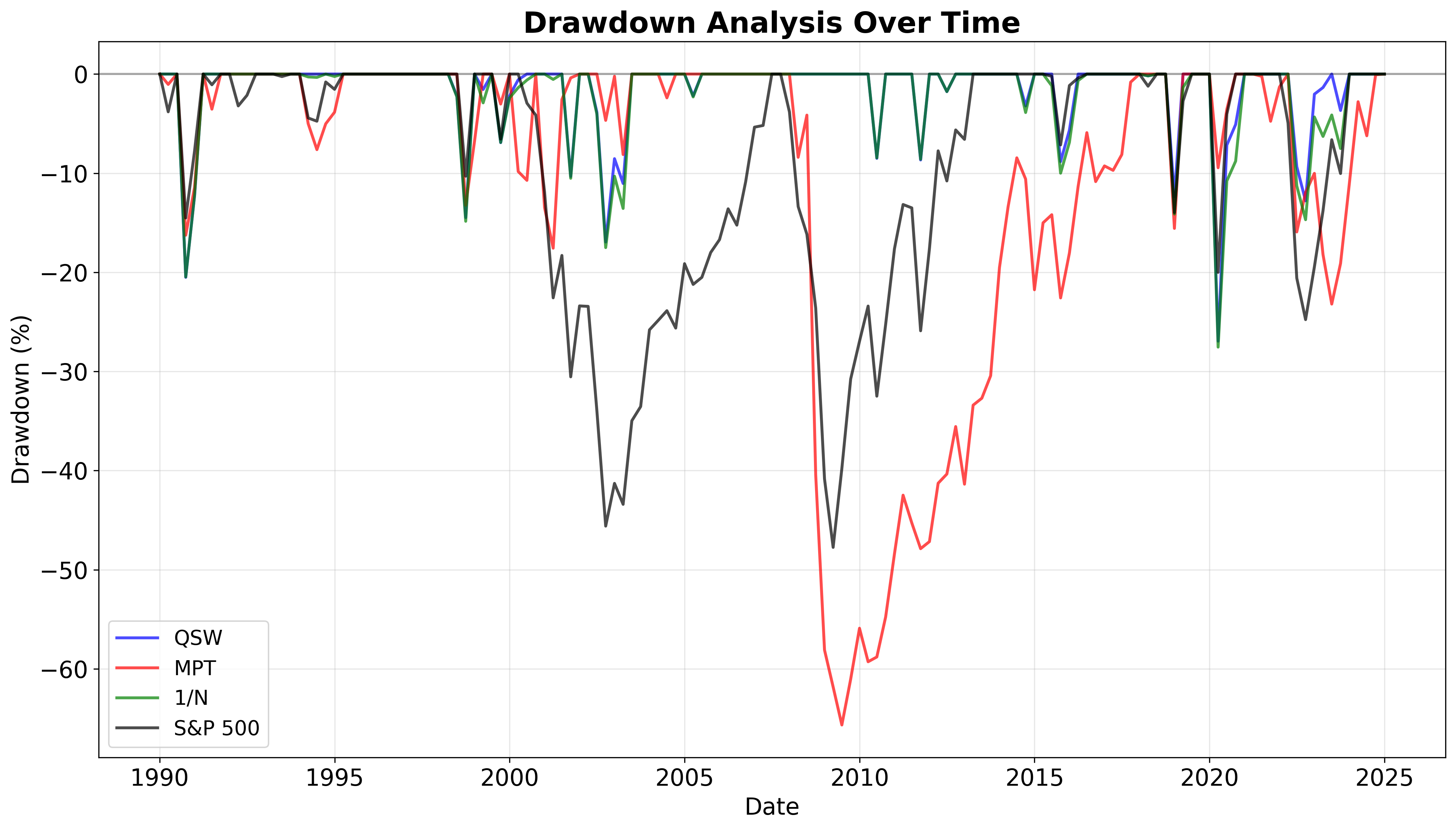}%
}
\hfill
\subfloat[Rolling volatility (8-quarter window)]{%
    \includegraphics[width=0.48\textwidth]{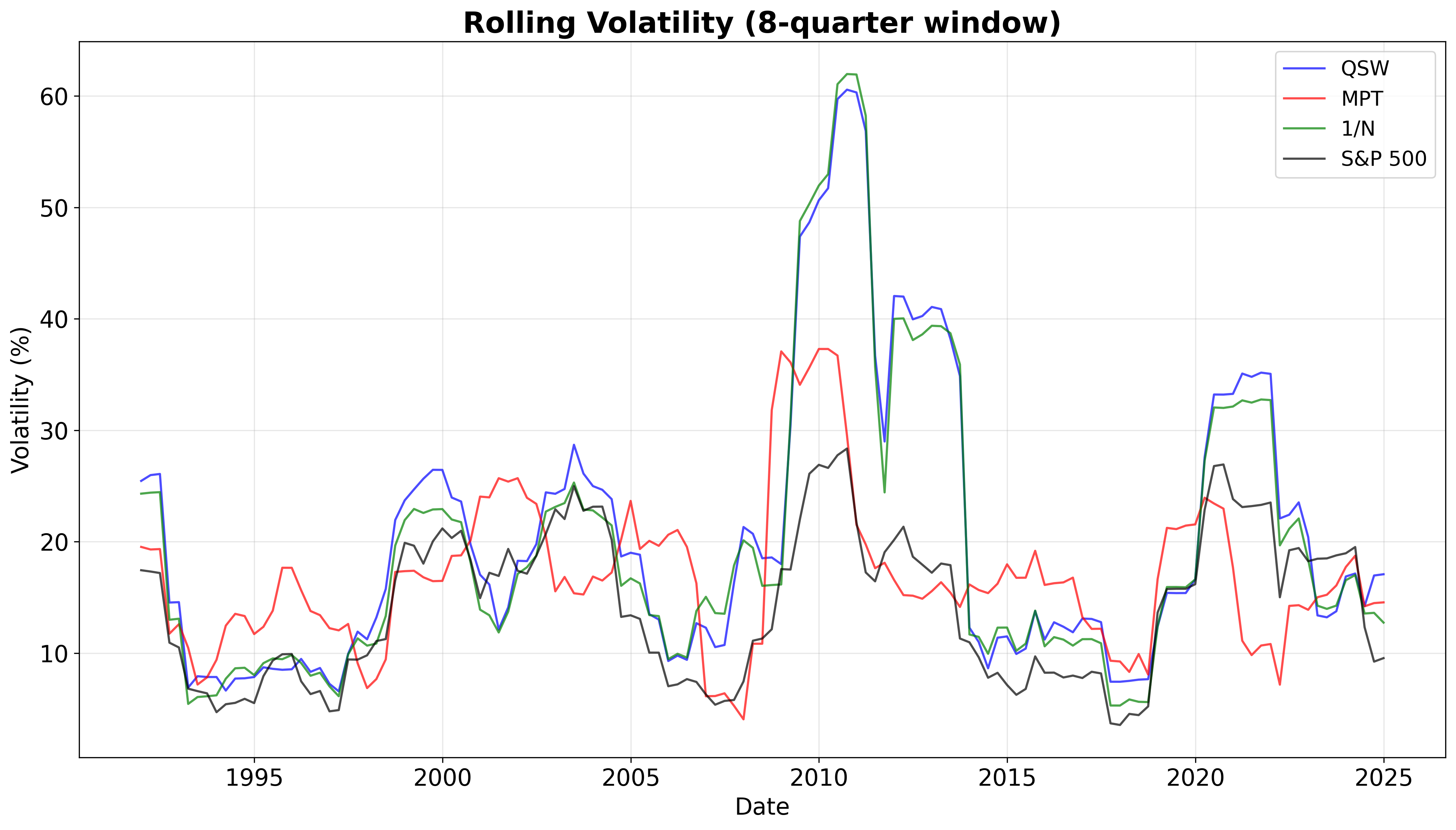}%
}
\end{figure}

Figure~\ref{fig:exp3_case_study_risk} further shows that QSW experiences shallower and faster-recovering drawdowns than MPT and the S\&P~500, while maintaining volatility in a similar range to the 1/N benchmark. The QSW optimizer therefore does not achieve its higher CAGR by leaning into extreme tail risk; instead, it improves the balance between return and drawdown over the full 34-year horizon. For brevity, we only report a representative subset of the 30-universe results and one detailed case study in this section. Additional diagnostic figures—including per-experiment performance panels for all 30 universes (final portfolio values, Sharpe ratios, CAGRs, Calmar ratios, maximum drawdowns, and annual turnover), as well as extended single-universe comparisons—are provided in an online supplementary image archive at \url{https://github.com/aceest/quantum-stochastic-walk-for-Portfolio-Optimization-Theory-and-Implementation-on-Financial-Networks}, in the \texttt{30\_experiments} subdirectory. These supplementary figures are fully consistent with the results reported here and allow interested readers to visually inspect the behavior of the QSW strategy across all individual universes and metrics.

\section{Discussion}

This section synthesizes the empirical findings from the Results section, discusses their implications for portfolio construction, and outlines limitations and directions for future work. Our experiments were designed to answer three core questions: (i) whether QSW-based portfolios can improve on classical MPT and naive 1/N benchmarks; (ii) how sensitive the QSW performance is to its hyperparameters; and (iii) whether these properties persist over long horizons and across different equity universes.

Across all Phase~1 experiments on the top-100 S\&P 500 universe (2018–2024), QSW-based portfolios consistently exhibited risk and diversification characteristics closely aligned with the naive 1/N benchmark. In Experiment~1, all six QSW strategy presets achieved Sharpe ratios around $0.96$–$0.98$, volatilities near $17\%$, and HHI values $\approx 0.01$, nearly identical to 1/N, while systematically outperforming the MPT maximum-Sharpe portfolio in the 1-year setting. Beyond raw risk-adjusted returns, the Monte Carlo shock analysis in Experiment~1 reveals a critical structural advantage of the QSW framework: intrinsic resistance to the "error maximization" that plagues classical optimization. While the MPT benchmark exhibited extreme fragility at the micro level—amplifying minor input perturbations into massive weight fluctuations with a standard deviation of nearly 8\%—the QSW allocations remained remarkably stable. This stability is architectural rather than accidental; the QSW parameters act as tunable dampers, where the diversification penalty ($\beta$) and holding coefficient ($\lambda$) effectively insulate the portfolio from input noise. Unlike classical solvers that require artificial constraints to prevent extreme rebalancing in response to estimation error, the QSW dynamics naturally suppress unforced turnover, providing a transparent mechanism to control the model's sensitivity to market noise. Experiment~2 generalized this behavior: the full 625-point grid search showed that the entire QSW parameter space yields HHI in the tight range $[0.010, 0.014]$, more than $20\times$ more diversified than the MPT benchmark (HHI $\approx 0.26$), and Sharpe ratios that are both high and remarkably stable. Taken together, these results indicate that the QSW dynamics are structurally anchored to a highly diversified, equal-weight-like regime, while still allowing for mild, data-driven tilts.

The grid search in Experiment~2 revealed that the Top-10 configurations (by Sharpe) all converge to the same pattern: minimal return-chasing ($\alpha = 0.1$) and maximal diversification pressure ($\beta = 500$), with low-to-moderate $\lambda$ and $\omega$. Heatmaps and correlation analyzes showed that $\beta$ acts as a structural “brake’’ that pins HHI at the 1/N floor across large regions of the grid, while $\omega$ and $\lambda$ act as behavioral knobs that trade off turnover and cost-efficiency without destabilizing performance. Importantly, the Sharpe surface is flat and robust: even the worst QSW configuration in the grid has a Sharpe comparable to, or better than, MPT under a 1-year window, and the median QSW Sharpe is substantially higher. This suggests that the QSW framework does not rely on a single fine-tuned parameter vector, but instead defines a broad, robust region of “good’’ configurations.

Phase~2 (Experiment~3) stress-tested the framework as a dynamic QSW optimizer from 1990 to 2024, with quarterly re-optimization over rolling 2-year windows and 30 independent 100-stock universes randomly sampled from point-in-time S\&P 500 constituents. Over 34 years and 30 universes, QSW remains a high-Sharpe, low-fragility strategy. On average, QSW achieved a CAGR of about $23\%$, a Sharpe ratio near $1.0$, and a Calmar ratio materially above those of MPT, 1/N, and the S\&P 500, while maintaining volatility between 1/N and MPT, and maximum drawdowns smaller than or comparable to 1/N and well below MPT and the index. The dynamic optimizer accepts a higher average turnover (around $200\%$ per year) than in the static Phase~1 experiments (typically $20$–$80\%$ for the best presets and about $40\%$ on average in Experiment~2) as the cost of adaptivity, but still trades substantially less than MPT’s $300$–$400\%$ levels. Distributional plots, win-rate charts in Figure~\ref{fig:exp3_win_scatter}a, and risk–return scatter plots in Figure~\ref{fig:exp3_win_scatter}b all support the same conclusion: QSW shifts the entire risk–return cloud upward relative to the benchmarks, rather than merely producing a few lucky outliers.

The single-universe case study in Experiment~3 shows how the rolling grid search behaves through the Dot-com bubble, the 2008 crisis, and the COVID shock. The parameter paths exhibit a recurring pattern: $\alpha$ stays low, $\beta$ is often at or near its upper bound, and $\omega$ toggles between more classical and more quantum regimes as volatility and market stress change. The dynamic hybrid optimizer repeatedly re-discovers the smart 1/N regime. The resulting portfolios retain a near-1/N diversification profile but adjust turnover and minor tilts to match the current regime. This behavior is consistent with the design rules inferred in Experiment~2, confirming that the same “smart 1/N’’ logic persists when the optimizer is allowed to adjust itself over three decades.

The empirical results have several implications for practical portfolio construction and for the role of quantum-inspired methods in finance. First, equal-weight remains a powerful baseline, but can be improved in a structured way. Our findings reinforce existing evidence that naive 1/N is a surprisingly effective benchmark in the presence of estimation error. However, QSW shows that it is possible to systematically improve on 1/N without sacrificing its key virtues. By embedding 1/N as a structural attractor (via diversification penalties and the QSW dynamics) and then allowing small, data-driven deviations on top, the QSW framework operationalizes the idea of a “smart 1/N’’: a portfolio that behaves like equal-weight in bad regimes and noisy environments, but tilts modestly when and where the data are reliable.

Second, quantum-inspired dynamics provide structure, not “magic’’ alpha. The advantage of the QSW framework does not come from exotic quantum effects per se, but from the structure imposed by the quantum-walk-inspired evolution: probability mass spreads over a graph in a way that naturally favors diversified, low-concentration states, while the decoherence and mixing parameters control how aggressively the system can deviate from this baseline. In practice, these dynamics act as a regularization mechanism that prevents the optimizer from collapsing into fragile, concentrated portfolios—a failure mode that classical MPT is particularly prone to, especially under short or noisy training windows.

Third, turnover and implementation constraints must be treated as first-class citizens. A recurring theme in our experiments is that nominal Sharpe or CAGR numbers can be deeply misleading if turnover and implementation costs are ignored. Classical MPT often wins on “paper Sharpe’’ but loses in any realistic cost-adjusted comparison due to extreme turnover and concentration. By contrast, the QSW design rules naturally push the framework into regions of the parameter space where efficiency (Sharpe per unit of turnover) is high and diversification is maintained. The comparison between Phase~1 and Phase~2 makes this trade-off explicit. In the static 2018–2024 setting of Experiments~1 and~2, the best QSW presets typically operate with annual turnover in the 20–80\% range (around 40\% on average), reflecting a low-footprint, “smart 1/N’’ implementation. In the dynamic robustness experiment (Experiment~3), the QSW optimizer accepts a higher average turnover of roughly 200\% per year in exchange for quarterly re-optimization of the classical and quantum channels over a rolling 2-year window. Even so, QSW still trades substantially less than the MPT maximum-Sharpe benchmark (about 330\% on average) while delivering markedly higher Sharpe and Calmar ratios. In this sense, turnover is not eliminated but budgeted: the extra trading capacity is allocated to regime adaptation rather than to fragile, high-frequency re-optimization of a highly concentrated portfolio.

Fourth, robustness to estimation error is more valuable than local optimality. From the Monte Carlo shock tests in Experiment~1 to the 625-point grid search and the 30-universe robustness study, a consistent pattern emerges: QSW prioritizes stability and robustness over extreme local optimality. In many institutional contexts, especially where mandates are sizeable and capital is sticky, such robustness is more valuable than squeezing out a few extra basis points of back-tested Sharpe from a highly fragile solution. The QSW framework therefore appears particularly well-suited to applications where estimation error, regime shifts, and market frictions are unavoidable.

Despite the encouraging results, several limitations of the present study should be acknowledged. First, our experiments focus on U.S. large-cap equities (S\&P 500 constituents) and a specific 34-year historical period. While this horizon includes multiple crises and structural shifts, the conclusions may not directly transfer to other asset classes (e.g., credit, commodities, FX), to strongly illiquid markets, or to regimes with very different microstructure and regulation. Extending the analysis to multi-asset portfolios and to non-U.S. markets is a natural next step. Second, implementation costs are treated via stylized turnover-based approximations (e.g., a constant round-trip rate). In practice, transaction costs depend on trade size, volatility, depth, and contemporaneous order flow, among other factors. A more detailed cost model, calibrated to actual execution data, could refine the comparison between QSW and MPT and may alter the optimal choice of $\alpha$, $\beta$, $\lambda$ and $\omega$ in the optimizer. Third, the 625-point grid search provides a transparent way to explore the QSW parameter space, but is not necessarily optimal from a computational or statistical standpoint. More sophisticated search methods (Bayesian optimization, bandit-style exploration, or meta-learning across universes) could lead to better parameter selection with fewer evaluations. Moreover, we have not explicitly optimized the re-optimization frequency; quarterly updates are a pragmatic choice, not a theoretically justified optimum. Fourth, the particular QSW formulation used here is only one of many possible quantum-inspired stochastic processes. Other variants—including different graph constructions, alternative decoherence models, or multi-particle walks—may yield different behavior. Similarly, we have not yet incorporated factor structures, sector constraints, or more complex risk models into the QSW dynamics. These design choices could be important in practical settings and warrant further exploration in future work.

The framework introduced in this paper opens several directions for further research. A natural extension is to apply QSW-based optimization to multi-asset portfolios that combine equities, fixed income, commodities, and alternative investments, potentially with a factor-based representation (value, momentum, quality, macro factors, etc.). In such settings, the graph structure could encode both security-level and factor-level relationships, allowing the quantum-walk dynamics to propagate information across multiple layers. Future work could also integrate more realistic transaction-cost models, liquidity constraints, and regulatory or mandate-specific constraints (e.g., tracking error, sector caps, ESG limits) directly into the QSW optimization. This would allow a joint design of the quantum and classical channels that explicitly targets cost-adjusted utility or drawdown-sensitive objectives. In particular, replacing the pure Sharpe objective used in the quarterly grid search of Experiment~3 with a cost-aware criterion—for example, maximizing the efficiency score $\mathcal{E}$ or Sharpe subject to an explicit turnover cap—could reduce the average turnover from its current $\sim 200\%$ level while preserving most of the adaptive benefit. Another promising direction is to embed QSW within an online-learning or change-point detection framework, allowing it to respond more intelligently to sudden regime shifts—for example, by adjusting the re-optimization frequency or temporarily increasing diversification pressure when uncertainty is high. Finally, although our implementation is entirely classical and quantum-inspired, the QSW framework is designed to be compatible with future quantum hardware. Exploring hardware-efficient encodings of the QSW dynamics and co-designing algorithms that leverage early quantum devices for the most computationally intensive subroutines (e.g., sampling, large-scale graph propagation) is an interesting avenue as quantum technologies mature.

In conclusion, this work introduces and empirically validates a QSW-based framework for portfolio construction that combines the robustness of naive 1/N with the flexibility of a parameterized, hybrid quantum–classical optimizer. Across multiple experiments, time horizons, and equity universes, the QSW model consistently delivers higher risk-adjusted returns than classical MPT and naive 1/N, while maintaining strong diversification and moderate turnover. The central message is not that quantum-inspired models magically generate alpha, but that they can enforce desirable structural properties—in particular, diversification and robustness to estimation error—in a principled way. As financial markets continue to grapple with noisy data, regime uncertainty, and implementation frictions, such structurally robust, “smart 1/N’’ approaches may offer a valuable alternative to both naive diversification and fragile, high-dimensional optimization. We hope that the QSW framework presented here can serve as a foundation for further work at the intersection of quantum-inspired algorithms and practical portfolio management.

\section{Methods}

\subsection*{QSW formulation on financial graphs}

Quantum stochastic walks (QSWs) extend classical random walks by blending coherent quantum evolution with controlled decoherence on the same graph. The coherent channel, driven by a Hamiltonian, allows superposition and interference across neighboring nodes, while the stochastic channel, implemented via a Lindblad dissipator, guarantees ergodicity and convergence to a unique stationary state. This dual-channel design overcomes the shortcomings of purely classical diffusion (no interference, fragile reliance on local correlations) and purely quantum walks (no stationary distribution), making QSWs a natural engine for portfolio allocation on financial networks.

Regarding classical random walks and the PageRank component, consider a graph $G = (V,E)$ with $n$ nodes representing financial assets. A simple continuous-time random walk can be described by the master equation
\begin{equation}
\frac{d\vec{p}}{dt} = (P - I)\,\vec{p},
\label{eq:classical_master}
\end{equation}
where $\vec{p} = (p_1,\dots,p_n)^\top$ is the probability distribution over nodes, $P$ is a row-stochastic transition matrix, and $(P-I)$ plays the role of an infinitesimal generator. The PageRank algorithm~\cite{brin1998anatomy} enhances this framework by introducing a “teleportation’’ step:
\begin{equation}
\mathcal{G} = \alpha_{\text{damp}} P + (1 - \alpha_{\text{damp}}) \frac{\mathbf{1}\mathbf{1}^\top}{n},
\label{eq:google_impl}
\end{equation}
where $\alpha_{\text{damp}} \in (0,1)$ is the damping parameter (typically $\approx 0.85$) and $\mathbf{1}\mathbf{1}^\top$ is the $n\times n$ all-ones matrix. This construction mixes graph-based transitions with a uniform teleportation step, ensuring that $\mathcal{G}$ is irreducible and aperiodic and therefore admits a unique stationary distribution even for directed graphs with dangling nodes.

A continuous-time quantum walk (CTQW) replaces the classical distribution $\vec{p}$ with a quantum state $|\psi\rangle$ evolving under the Schrödinger equation~\cite{farhi1998quantum,kempe2003quantum}
\begin{equation}
i\hbar\, \frac{d|\psi\rangle}{dt} = H\,|\psi\rangle,
\label{eq:schrodinger_ctqw}
\end{equation}
where $H$ is a Hermitian Hamiltonian. The state evolves as $|\psi(t)\rangle = e^{-iHt/\hbar}|\psi(0)\rangle$, and the probability of finding the walker at node $j$ is $P_j(t) = |\langle j|\psi(t)\rangle|^2$. While CTQWs can achieve speedups on certain graph problems~\cite{childs2003exponential}, they lack a stationary state and are highly sensitive to decoherence.

QSWs interpolate between the classical master equation \eqref{eq:classical_master} and the Schrödinger-type CTQW \eqref{eq:schrodinger_ctqw} via the Gorini–Kossakowski–Lindblad–Sudarshan (GKLS) master equation~\cite{chruscinski2017gkls}:
\begin{equation}
\frac{d\rho}{dt} = -i(1-\omega)[H,\rho] + \omega\sum_{i,j} c_{ij}\!\left( L_{ij}\rho L_{ij}^{\dagger} - \tfrac12\{L_{ij}^{\dagger}L_{ij},\rho\}\right),
\label{eq:qsw_master_final}
\end{equation}
where $\rho$ is the density matrix, $\omega\in[0,1]$ is the quantum–classical mixing parameter, $H$ is the Hamiltonian, $L_{ij} = |i\rangle\langle j|$ are Lindblad jump operators representing transitions from node $j$ to node $i$, and $c_{ij}\ge 0$ are transition rates. The generator consists of two competing terms: a coherent term $-i(1-\omega)[H,\rho]$ that drives unitary evolution, and a dissipative term $\omega\sum_{i,j} c_{ij}(\cdot)$ that introduces decoherence and classical jumps. The mixing parameter $\omega$ controls the balance: $\omega=0$ recovers a pure CTQW, $\omega=1$ a classical stochastic process, and $0<\omega<1$ a hybrid dynamics. For optimization applications we require convergence to a unique stationary state. Following~\cite{spohn1977algebraic}, if we set $c_{ij} = \mathcal{G}_{ij}$ where $\mathcal{G}$ is the Google matrix~\eqref{eq:google_impl}, then for any $\omega>0$ the quantum dynamical semigroup generated by Eq.~\eqref{eq:qsw_master_final} is primitive: a unique, full-rank stationary state $\rho_\infty$ exists and $\rho(t)\to\rho_\infty$ exponentially fast. We define the final portfolio weights as $w_i = \rho_{\infty,ii}$, i.e. the diagonal of the stationary density matrix.

\begin{figure}[htbp]
  \centering
  \includegraphics[width=0.9\textwidth]{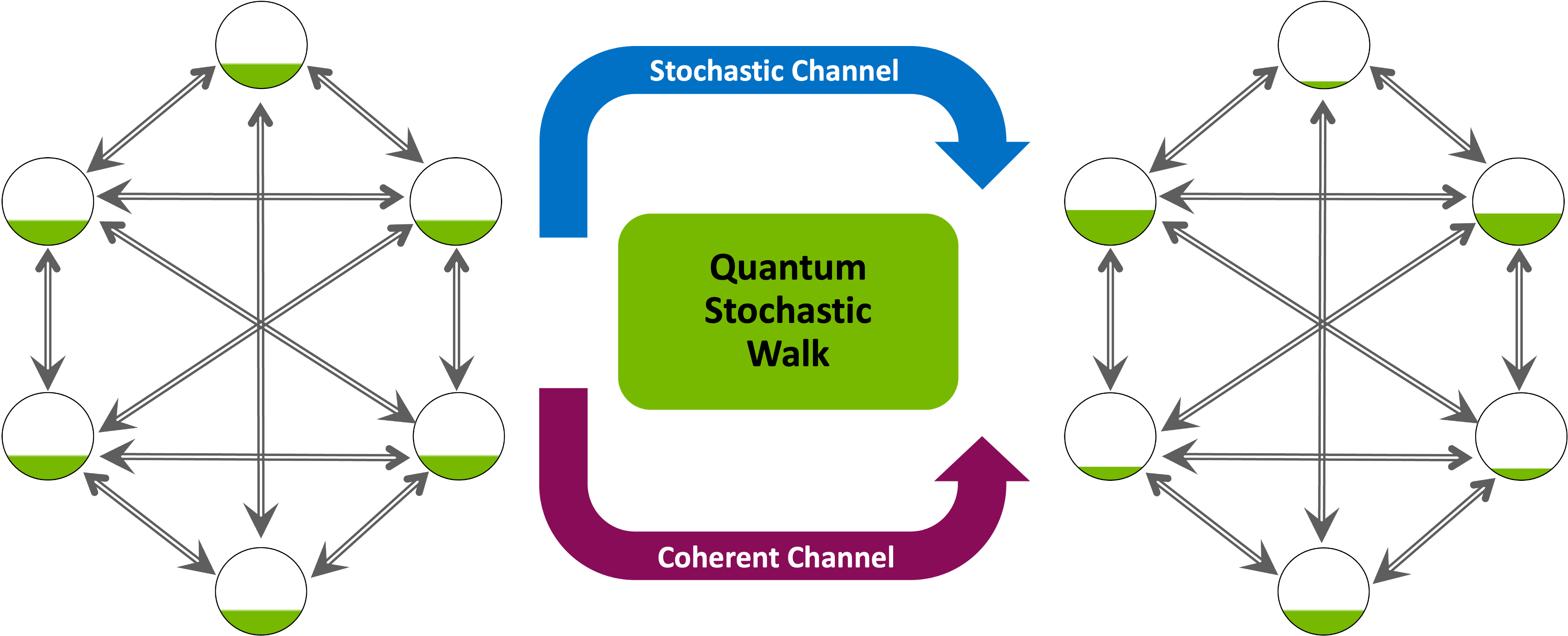}
  \caption{Conceptual illustration of the quantum stochastic walk (QSW) as a dual-channel engine on a financial graph: a coherent channel (purple) explores correlated clusters through Hamiltonian evolution, while a stochastic channel (blue) implements Google-type jumps that enforce diversification and guarantee convergence to a unique stationary state.}
  \label{fig:graph_map_demo}
\end{figure}

We encode financial data on the graph by taking $n$ financial assets with daily returns $\{r_{i,t}\}$ over a training window of length $T$. Each node $i\in\{1,\dots,n\}$ is characterized by its mean return $\mu_i = \frac{1}{T}\sum_{t=1}^T r_{i,t}$, its mean excess return $\mu_i^{(\mathrm{ex})} = \mu_i - r_f^{(d)}$, its volatility $\sigma_i$, and its daily Sharpe ratio $SR_i = \mu_i^{(\mathrm{ex})}/\sigma_i$, where $r_f^{(d)}$ is the daily risk-free rate. The graph weight matrix $W$ encodes both individual asset quality and pairwise relationships:
\begin{equation}
W_{ij} = \begin{cases}
\exp\bigl(\alpha\, SR_j - \beta\, \Sigma_{ij}\bigr) & \text{if } i \neq j, \\
\exp\bigl(\lambda\, SR_i\bigr)                    & \text{if } i = j,
\end{cases}
\label{eq:weight_matrix_impl}
\end{equation}
where $\alpha>0$ controls the preference for high-Sharpe destinations, $\beta>0$ penalizes transitions between highly correlated assets, $\Sigma_{ij}$ is the sample covariance between assets $i$ and $j$ over the same window, and $\lambda$ is the holding coefficient for self-loops. Row-normalizing $W$ yields a stochastic transition matrix
\begin{equation}
P_{ij} = \frac{W_{ij}}{\sum_k W_{ik}},
\label{eq:transition_matrix}
\end{equation}
which is then embedded into the Google matrix $\mathcal{G}$ in Eq.~\eqref{eq:google_impl}. The same financial information is encoded in the Hamiltonian via
\begin{equation}
H_{ij} = \begin{cases}
-\gamma_1\, SR_i & \text{if } i = j, \\
\gamma_2\, \hat{\Sigma}_{ij} & \text{if } i \neq j,
\end{cases}
\label{eq:hamiltonian_impl}
\end{equation}
where $\hat{\Sigma}_{ij}$ is a normalized covariance $\hat{\Sigma}_{ij} = (\Sigma_{ij}-\Sigma_{\min}) / (\Sigma_{\max}-\Sigma_{\min})$ and $\gamma_1,\gamma_2$ are scaling factors chosen for numerical stability. This construction yields a dual-channel encoding of financial data: the coherent channel explores highly correlated clusters through $H$, while the stochastic channel enforces diversification through $\mathcal{G}$ and the rates $c_{ij} = \mathcal{G}_{ij}$. The steady-state diagonal of $\rho_\infty$ then defines a well-behaved, graph-based mapping from return/covariance inputs to portfolio weights. Figure~\ref{fig:graph_map_demo} summarizes how the QSW operates on the financial graph. At each step, the portfolio state (represented by node occupancies on the left-hand graph) is updated by two complementary channels acting on the same network. The coherent channel (purple arrow) applies Hamiltonian evolution, allowing the walk to tunnel within highly correlated clusters and to explore multiple paths in superposition. This captures nonlinear dependency structure that is invisible to purely classical diffusion. The stochastic channel (blue arrow) then applies a Google-type update based on the weighted transition matrix, which damps probability mass inside tight clusters and redistributes it toward under-represented nodes. Iterating these two channels in alternation drives the system toward a unique stationary density matrix, whose diagonal elements define the final portfolio weights on the right-hand graph. In this way, the QSW framework implements portfolio allocation as a dual-channel process that jointly encodes return, risk and correlation directly in the dynamics of the walk.

\subsection*{QSW algorithm implementation}

\begin{figure}[htbp]
  \centering
  \includegraphics[width=\linewidth]{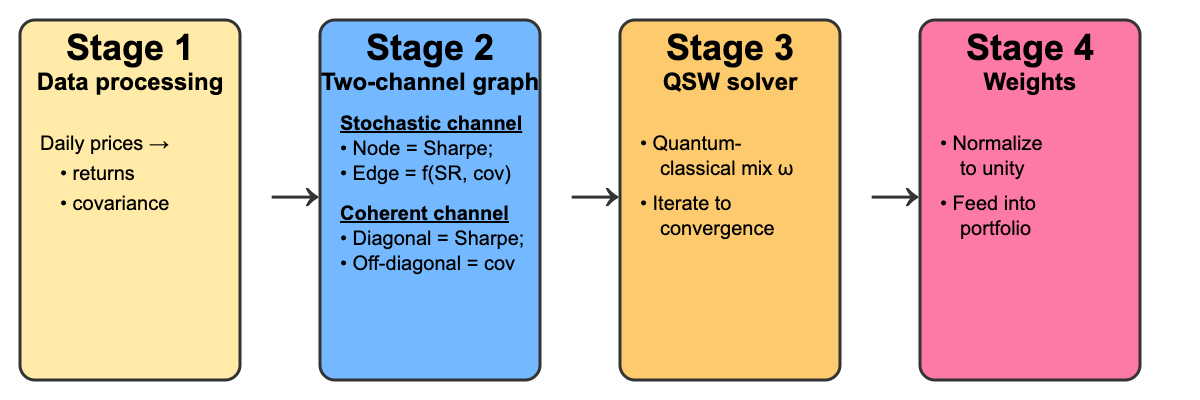}
  \caption{Four-stage quantum–stochastic-walk (QSW) portfolio framework employed in all subsequent experiments.}
  \label{fig:qsw_framework}
\end{figure}

We implement the QSW-based portfolio optimizer in four stages, summarized in Figure~\ref{fig:qsw_framework}.
\begin{enumerate}[label=\textbf{Stage~\arabic*:},leftmargin=*,itemsep=0.2em]
 \item Data processing – Ingest daily adjusted close prices. For each rebalancing date, compute inputs from a rolling window of $T_w$ trading days of simple returns $r_{i,t} = P_{i,t}/P_{i,t-1} - 1$:
       \begin{itemize}[leftmargin=*]
         \item Daily statistics: sample mean $\mu_i$, sample standard deviation $\sigma_i$, and the $N \times N$ sample covariance matrix $\boldsymbol{\Sigma}$ over the window.
         \item Sharpe ratios: daily Sharpe for each asset
               \[
               \mathrm{SR}_i = \frac{\mu_i - r_f^{(d)}}{\sigma_i},
               \]
               where we assume an annual risk-free rate of 3\% and set $r_f^{(d)} = 0.03/252$.
       \end{itemize}

 \item Dual-channel graph – Embed these statistics in a dual network:
       \begin{itemize}[leftmargin=*]
         \item Stochastic channel: construct the weight matrix $W_{ij} \propto \exp\bigl(\alpha\,\mathrm{SR}_j - \beta\,\Sigma_{ij}\bigr)$, row-normalize it to obtain a transition matrix $P$, and form the Google matrix $\mathcal{G}$ as in Eq.~\eqref{eq:google_impl}; this favors moves toward high-Sharpe, low-covariance destinations.
         \item Coherent channel: build a Hamiltonian with $H_{ii} \propto \mathrm{SR}_i$ and $H_{ij} \propto \hat\Sigma_{ij}$ for $i \neq j$, where $\hat\Sigma_{ij}$ is the normalized covariance, capturing quantum interference across correlated assets.
       \end{itemize}

 \item QSW solver – Evolve the density operator under the quantum–classical mix~$\omega$ (GKLS generator) until convergence to the stationary state $\rho_\infty$.
 \item Weights – Extract the diagonal of $\rho_\infty$ and normalize it to obtain fully invested portfolio weights $\mathbf{w}$ with $\sum_i w_i = 1$, which are then fed into the back-test loop.
\end{enumerate}

We use efficient Kraus-operator evolution to integrate the GKLS master equation \eqref{eq:qsw_master_final} because state-vector unraveling increases the state dimension from $N$ to $N^2$ and is computationally expensive. To accelerate the simulation, we map the Hamiltonian and Lindblad operators $\{H,L_{ij}\}$ to a set of Kraus operators $\{K_0,K_{ij}\}$ and implement the quantum channel in operator-sum form:
\begin{equation}
\rho(t + \Delta t) = K_0(\Delta t)\,\rho(t)\,K_0^\dagger(\Delta t) + \sum_{i,j} K_{ij}(\Delta t)\,\rho(t)\,K_{ij}^\dagger(\Delta t),
\label{eq:kraus_sum}
\end{equation}
with the trace-preserving condition
\begin{equation}
K_0^\dagger K_0 + \sum_{i,j} K_{ij}^\dagger K_{ij} = I.
\label{eq:trace_preservation}
\end{equation}
Discretizing Eq.~\eqref{eq:qsw_master_final} yields
\begin{align}
\rho(t + \Delta t) &= \rho(t) - i (1-\omega) [H, \rho(t)] \Delta t
 + \omega \sum_{i,j} c_{ij} \left( L_{ij} \rho(t) L_{ij}^\dagger
 - \tfrac{1}{2} \{ L_{ij}^\dagger L_{ij}, \rho(t) \} \right) \Delta t
 + \mathcal{O}(\Delta t^2).
\label{eq:lindblad_discretized}
\end{align}
Comparing Eqs.~\eqref{eq:lindblad_discretized} and \eqref{eq:kraus_sum} leads to the approximate Kraus operators
\begin{equation}
K_{ij} = \sqrt{1 - e^{-\omega c_{ij} \Delta t}}\, L_{ij}, \qquad
K_0 = \sqrt{ I - \sum_{i,j} K_{ij}^\dagger K_{ij} }\, e^{-i (1 - \omega) H \Delta t}.
\label{eq:kraus_ops}
\end{equation}
In our setting each $K_{ij}$ has the form $\gamma_{ij}\,|i\rangle\langle j|$ and is therefore extremely sparse. The double-sided product simplifies to
\begin{equation}
K_{ij} \rho K_{ij}^\dagger = \gamma_{ij}^2\, \rho_{jj}\, |i\rangle \langle i|,
\label{eq:kraus_diag}
\end{equation}
so the dissipative term only involves the diagonal of $\rho$. Defining $P = (\rho_{11},\rho_{22},\dots,\rho_{NN})^\top$ and a matrix $\Gamma$ with entries $\Gamma_{ij} = \gamma_{ij}^2$, we obtain
\begin{equation}
\sum_{i,j} K_{ij}\rho K_{ij}^\dagger = \Gamma P,
\label{eq:sum_gamma_p}
\end{equation}
and the operator-sum update can be written as
\begin{equation}
\rho' = K_0 \rho K_0^\dagger + \Gamma P.
\label{eq:final_op_sum}
\end{equation}
Thus the expensive sum of many matrix–matrix products reduces to a single matrix–vector multiplication.

An iterative solver initializes the density matrix at the maximally mixed state $\rho^{(0)} = I/n$, chooses a time step $\Delta t$ and tolerance $\epsilon$, and iterates Eq.~\eqref{eq:final_op_sum} until convergence. Algorithm~\ref{alg:qsw_implementation} summarizes the implementation used in our experiments.

\begin{algorithm}[H]
\caption{GPU-accelerated QSW portfolio optimizer}
\label{alg:qsw_implementation}
\begin{algorithmic}[1]
\Require Historical returns, QSW parameters $\{\omega, \alpha, \beta, \lambda\}$
\Ensure Portfolio weights $\mathbf{w}$
\State Compute $\mu_i, \sigma_i, SR_i, \Sigma_{ij}$ from the training window
\State Construct weight matrix $W$ and Hamiltonian $H$; form Google matrix $\mathcal{G}$
\State Initialize $\rho^{(0)} = I/n$, choose $\Delta t$, tolerance $\epsilon$
\Repeat
    \State Build Kraus operators $K_0, K_{ij}$ from Eq.~\eqref{eq:kraus_ops}
    \State $\tilde{\rho} \leftarrow K_0 \rho^{(t)} K_0^\dagger$
    \State $P \leftarrow \operatorname{diag}(\tilde{\rho})$
    \State $\rho^{(t+1)} \leftarrow \tilde{\rho} - \operatorname{diag}(P) + \operatorname{diag}(\Gamma P)$
    \State Normalize $\rho^{(t+1)} \leftarrow \rho^{(t+1)}/\operatorname{Tr}(\rho^{(t+1)})$
\Until{$\|\rho^{(t+1)} - \rho^{(t)}\|_1 < \epsilon$ or $t$ reaches a maximum}
\State Set $w_i = \rho^{(\infty)}_{ii} / \sum_j \rho^{(\infty)}_{jj}$ for $i=1,\dots,n$
\State \Return $\mathbf{w}$
\end{algorithmic}
\end{algorithm}

The computational efficiency is ensured by GPU acceleration. The implementation leverages \texttt{cuPyNumeric}~\cite{cupynumeric} for linear-algebra kernels, an NVIDIA-developed library that serves as a drop-in replacement for CPU-based NumPy. It accelerates all core primitives in the QSW solver, including matrix exponentials for unitary evolution, element-wise operations for transition rates, and diagonal extraction and reconstruction. In practice, a single NVIDIA A100 GPU can evaluate the full 625-parameter grid of Experiment~2 in roughly three hours of wall-clock time (about 17 seconds per configuration), including data loading, graph construction, QSW convergence, and metric computation. This enables real-time or near-real-time portfolio optimization for universes of order $10^2$–$10^3$ assets on modern GPU hardware.

\subsection*{Data and experimental setup}

For the asset universe and data collection, we consider two types of equity universes. In Phase~1 (parameter exploration) we use the 100 largest U.S. stocks by market capitalisation (“top-100” universe), including technology names (e.g. AAPL, MSFT, GOOGL, AMZN, NVDA), financials (e.g. JPM, BAC, WFC), healthcare (e.g. JNJ, UNH, PFE), and other sectors. This universe provides sufficient cross-sectional diversity while keeping the QSW simulations computationally tractable. To assess robustness to stock selection, Phase~2 employs a multi-universe design based on 30 random subsets of the S\&P~500. Each trial defines a distinct 100-stock universe that is maintained in a point-in-time fashion to avoid survivorship bias: at the initial date (1990Q1) we draw 100 constituents from the S\&P~500 members active at that time; at each quarterly rebalance we remove any delisted names (e.g. due to mergers or bankruptcies) and replace them by randomly sampling from the S\&P~500 constituents active on that date. Historical daily adjusted close prices from 1988-01-01 to 2024-12-31 are obtained from Yahoo Finance.

All strategies are evaluated in a common backtesting framework with quarterly rebalancing. The specific backtest horizon and training window depend on the experiment:

\begin{itemize}
  \item \emph{Phase~1: parameter exploration (2018–2024).}
        \begin{itemize}[leftmargin=*,label=--]
          \item Backtest period: 2018-01-02 to 2024-12-31 on the top-100 universe.
          \item Training window: rolling 1–2 years of daily data (252, 378, or 504 trading days) to study the impact of lookback length.
        \end{itemize}
  \item \emph{Phase~2: multi-universe robustness (1990–2024).}
        \begin{itemize}[leftmargin=*,label=--]
          \item Backtest period: 1990-01-02 to 2024-12-31.
          \item Training window: fixed 2-year (504 trading days) rolling window.
        \end{itemize}
\end{itemize}

In both phases the rebalancing frequency is quarterly (every three months), and transaction costs are incorporated indirectly via portfolio turnover statistics. This frequency reflects common institutional practice, balancing responsiveness to changing market conditions against implementation costs.

Regarding computational performance, all QSW simulations and grid searches are implemented in Python using GPU-accelerated linear algebra (see above). On a single NVIDIA A100 GPU, the full 625-parameter grid of Phase~1, Experiment~2 can be evaluated in roughly three hours of wall-clock time (about 17 seconds per configuration), including data loading, graph construction, QSW convergence, and performance metric computation.

In Phase~1: parameter exploration, we use two complementary approaches on the top-100 universe over 2018–2024. First, we define six QSW strategy presets that capture different investment styles (Table~\ref{tab:scenario_parameters}) by fixing $(\alpha,\beta,\lambda)$:


Each preset is combined with $\omega \in \{0.2, 0.4, 0.6, 0.8, 1.0\}$ to explore the quantum–classical spectrum. Second, we perform a comprehensive grid search over the four QSW parameters: $\alpha,\beta,\lambda \in \{0.1, 5, 50, 100, 500\}$ and $\omega \in \{0.2, 0.4, 0.6, 0.8, 1.0\}$, yielding 625 parameter combinations per training-window choice. This grid is used to identify robust high-performing regions of the parameter space.

In Phase~2: multi-universe robustness, to ensure that our findings are not artefacts of a particular stock universe, we run the full 625-point grid search on each of the 30 point-in-time 100-stock universes over 1990–2024. For each trial we (a) construct the initial universe at 1990Q1 from contemporaneous S\&P~500 constituents, (b) maintain the universe through quarterly replacement of delisted names, and (c) backtest all QSW configurations and benchmarks. We then analyze the cross-universe distribution of performance and risk metrics. This multi-universe design mitigates survivorship and selection bias and demonstrates that the observed “smart 1/N’’ behavior and robustness of the QSW optimizer are not driven by a single favorable sample.

\subsection*{Performance metrics and evaluation}

We evaluate portfolio performance using a set of complementary metrics that capture returns, risk, drawdown behavior, concentration, and trading efficiency.

For return and risk metrics, we track both level and risk-adjusted returns.

\emph{1. Cumulative return.}
\begin{equation}
V_T = V_0 \prod_{t=1}^{T} (1 + r_{p,t}),
\end{equation}
where $r_{p,t} = \sum_{i=1}^{n} w_{i,t}\, r_{i,t}$ is the portfolio return at time $t$.

\emph{2. Compound annual growth rate (CAGR).}
Let $Y$ denote the investment horizon in years. The CAGR is defined as
\begin{equation}
\text{CAGR} = \left(\frac{V_T}{V_0}\right)^{1/Y} - 1,
\end{equation}
which measures the long-run average growth rate of portfolio wealth.

\emph{3. Annualized Sharpe ratio.}
We compute the Sharpe ratio using daily excess returns and annualize it with 252 trading days:
\begin{equation}
SR = \frac{\mu_{p}^{\text{(excess)}}}{\sigma_p} \sqrt{252},
\end{equation}
where $\mu_{p}^{\text{(excess)}}$ and $\sigma_p$ are the mean and standard deviation of daily portfolio excess returns (portfolio return minus the daily risk-free rate). In all experiments we assume a constant annual risk-free rate of $3\%$.

\emph{4. Maximum drawdown (MDD).}
\begin{equation}
\text{MDD} = \max_{t \in [0,T]} \frac{\max_{s \in [0,t]} V_s - V_t}{\max_{s \in [0,t]} V_s},
\end{equation}
which captures the worst peak-to-trough loss in percentage terms over the evaluation period.

\emph{5. Calmar ratio.}
To relate long-run growth to downside risk, we use the Calmar ratio,
\begin{equation}
\text{Calmar} = \frac{\text{CAGR}}{|\text{MDD}|},
\end{equation}
which measures annualized return per unit of worst drawdown. Unless otherwise noted, $\text{MDD}$ is computed over the full evaluation horizon of the experiment (e.g. 1990–2024 for Experiment~3); rolling Calmar ratios in the figures are defined analogously using window-specific MDD values. This metric is used extensively in Experiment~3 to compare strategies on a drawdown-adjusted basis.

For portfolio concentration metrics, we quantify how capital is distributed across names using three standard measures.

\emph{1. Herfindahl–Hirschman Index (HHI).}
\begin{equation}
\text{HHI} = \sum_{i=1}^{n} w_i^2,
\end{equation}
where $w_i$ is the weight of asset $i$. A higher HHI indicates greater concentration.

\emph{2. Effective number of stocks.}
\begin{equation}
N_{\text{eff}} = \frac{1}{\text{HHI}},
\end{equation}
which can be interpreted as the number of equally weighted assets that would yield the same HHI.

\emph{3. Top-5 concentration.}
\begin{equation}
C_5 = \sum_{i=1}^{5} w_{[i]},
\end{equation}
where $w_{[i]}$ denotes the $i$-th largest portfolio weight. $C_5$ measures how much of the portfolio is concentrated in the five largest positions.

We also track trading and efficiency metrics. Portfolio turnover represents the frequency of rebalancing activity and directly impacts implementation costs, market impact, and strategy scalability. High-turnover strategies face substantial transaction costs, price impact during execution, and capacity constraints that can erode theoretical performance gains in practice~\cite{almgren2001optimal,grinold1999active}.

\emph{1. Average annualized turnover.}
\begin{equation}
\overline{\text{TO}}_{\text{ann}} = 4\;\times\;\frac{1}{N_{\text{quarters}}} \sum_{q=1}^{N_{\text{quarters}}} \sum_{i=1}^{n}\bigl|w_{i,q}-w_{i,q-1}\bigr|,
\label{eq:annual_turnover}
\end{equation}
which expresses the proportion of portfolio value traded \emph{per year}. In buy-side practice, annual turnover above roughly $200\%$ is often considered prohibitively expensive because of spread and market-impact costs, whereas turnover below about $50\%$ is usually viewed as readily implementable~\cite{kissell2013science}. Classical mean–variance optimization frequently produces $300$–$500\%$ annual turnover, posing significant implementation challenges~\cite{demiguel2009optimal,best1991sensitivity}.

\emph{2. Sharpe–turnover efficiency ratio.}
\begin{equation}
\mathcal{E} = \frac{SR}{\overline{\text{TO}}_{\text{ann}} + 0.01},
\end{equation}
which captures the trade-off between risk-adjusted returns and trading intensity by measuring how much Sharpe ratio is achieved per unit of annual turnover~\cite{grinold1999active}. The small constant $0.01$ prevents division by zero for extremely low-turnover strategies. Higher efficiency ratios indicate strategies that deliver superior risk-adjusted performance without excessive trading activity—a critical requirement for real-world implementation, where transaction costs, market impact, and capacity constraints dominate practical considerations~\cite{almgren2001optimal,perold1988implementation}. For example, a strategy achieving a Sharpe ratio of $1.0$ with $20\%$ annual turnover ($\mathcal{E} \approx 5.0$) is significantly more valuable than one achieving $1.2$ with $400\%$ turnover ($\mathcal{E} \approx 0.3$) once implementation frictions are taken into account~\cite{kissell2013science}.

\subsection*{Benchmark strategies}

We compare QSW-based portfolios against three primary benchmarks.

The classical benchmark is the maximum-Sharpe (MSR) portfolio based on Modern Portfolio Theory (MPT). While the MSR is defined as a non-convex ratio, it can be computed via an equivalent convex quadratic program. Since our inputs $\boldsymbol{\mu}$ are excess returns (over the risk-free rate), we solve~\cite{boyd2004convex}
\begin{equation}
\mathbf{y}^{\star} = \arg\min_{\mathbf{y}} \frac{1}{2}\mathbf{y}^T\boldsymbol{\Sigma}\mathbf{y}
\end{equation}
subject to $\mathbf{y}^T\boldsymbol{\mu} = 1$ and $\mathbf{y} \geq \mathbf{0}$. This finds the portfolio that minimizes variance for a fixed unit of excess return. The fully invested MPT weights are then $\mathbf{w}_{\text{MPT}} = \mathbf{y}^{\star} / \sum_i y^{\star}_i$. We implement this using the \texttt{PyPortfolioOpt} library, feeding it the same sample mean $\boldsymbol{\mu}$ and sample covariance $\boldsymbol{\Sigma}$ estimated from the rolling window as for the QSW optimizer.

The naive diversification (1/N) strategy assigns an equal weight $w_i = 1/N$ to each of the $N$ available assets at every rebalance. It completely avoids estimating $\boldsymbol{\mu}$ and $\boldsymbol{\Sigma}$ and is therefore immune to the estimation errors that plague MVO, especially in high dimensions~\cite{demiguel2009optimal}. Despite its simplicity, 1/N is a strong benchmark and serves as the natural reference for evaluating whether QSW provides a genuinely robust improvement.

Finally, the capitalization-weighted S\&P 500 index serves as a passive market benchmark, representing the opportunity cost of active management in large-cap U.S. equities.

\subsection*{Analysis framework}

We analyze QSW performance along several complementary dimensions. First, we study parameter sensitivity by analyzing correlations between QSW parameters $(\alpha,\beta,\lambda,\omega)$ and key performance metrics (Sharpe, Calmar, turnover, concentration), and identify robust regions of the parameter space. Second, we examine concentration–performance trade-offs by analyzing the efficient frontier in Sharpe–HHI space, comparing concentration levels versus MPT and 1/N, and assess how effective diversification varies under different parameter settings.

To ensure robust conclusions, we use simple but informative diagnostics. We perform a win-rate analysis by computing the percentage of QSW configurations that outperform each benchmark on a given metric (e.g. Sharpe, Calmar, final value). We also examine cross-universe consistency. In the multi-universe setting (Experiment~3), we examine the distribution of performance across all 30 random 100-stock universes, focusing on how often and by how much QSW dominates MPT, 1/N, and the index on risk and return metrics.

\subsection*{Summary of methods}

The implementation and evaluation methodology is designed to bridge the gap between theoretical innovation and realistic portfolio practice. In particular, it:

\begin{enumerate}[leftmargin=*]
  \item adopts a two-phase design: a short-horizon parameter exploration (2018–2024, top-100 S\&P 500) and a long-horizon robustness phase (1990–2024, rolling 2-year windows);
  \item combines three complementary experiment types: fixed strategy presets, a full 625-point hyper-parameter grid search, and a dynamic, multi-universe study;
  \item uses quarterly rebalancing and Sharpe ratios computed on excess returns over a 3\% annual risk-free rate, in line with institutional practice;
  \item evaluates portfolios with a rich set of metrics, including Sharpe, Calmar, maximum drawdown, turnover, and multiple concentration measures (HHI, effective number of stocks, top-5 weights);
  \item benchmarks QSW against maximum-Sharpe MPT portfolios, naive 1/N portfolios on the same universe, and the S\&P 500 index;
  \item validates robustness through extensive parameter-sensitivity analysis and a 34-year, 30-universe experiment based on point-in-time S\&P 500 constituents.
\end{enumerate}

Together, these elements provide a rigorous framework for evaluating the QSW framework as a quantum-inspired portfolio construction tool, ensuring that the reported results are not an artefact of a particular sample, universe, or parameter choice but reflect stable behavior under realistic market and implementation conditions.

\section*{Data availability}

All numerical results in this study are generated by simulating the QSW optimizer on equity return series as described in the Methods section. Daily adjusted close prices for S\&P 500 constituents over 1988–2024 were obtained from Yahoo Finance (\url{https://finance.yahoo.com}); no proprietary or restricted datasets were used. The raw price data can be re-downloaded from the same public source, and all preprocessing steps (return calculation, training windows, universe construction) are fully specified in the text. Aggregated backtest outputs (CAGR, Sharpe, Calmar, drawdown, turnover, and concentration metrics for all configurations and universes), as well as the per-universe diagnostic figures shown in the Results and Supplementary Information, are available from the corresponding author upon reasonable request.

\section*{Code availability}

All algorithms and parameter settings required to reproduce the results are described in the Methods section. The custom code used to implement the QSW optimizer and run the backtests is not publicly archived at this time, as it is part of an ongoing research project. Results from specific simulations, as well as pseudo-code and representative implementation details, are available from the corresponding author upon reasonable request for academic and non-commercial purposes.

\section*{ACKNOWLEDGMENTS}

We gratefully acknowledge support from the National Science and Technology Council (NSTC), Taiwan \textbf{NSTC 114-2112-M-033-010-MY3} \emph{(Applications of Quantum Computing in Financial Market Simulation and Optimization)} and Grants \textbf{NSTC 114-2119-M-033-001} \emph{(Applications of Quantum Computing in Optimization and Finance)}. We also thank the NVAITC and the NVIDIA Quantum team for their continued technical guidance. Our gratitude further extends to the NVIDIA Academic Grant Program for providing the GPU resources essential to this research, and to the NVIDIA Strategic Researcher Engagement team for their steadfast support.

\section*{Author contributions}

Yen Jui Chang conceived the study, designed and executed the experiments, and wrote the manuscript. Wei-Ting Wang developed the QSW software. Yun-Yuan Wang implemented GPU acceleration for the QSW framework. Chen-Yu Liu and Kuan-Cheng Chen analyzed and visualized the experimental results and contributed to their interpretation. Ching-Ray Chang provided critical discussion of the results and reviewed the manuscript. All authors reviewed and approved the final version of the manuscript.

\section*{Competing interests}

The authors declare no competing interests.

\bibliography{sn-bibliography}

@article{Eduardo2012quantum,
  title = {Quantum Navigation and Ranking in Complex Networks},
  author = {S{\'a}nchez-Burillo, Eduardo and Duch, Jordi and G{\'o}mez-Garde{\~n}es, Jes{\'u}s and Zueco, David},
  journal = {Scientific Reports},
  volume = {2},
  pages = {605},
  year = {2012},
  doi = {10.1038/srep00605}
}

@article{whitfield2010quantum,
  title = {Quantum stochastic walks: A generalization of classical random walks and quantum walks},
  author = {Whitfield, James D. and Rodriguez-Rosario, Cesar A. and Aspuru-Guzik, Al{\'a}n},
  journal = {Physical Review A},
  volume = {81},
  number = {2},
  pages = {022323},
  year = {2010},
  doi = {10.1103/PhysRevA.81.022323}
}

@article{attal2012open,
  title = {Open Quantum Random Walks},
  author = {Attal, St{\'e}phane and Petruccione, Francesco and Sabot, Christophe and Sinayskiy, Ilya},
  journal = {Journal of Statistical Physics},
  volume = {147},
  number = {4},
  pages = {832--852},
  year = {2012},
  doi = {10.1007/s10955-012-0491-0}
}

@article{paparo2013quantum,
  title = {Quantum {Google} in a Complex Network},
  author = {Paparo, Giuseppe Davide and M{\"u}ller, Markus and Comellas, Francesc and Martin-Delgado, Miguel Angel},
  journal = {Scientific Reports},
  volume = {2},
  pages = {444},
  year = {2013},
  doi = {10.1038/srep02773}
}

@article{wang2022continuous,
  title = {Continuous-time quantum walk based centrality testing on weighted graphs},
  author = {Wang, Yang and Xue, Shichuan and Wu, Junjie and Xu, Ping},
  journal = {Scientific Reports},
  volume = {12},
  number = {1},
  pages = {6001},
  year = {2022},
  publisher = {Nature Publishing Group},
  doi = {10.1038/s41598-022-09915-1}
}

@article{farhi1998quantum,
  title = {Quantum computation and decision trees},
  author = {Farhi, Edward and Gutmann, Sam},
  journal = {Physical Review A},
  volume = {58},
  number = {2},
  pages = {915},
  year = {1998},
  doi = {10.1103/PhysRevA.58.915}
}

@article{kempe2003quantum,
  title = {Quantum random walks: An introductory overview},
  author = {Kempe, Julia},
  journal = {Contemporary Physics},
  volume = {44},
  number = {4},
  pages = {307--327},
  year = {2003},
  doi = {10.1080/00107151031000110776}
}

@article{childs2003exponential,
  title = {Exponential algorithmic speedup by a quantum walk},
  author = {Childs, Andrew M. and Cleve, Richard and Deotto, Enrico and Farhi, Edward and Gutmann, Sam and Spielman, Daniel A.},
  journal = {Proceedings of the 35th Annual ACM Symposium on Theory of Computing},
  pages = {59--68},
  year = {2003},
  doi = {10.1145/780542.780552}
}

@article{chruscinski2017gkls,
  title = {A Brief History of the {GKLS} Equation},
  author = {Chru{\'s}ci{\'n}ski, Dariusz and Pascazio, Saverio},
  journal = {Open Systems \& Information Dynamics},
  volume = {24},
  number = {03},
  pages = {1740001},
  year = {2017},
  doi = {10.1142/S1230161217400017}
}

@article{zhang2021deep,
  title = {Deep learning for portfolio optimization},
  author = {Zhang, Zihao and Zohren, Stefan and Roberts, Stephen},
  journal = {The Journal of Financial Data Science},
  volume = {2},
  number = {4},
  pages = {8--20},
  year = {2021},
  doi = {10.3905/jfds.2020.1.042}
}

@article{gu2020empirical,
  title = {Empirical asset pricing via machine learning},
  author = {Gu, Shihao and Kelly, Bryan and Xiu, Dacheng},
  journal = {The Review of Financial Studies},
  volume = {33},
  number = {5},
  pages = {2223--2273},
  year = {2020},
  doi = {10.1093/rfs/hhaa009}
}

@article{han2023risk,
  title = {Risk budgeting portfolio optimization with deep reinforcement learning},
  author = {Han, Seungwoo},
  journal = {The Journal of Financial Data Science},
  volume = {5},
  number = {4},
  pages = {86--101},
  year = {2023},
  doi = {10.3905/jfds.2023.1.137}
}

@article{jiang2017deep,
  title = {A deep reinforcement learning framework for the financial portfolio management problem},
  author = {Jiang, Zhengyao and Xu, Dixing and Liang, Jinjun},
  journal = {arXiv preprint arXiv:1706.10059},
  year = {2017},
  doi = {10.48550/arXiv.1706.10059}
}

@article{arrieta2020explainable,
  title = {Explainable Artificial Intelligence ({XAI}): Concepts, taxonomies, opportunities and challenges},
  author = {Barredo Arrieta, Alejandro and D{\'\i}az-Rodr{\'\i}guez, Natalia and Del Ser, Javier and Bennetot, Adrien and Tabik, Siham and Pastore, Alberto and Gtr, Inma and De, S. and Herrera, Francisco},
  journal = {Information Fusion},
  volume = {58},
  pages = {82--115},
  year = {2020},
  doi = {10.1016/j.inffus.2019.12.012}
}

@misc{cupynumeric,
  title = {{cuPyNumeric}: A {NumPy}-Compatible Library for {NVIDIA} {GPU} Calculations},
  author = {{NVIDIA Corporation}},
  year = {2024},
  howpublished = {\url{https://developer.nvidia.com/cupynumeric}},
  note = {Accessed: 2024-06-30}
}

@article{markowitz1952portfolio,
  title = {Portfolio Selection},
  author = {Markowitz, Harry},
  journal = {The Journal of Finance},
  volume = {7},
  number = {1},
  pages = {77--91},
  year = {1952},
  doi = {10.1111/j.1540-6261.1952.tb01525.x}
}

@article{sharpe1964capital,
  title = {Capital Asset Prices: A Theory of Market Equilibrium Under Conditions of Risk},
  author = {Sharpe, William F.},
  journal = {The Journal of Finance},
  volume = {19},
  number = {3},
  pages = {425--442},
  year = {1964},
  doi = {10.1111/j.1540-6261.1964.tb02865.x}
}

@article{tobin1958liquidity,
  title = {Liquidity Preference as Behavior Towards Risk},
  author = {Tobin, James},
  journal = {The Review of Economic Studies},
  volume = {25},
  number = {2},
  pages = {65--86},
  year = {1958},
  doi = {10.2307/2296205}
}

@article{merton1972analytic,
  title = {An Analytic Derivation of the Efficient Portfolio Frontier},
  author = {Merton, Robert C.},
  journal = {Journal of Financial and Quantitative Analysis},
  volume = {7},
  number = {4},
  pages = {1851--1872},
  year = {1972},
  doi = {10.2307/2329621}
}

@book{fabozzi2007robust,
  title = {Robust Portfolio Optimization and Management},
  author = {Fabozzi, Frank J. and Kolm, Petter N. and Pachamanova, Despina A. and Focardi, Sergio M.},
  publisher = {John Wiley \& Sons},
  address = {Hoboken, NJ, USA},
  year = {2007},
  doi = {10.1002/9781119202172}
}

@article{black1992global,
  title = {Global Portfolio Optimization},
  author = {Black, Fischer and Litterman, Robert},
  journal = {Financial Analysts Journal},
  volume = {48},
  number = {5},
  pages = {28--43},
  year = {1992},
  doi = {10.2469/faj.v48.n5.28}
}

@book{mantegna1999introduction,
  title = {Introduction to Econophysics: Correlations and Complexity in Finance},
  author = {Mantegna, Rosario N. and Stanley, H. Eugene},
  publisher = {Cambridge University Press},
  address = {Cambridge, UK},
  year = {1999},
  doi = {10.1017/CBO9780511755767}
}

@article{jegadeesh1993returns,
  title = {Returns to Buying Winners and Selling Losers: Implications for Stock Market Efficiency},
  author = {Jegadeesh, Narasimhan and Titman, Sheridan},
  journal = {The Journal of Finance},
  volume = {48},
  number = {1},
  pages = {65--91},
  year = {1993},
  doi = {10.1111/j.1540-6261.1993.tb04702.x}
}

@book{fletcher2013practical,
  title = {Practical Methods of Optimization},
  author = {Fletcher, R.},
  edition = {2nd},
  publisher = {John Wiley \& Sons},
  address = {Hoboken, NJ, USA},
  year = {2013},
  doi = {10.1002/9781118723203}
}

@article{mantegna1999hierarchical,
  title = {Hierarchical structure in financial markets},
  author = {Mantegna, Rosario N.},
  journal = {The European Physical Journal B},
  volume = {11},
  pages = {193--197},
  year = {1999},
  doi = {10.1007/s100510050929}
}

@article{TUMMINELLO201040,
  title = {Correlation, hierarchies, and networks in financial markets},
  author = {Tumminello, Michele and Lillo, Fabrizio and Mantegna, Rosario N.},
  journal = {Journal of Economic Behavior \& Organization},
  volume = {75},
  number = {1},
  pages = {40--58},
  year = {2010},
  doi = {10.1016/j.jebo.2010.01.004}
}

@article{pozzi2013spread,
  title = {Spread of Risk Across Financial Markets: Better to Invest in the Peripheries},
  author = {Pozzi, F. and Di Matteo, T. and Aste, T.},
  journal = {Scientific Reports},
  volume = {3},
  pages = {1665},
  year = {2013},
  doi = {10.1038/srep01665}
}

@article{kenett2012networks,
  title = {Dependency Network and Node Influence: Application to the Study of Financial Markets},
  author = {Kenett, Dror Y. and Preis, Tobias and Gur-Gershgoren, Gitit and Ben-Jacob, Eshel},
  journal = {International Journal of Bifurcation and Chaos},
  volume = {22},
  number = {07},
  pages = {1250181},
  year = {2012},
  doi = {10.1142/S0218127412501817}
}

@article{onnela2003dynamic,
  title = {Dynamic asset trees and Black Monday},
  author = {Onnela, Jukka-Pekka and Chakraborti, Anirban and Kaski, Kimmo and Kert{\'e}sz, J{\'a}nos},
  journal = {Physica A: Statistical Mechanics and its Applications},
  volume = {324},
  number = {1-2},
  pages = {247--252},
  year = {2003},
  doi = {10.1016/S0378-4371(02)01882-4}
}

@article{goldfarb2003robust,
  title = {Robust Portfolio Selection Problems},
  author = {Goldfarb, Donald and Iyengar, Garud},
  journal = {Mathematics of Operations Research},
  volume = {28},
  number = {1},
  pages = {1--38},
  year = {2003},
  doi = {10.1287/moor.28.1.1.14260}
}

@article{garlappi2007portfolio,
  title = {Portfolio Selection with Parameter and Model Uncertainty: A Multi-Prior Approach},
  author = {Garlappi, Lorenzo and Uppal, Raman and Wang, Tan},
  journal = {The Review of Financial Studies},
  volume = {20},
  number = {1},
  pages = {41--81},
  year = {2007},
  doi = {10.1093/rfs/hhl003}
}

@article{artzner1999coherent,
  title = {Coherent Measures of Risk},
  author = {Artzner, Philippe and Delbaen, Freddy and Eber, Jean-Marc and Heath, David},
  journal = {Mathematical Finance},
  volume = {9},
  number = {3},
  pages = {203--228},
  year = {1999},
  doi = {10.1111/1467-9965.00068}
}

@article{rockafellar2000optimization,
  title = {Optimization of Conditional Value-at-Risk},
  author = {Rockafellar, R. Tyrrell and Uryasev, Stanislav},
  journal = {Journal of Risk},
  volume = {2},
  pages = {21--42},
  year = {2000},
  doi = {10.21314/JOR.2000.038}
}

@article{demiguel2009optimal,
  title = {Optimal Versus Naive Diversification: How Inefficient is the 1/{N} Portfolio Strategy?},
  author = {DeMiguel, Victor and Garlappi, Lorenzo and Uppal, Raman},
  journal = {The Review of Financial Studies},
  volume = {22},
  number = {5},
  pages = {1915--1953},
  year = {2009},
  doi = {10.1093/rfs/hhm075}
}

@book{boyd2004convex,
  title = {Convex Optimization},
  author = {Boyd, Stephen and Vandenberghe, Lieven},
  year = {2004},
  publisher = {Cambridge University Press},
  address = {Cambridge, UK},
  doi = {10.1017/CBO9780511804441}
}

@article{longin2001extreme,
  title = {Extreme Correlation of International Equity Markets},
  author = {Longin, Fran{\c{c}}ois and Solnik, Bruno},
  journal = {The Journal of Finance},
  volume = {56},
  number = {2},
  pages = {649--676},
  year = {2001},
  doi = {10.1111/0022-1082.00340}
}

@article{ang2002asymmetric,
  title = {Asymmetric Correlations of Equity Portfolios},
  author = {Ang, Andrew and Chen, Joseph},
  journal = {Journal of Financial Economics},
  volume = {63},
  number = {3},
  pages = {443--494},
  year = {2002},
  doi = {10.1016/S0304-405X(02)00068-5}
}

@article{cont2001empirical,
  title = {Empirical Properties of Asset Returns: Stylized Facts and Statistical Issues},
  author = {Cont, Rama},
  journal = {Quantitative Finance},
  volume = {1},
  number = {2},
  pages = {223--236},
  year = {2001},
  doi = {10.1080/713665670}
}

@article{chang2000heuristics,
  title = {Heuristics for Cardinality Constrained Portfolio Optimisation},
  author = {Chang, Tun-Jen and Meade, Nigel and Beasley, John E. and Sharaiha, Yazid M.},
  journal = {Computers \& Operations Research},
  volume = {27},
  number = {13},
  pages = {1271--1302},
  year = {2000},
  doi = {10.1016/S0305-0548(99)00074-X}
}

@article{almgren2001optimal,
  title = {Optimal Execution of Portfolio Transactions},
  author = {Almgren, Robert and Chriss, Neil},
  journal = {Journal of Risk},
  volume = {3},
  pages = {5--40},
  year = {2001},
  doi = {10.21314/JOR.2001.041}
}

@book{grinold1999active,
  title = {Active Portfolio Management: A Quantitative Approach for Producing Superior Returns and Controlling Risk},
  author = {Grinold, Richard C. and Kahn, Ronald N.},
  publisher = {McGraw-Hill},
  address = {New York, NY},
  edition = {2nd},
  year = {1999},
  isbn = {978-0070248823}
}

@article{best1991sensitivity,
  title = {On the Sensitivity of Mean-Variance-Efficient Portfolios to Changes in Asset Means: Some Analytical and Computational Results},
  author = {Best, Michael J. and Grauer, Robert R.},
  journal = {The Review of Financial Studies},
  volume = {4},
  number = {2},
  pages = {315--342},
  year = {1991},
  doi = {10.1093/rfs/4.2.315}
}

@article{chopra1993effect,
  title = {The Effect of Errors in Means, Variances, and Covariances on Optimal Portfolio Choice},
  author = {Chopra, Vijay K. and Ziemba, William T.},
  journal = {Journal of Portfolio Management},
  volume = {19},
  number = {2},
  pages = {6--11},
  year = {1993},
  doi = {10.3905/jpm.1993.409440}
}

@article{perold1988implementation,
  title = {The implementation shortfall: Paper versus reality},
  author = {Perold, Andre F.},
  journal = {The Journal of Portfolio Management},
  volume = {14},
  number = {3},
  pages = {4--9},
  year = {1988},
  doi = {10.3905/jpm.1988.409150}
}

@techreport{frazzini2018trading,
  title = {Trading Costs},
  author = {Frazzini, Andrea and Israel, Ronen and Moskowitz, Tobias J.},
  institution = {SSRN},
  year = {2018},
  number = {3229719},
  doi = {10.2139/ssrn.3229719}
}

@book{kissell2013science,
  title = {The Science of Algorithmic Trading and Portfolio Management},
  author = {Kissell, Robert},
  year = {2013},
  publisher = {Academic Press},
  address = {Boston, MA},
  doi = {10.1016/C2012-0-00818-6}
}

@article{brin1998anatomy,
  title = {The anatomy of a large-scale hypertextual web search engine},
  author = {Brin, Sergey and Page, Lawrence},
  journal = {Computer Networks and ISDN Systems},
  volume = {30},
  number = {1-7},
  pages = {107--117},
  year = {1998},
  doi = {10.1016/S0169-7552(98)00110-X}
}

@article{spohn1977algebraic,
  title = {Algebraic Conditions for the Approach to Equilibrium of an Open $N$-Level System},
  author = {Spohn, Herbert},
  journal = {Letters in Mathematical Physics},
  volume = {2},
  pages = {33--38},
  year = {1977},
  doi = {10.1007/BF00420668}
}

\end{document}